\documentclass[prd,aps,twocolumn,a4paper,floatfix,nofootinbib,showpacs]{revtex4}
   
\usepackage{graphicx,psfrag}
\usepackage{mathrsfs}
\usepackage{amsmath,amsfonts,amssymb}
\usepackage{multirow}
\usepackage{comment}
\usepackage{hyperref}
\usepackage{bbm}
\usepackage{placeins}
\usepackage[normalem]{ulem}

\usepackage{pifont}% http://ctan.org/pkg/pifont
\newcommand{\cmark}{\ding{51}}%
\newcommand{\xmark}{\ding{55}}%

\def\ul{\underline}
\def\p{\partial}
\def\Lie{{\cal L}}

\def\perpdsq{{}^{\mathbbmss{q}}\!\!\!\perp}

\def\gamN{{}^{\textrm{\tiny{(N)}}}\!\gamma}

\def\slN{{}^{\textrm{\tiny{(N)}}}b}

\def\gBN{{}^{\textrm{\tiny{(N)}}}\!\mathbbmss{g}}

\def\qN{{}^{\textrm{\tiny{(N)}}}\!q}
\def\bN{{}^{\textrm{\tiny{(N)}}}\!b}

\def\DN{{}^{\textrm{\tiny{(N)}}}\!D}

\def\KN{{}^{\textrm{\tiny{(N)}}}\!K}

\def\HN{{}^{\textrm{\tiny{(N)}}}\!H}
\def\MN{{}^{\textrm{\tiny{(N)}}}\!M}

\usepackage{color}

\newcommand{\RM}[1]{\MakeUppercase{\romannumeral #1{}}}

\begin{document}

\title{The evolution of hyperboloidal data with the dual foliation
formalism: \newline Mathematical analysis and wave equation tests} 

\author{David \surname{Hilditch}, Enno \surname{Harms},
Marcus \surname{Bugner}, Hannes \surname{R\"uter} and
Bernd \surname{Br\"ugmann}}

\affiliation{Theoretical Physics Institute, University of Jena, 
07743 Jena, Germany}

\date{\today}

\begin{abstract}
A long-standing problem in numerical relativity is the satisfactory treatment of 
future null-infinity. We propose an approach for the evolution of hyperboloidal 
initial data in which the outer boundary of the computational domain is placed at infinity. The main idea is to apply 
the `dual foliation' formalism in combination with hyperboloidal coordinates and the generalized harmonic gauge formulation.
The strength of the present approach is that, following the ideas of Zengino\u{g}lu, a hyperboloidal layer can be naturally 
attached to a central region using standard coordinates of numerical relativity applications. Employing a generalization of the 
standard hyperboloidal slices, developed by Calabrese~et.~al., we find that all formally singular terms take a trivial 
limit as we head to null-infinity. A byproduct is a numerical approach for hyperboloidal evolution of nonlinear wave equations 
violating the null-condition. The height-function method, used often for fixed background spacetimes, is generalized in such a 
way that the slices can be dynamically `waggled' to maintain the desired outgoing coordinate lightspeed precisely. This is 
achieved by dynamically solving the eikonal equation. As a first numerical test of the new approach we solve the 3D flat space 
scalar wave equation. The simulations, performed with the pseudospectral~\texttt{bamps} code, show that outgoing waves are 
cleanly absorbed at null-infinity and that errors converge away rapidly as resolution is increased. 
\end{abstract}

\pacs{
  04.25.D-,   % numerical relativity
  95.30.Sf   % relativity and gravitation
}

\maketitle

\tableofcontents

%%%%%%%%%%%%%%%%%%%%%%%%%%%%%%%%%%%%%%%%%%%%%%%%%%%%%%%%%%%%%%%%%%%%%%%%%%%%%%%%%%%%%%%
\section{Introduction}\label{Section:Introduction}
%%%%%%%%%%%%%%%%%%%%%%%%%%%%%%%%%%%%%%%%%%%%%%%%%%%%%%%%%%%%%%%%%%%%%%%%%%%%%%%%%%%%%%%

The ultimate aim of numerical relativity (NR) could be considered the ability to solve the 
field equations of general relativity (GR) to arbitrary accuracy in any desired configuration. 
But more extreme initial data systematically become more difficult to treat, and thus require 
more sophisticated numerical and analytical techniques. Extreme data of particular interest 
are strong-field gravitational waves (GWs), which may either collapse to form a blackhole or 
disperse to infinity. Tuning the strength of such initial data so that we approach the 
{\it threshold of blackhole formation} is expected to reveal interesting 
phenomena~\cite{AbrEva92,GunGar07}. This regime is of great relevance to the weak cosmic 
censorship conjecture, which, roughly speaking, says that future null-infinity is 
generically complete. Continuing our research 
programme~\cite{HilBauWey13,HilWeyBru15,Hil15} in this direction, a grand goal is to 
include infinity in the computational domain so that we may {\it see} 
conclusively the manner in which null-infinity is terminated, should it do so. This is 
but one motivation for the explicit treatment of null-infinity, the most obvious 
alternative being the unambiguous extraction of GWs from compact binaries. 
In the latter case the current method of Cauchy-characteristic-extraction (CCE), 
in which a characteristic domain is attached to a timelike worldtube in the interior of 
the computational domain of a standard NR code, and fed data which is then integrated out 
to infinity to obtain the wave signal there, appears 
sufficient~\cite{ReiBisPol09,ReiBisPol12,TayBoyRei13,HanSzi14,HanSziWin15,HanSziWin16}. One should 
however be aware of the principle weakness of CCE that the weak-field characteristic domain 
does not couple back onto the central Cauchy domain.

Several alternative strategies have been suggested as proper numerical treatments of null-infinity. 
The first is the big sister of CCE, namely Cauchy-Characteristic-Matching (CCM)~\cite{Win12}, 
in which the coupling from the characteristic domain is properly taken care of so that 
the outer boundary of the Cauchy region can be taken as the inner boundary of the 
characteristic domain directly. The second such approach, to which we adhere, is the use 
of initial data which is specified everywhere on a spacelike surface, but in which the 
surface rises up in a spacetime diagram in such a way as to intersect with future 
null-infinity. Such data can be naturally combined with a radial compactification so that 
null-infinity is drawn to a finite coordinate radius. This combination is often referred to as 
a {\it hyperboloidal initial data set}. One cannot simply evolve hyperboloidal data with standard 
methods because they turn out to be formally singular at null-infinity. Therefore some kind 
of regularization is needed. Hyperboloidal data can be naturally viewed within the conformal 
approach pioneered by Penrose~\cite{Pen64}. In this approach a conformal regularization 
may be performed, resulting in the conformal field equations of Friedrich~\cite{Fri81,Fri81a}.
Recently work towards treating the conformal field equations numerically was presented by Doulis 
and Frauendiener~\cite{DouFra16}. Another suggestion, due to Zengino\u{g}lu~\cite{Zen08}, is to use 
a conformal transformation in combination with the standard construction of the generalized harmonic 
gauge (GHG) formulation~\cite{Fri86,Gar02}. This does not result in completely regular field equations, 
since formally singular terms remain in the equations of motion. Nevertheless, with a suitable gauge 
these take a regular limit. This approach was taken by Moncrief and Rinne, for an alternative 
constrained formulation of the field equations; in~\cite{MonRin08} the necessary limits were 
evaluated explicitly and in~\cite{Rin09} used numerically. Again following this tack, 
Va\~n\'o-Vi\~nuales and collaborators~\cite{VanHusHil14,VanHus14,Van15}, working in spherical 
symmetry, performed the same type of partial regularization, but this time performed numerics 
with standard free-evolution formulations~\cite{BauSha98,ShiNak95,NakOohKoj87,BerHil09} and, as in~\cite{Zen07}, 
a staggered numerical grid so as to avoid explicitly dealing with the problematic terms. Yet another approach, similar 
to that of Moncrief and Rinne, is the tetrad formulation of Bardeen, Sarbach and Buchman~\cite{BarSarBuc11}. This formulation 
was recently succesfully tested in spherical symmetry by Morales and Sarbach~\cite{MorSar16}. Hyperboloidal slices are furthermore 
a main ingredient in the general approach of LeFloch and Yue~\cite{LeFYue14} to the proof of global existence for systems 
of nonlinear wave equations, and of particular relevance here, in their application to GR~\cite{LeFYue15}.

In this work we develop a new strategy for the aforementioned regularization that avoids the conformal 
decomposition. The idea is to use a global tensor basis for the representation of all tensors which is 
asymptotically flat, but nevertheless to use hyperboloidal coordinates. This procedure can be naturally put 
into practice within the dual-foliation (DF) formalism proposed in~\cite{Hil15}, hereafter 
referred to as `the DF paper'. The DF formalism is built on the idea of using two coordinate 
systems, one to determine a particular tensor basis and taking care of the required hyperbolicity,
and the other for covering the spacetime with coordinates; in our application here these labels are
chosen to be hyperboloidal coordinates. To see that this approach has a chance to work one need only
consider the Minkowski metric represented in a global inertial frame. Clearly the coordinate choice we
make on the spacetime is irrelevant to the regularity of the metric in that representation. Work is needed
however to establish whether or not the resulting regularization is partial, or if completely regular field
equations can be obtained. The necessary investigation follows in the first part of the paper. Our main
finding is that, with due care and reasonable assumptions on the data, divergent terms can indeed be
eliminated from the evolution equations all the way out to null-infinity. 
 
The article is structured as follows. In Sec~\ref{Section:DF_GHG} we give a 
geometric derivation of the first order DF GHG formulation.
Next, in Sec~\ref{Section:Hyperboloidal_Coordinates} we specialize this 
formulation to the relevant case of hyperboloidal coordinates. We then consider the 
asymptotics to convince ourselves of the regularity of the equations of motion. In 
Sec~\ref{Section:Numerical_Experiments} we present the 
first numerical results of the new DF approach with hyperboloidal coordinates.
We restrict the numerical experiments to the wave equation, which we take as a toy model for the full 
setup. Finally in Sec~\ref{Section:Conclusion} we make some concluding remarks. Geometric units are 
used throughout.

%%%%%%%%%%%%%%%%%%%%%%%%%%%%%%%%%%%%%%%%%%%%%%%%%%%%%%%%%%%%%%%%%%%%%%%%%%%%%%%%%%%%%%%
\section{Dual Foliation GHG}\label{Section:DF_GHG}
%%%%%%%%%%%%%%%%%%%%%%%%%%%%%%%%%%%%%%%%%%%%%%%%%%%%%%%%%%%%%%%%%%%%%%%%%%%%%%%%%%%%%%%

%------------------------------------------
% Fig: DF_slices
%------------------------------------------
\begin{figure}[t]
  \centering  
  \includegraphics[width=0.5\textwidth]{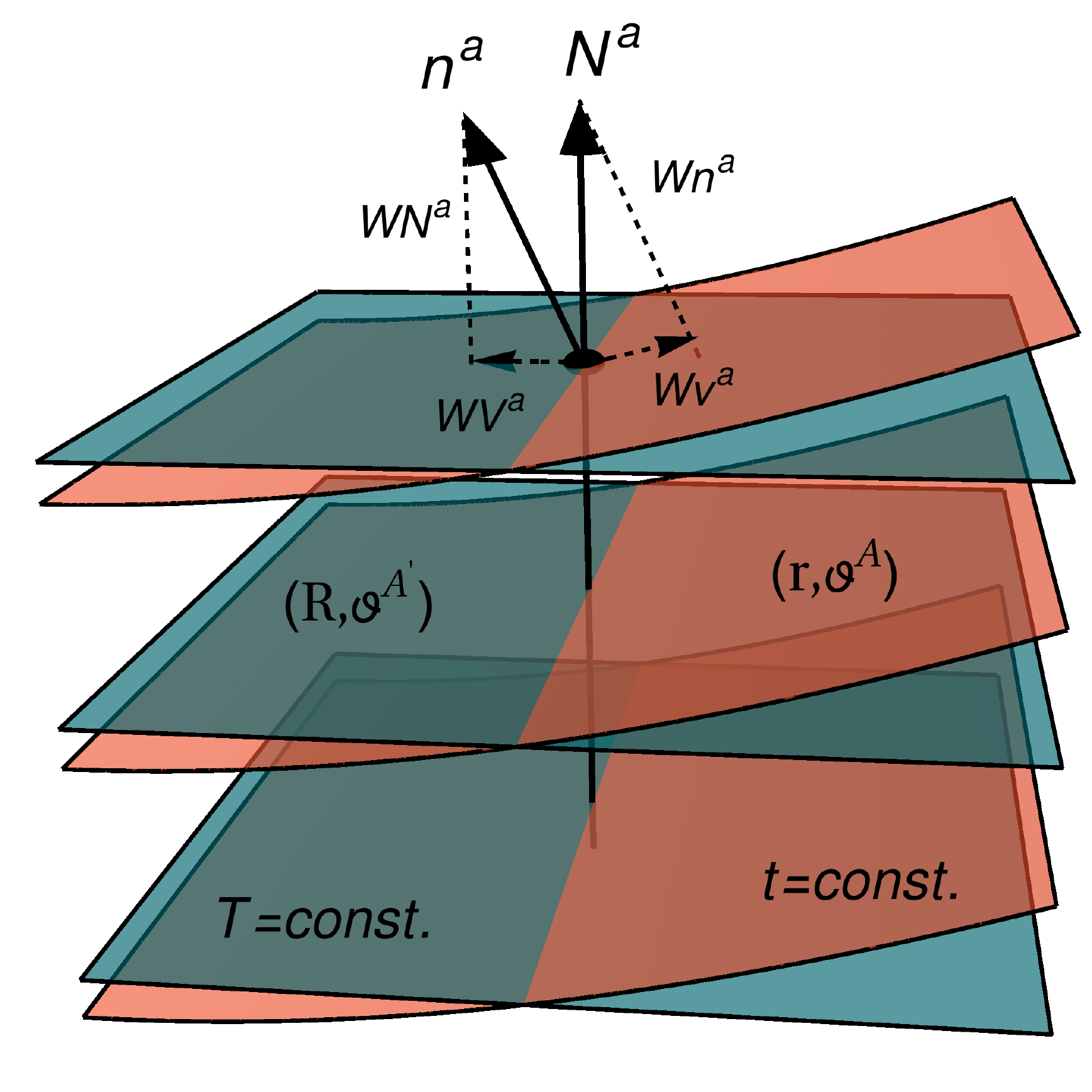} 
  \caption{
   Illustration of a spacetime with two different slicings as used
   in the `Dual Foliation' (DF) approach of~\cite{Hil15}. 
   We have two coordinate systems,
   i)~the lower case $x^\mu=(t,x^i)$, and
   ii)~the upper case $X^{\ul{\mu}}=(T,X^{\ul{i}})$.
   In the illustration we already anticipate the usage
   of spherical-like coordinates with a radius and two `angles',
   $x^i=(r,\vartheta^A)$ and $X^{\ul{i}'}=(R,\vartheta^{A'})$.
   The time-functions define the two slicings
   by $t=\rm{const.}$ and $T=\rm{const.}$ respectively.
   Associated with each slicing is the timelike unit normal,
   i.e.\ $n^a$ for the lower case slices
   and $N^a$ for the upper case slices.
   The inner product of the normal vectors
   is called the Lorentz factor, $W=-(N^a n_a)$.
   The mutual projections of the normal vectors
   into the other slicing, divided by $W$, are called the boost vectors, 
   $W v^a=\perp\!\!N^a$ and ${W V^a = {}^\textrm{\tiny{(N)}}\!\!\!\perp\!n^a}$.   
   In this work we apply the DF approach to evolve the scalar wave-equation on flat space
   along hyperboloidal slices of constant time $t$ and on a compactified radial grid in $r$.
     }
  \label{fig:DF_slices}
\end{figure}

In this section we compute the field equations of the first order generalized harmonic 
formulation with a general DF-setup~\cite{Hil15}. Later on, in 
Sec~\ref{Section:Hyperboloidal_Coordinates}, we will specify the two coordinate systems of the 
DF-formulation to be standard GHG coordinates on the one hand and hyperboloidal coordinates on 
the other hand. For the convenient application of hyperboloidal coordinates the used GHG 
coordinates have to be spherical-polar-like with certain asymptotic features, which means an intermediate 
step is required for transforming the Cartesian-like GHG formulation to what we call 
`shell-coordinates'. In the following section we give the associated GHG evolution equations in 
shell-coordinates and then employ the DF formalism proper.

%%%%%%%%%%%%%%%%%%%%%%%%%%%%%%%%%%%%%%%%%%%%%%%%%%%%%%%%%%%%
\subsection{Review of the DF paper}\label{subsection:DF_Review}
%%%%%%%%%%%%%%%%%%%%%%%%%%%%%%%%%%%%%%%%%%%%%%%%%%%%%%%%%%%%

The present work strongly relies on the `DF approach' as introduced in the DF~paper~\cite{Hil15}. We 
recommend reading~\cite{Hil15} but will give a short review on the essentials, collecting those parts 
that are imperative for understanding the discussion here.

%%%%%%%%%%%%%%%%%%%%%%%%%%%%%%%%%%%%%%%%%%%%%%%%%%%%%%%%%%%%
\paragraph*{Basics and notation:} The idea is to employ two different coordinate systems,~$X^{\ul{\mu}}$ 
and~$x^\mu$, and to exploit the good properties of each. More precisely, the ultimate goal of this work 
is to use the DF-approach to evolve the asymptotically regular standard GHG-variables along hyperboloidal 
slices that include future null infinity, see more in Sec~\ref{Section:Hyperboloidal_Coordinates}. We write 
the GHG coordinates, which are Cartesian and asymptotically Minkowskian, in upper case, $X^{\ul{\mu}}=(T,X^{\ul{i}})$.
Tensors are represented with respect to the basis~$(\p_{\ul{\mu}})^a$ and natural dual. Indices belonging to this basis 
are underlined. Note that Greek indices $\ul{\mu},\ul{\nu}$ go over space and time, whereas Latin 
indices~$\ul{i},\ul{j},\ul{k},\ul{l}$ are purely spatial. Latin indices~$a,b,c,d$ will be abstract. The 
future pointing unit normal vector to constant time~$T$ slices is~$N^a$. The geometric quantities associated 
to a~$3+1$-split along~$N^a$ are denoted by either upper case symbols, that is, for example, the lapse~$A$ and 
the shift $B^{\ul{i}}$, or by a preceding superscript $(N)$, e.g., the induced metric on 
the~$T$-slices~$\gamN_{\ul{\mu\nu}}$. The lower case coordinates $x^\mu$ will be understood as the hyperboloidal 
coordinates, with~$\mu,\nu$ going over space and time and Latin indices~$i,j,k,l$ denoting the spatial 
components. The future pointing unit normal vector to constant time~$t$ slices is~$n^a$. The geometric quantities 
associated to a $3+1$-split along $n^a$ are denoted by lower case Greek letters, that is the lapse~$\alpha$, the 
shift~$\beta^i$, and the induced $3$-metric on the~$t$-slices,~$\gamma_{\mu\nu}$.

%%%%%%%%%%%%%%%%%%%%%%%%%%%%%%%%%%%%%%%%%%%%%%%%%%%%%%%%%%%%
\paragraph*{Frequently used DF-quantities and important relations:} The following general relations between the 
normal vectors will be important throughout the paper,
\begin{align}
N^a=W(n^a+v^a)\,, \qquad n^a=W(N^a+V^a)\,,\label{eqn:N_ndecomp}
\end{align}
with Lorentz factor~$W=(1-v^iv_i)^{-1/2}=(1-V^{\ul{i}}V_{\ul{i}})^{-1/2}$ and boost vectors~$v^a,V^a$, 
which are spatial with respect to~$t$ and~$T$ respectively.
This is nothing but the decomposition of the normal vectors
in the respective other normal and traverse directions,
see Fig.~\ref{fig:DF_slices}.
In the DF formalism two further metrics naturally appear, 
namely the respective projections of the metrics onto the opposing slices.
These are called the {\it boost metrics},
\begin{align}
  \label{eq:boost_metrics_abstract}
  \mathbbmss{g}_{ab} &= {\gamma^c}_a \,  {\gamma^d}_b \,  \gamN_{cd}  \,, \qquad 
  {}^\textrm{\tiny{(N)}}\!\mathbbmss{g}_{ab} = {\gamN^c}_a \, {\gamN^d}_b \,  \gamma_{cd}\,.  
\end{align}
In adapted coordinates, using the Jacobians relating the two tensor bases, we have,
\begin{align}
  \mathbbmss{g}^{ij} &= {\varPhi^i}_{\ul{i}} \; {\varPhi^j}_{\ul{j}} \; \gamN^{\ul{ij}}\,, \qquad
  {}^\textrm{\tiny{(N)}}\!\mathbbmss{g}^{\ul{ij}} = 
                         {\varphi^{\ul{i}}}_i \,  {\varphi^{\ul{j}}}_j \, \gamma^{ij}\,,  
\end{align}
where we define the `projected Jacobians' as,
\begin{align}
 \label{eq:proj_jacobians}
 {\varPhi^i}_{\ul{i}} &:= {\gamma^i}_\mu  (J^{-1})^\mu{}_{\ul{i}} \,, \qquad
 {\varphi^{\ul{i}}}_{i} := {\gamN^{\ul{i}}}_{\ul{\mu}}  J^{\ul{\mu}}{}_{i}\,.
\end{align}
The respective inverse quantities, ${(\varPhi^{-1})^{\ul{i}}}_{i}$ and ${(\varphi^{-1})^{i}}_{\ul{i}}$,
can be straightforwardly computed and are given in the DF paper. For readers interested in repeating all calculations done
in the course of this paper it is vital to emphasize here that we use the up/down index notation
associated with the metrics $\gamma_{ab}, \gamN_{ab}$.
Therefore indices of the boost metrics written in the respective adapted coordinates
cannot be raised/lowered with the boost metrics themselves. This implies some noteworthy consequences, 
in particular~$\mathbbmss{g}^{ij} \equiv \gamma^{ik} \, \gamma^{jl} \, \mathbbmss{g}_{kl}
\neq ({\mathbbmss{g}^{-1}})^{ij}$. Explicitly one finds, 
\begin{align}
 \label{eq:boost_metric_explicit_formulas}
 \mathbbmss{g}_{ij}        &=\gamma_{ij}+W^2v_iv_j \; , \nonumber \\
 (\mathbbmss{g}^{-1})^{ij} &=\gamma^{ij} - v^i v^j \; ,
\end{align}
and analogous expressions for~${}^\textrm{\tiny{(N)}}\!\mathbbmss{g}_{\ul{ij}}$
and ${}^\textrm{\tiny{(N)}}(\mathbbmss{g}^{-1})^{\ul{ij}}$ in terms of $\gamN_{\ul{ij}}$ and $V_{\ul{i}}$, see~Sec~\RM{2} in the DF paper.
Finally, we also have,
\begin{align}
 \label{eq:upper_case_Pi_abbrev}
 \Pi^i &=W v^i - \alpha^{-1} W \beta^i\,.
\end{align}
The quantities introduced in this section will be used to abbreviate many equations below.

%%%%%%%%%%%%%%%%%%%%%%%%%%%%%%%%%%%%%%%%%%%%%%%%%%%%%%%%%%%%
\subsection{Shells adapted GHG}\label{subsection:Shells}
%%%%%%%%%%%%%%%%%%%%%%%%%%%%%%%%%%%%%%%%%%%%%%%%%%%%%%%%%%%%

%%%%%%%%%%%%%%%%%%%%%%%%%%%%%%%%%%%%%%%%%%%%%%%%%%%%%%%%%%%%
\paragraph*{The first order GHG formulation:} 
At the risk of repeating the presentation given in~\cite{HilWeyBru15},
let us start with a light modification of the first order GHG~system~\cite{LinSchKid05}.
The GHG evolution equations read,
\begin{align}
\p_Tg_{\ul{\mu\nu}}&=B^{\ul{i}}\p_{\ul{i}}g_{\ul{\mu\nu}}+A\,S^{(g)}_{\ul{\mu\nu}}\,,\nonumber\\
\p_T\Phi_{\ul{i\,\mu\nu}}&=B^{\ul{j}}\p_{\ul{j}}\Phi_{\ul{i\,\mu\nu}}-A\,\p_{\ul{i}}\Pi_{\ul{\mu\nu}}
+\gamma_2\,A\,\p_{\ul{i}}g_{\ul{\mu\nu}}+A\,\,S^{(\Phi)}_{\ul{i\,\mu\nu}}\,,\nonumber\\
\p_T\Pi_{\ul{\mu\nu}}&=B^{\ul{i}}\p_{\ul{i}}\Pi_{\ul{\mu\nu}}-A\,\gamN^{\ul{ij}}\,\p_{\ul{i}}\Phi_{\ul{j\,\mu\nu}}
+A\,S^{(\Pi)}_{\ul{\mu\nu}}\,,\label{eqn:GHG_11}
\end{align}
with shorthands for the source terms,
\begin{align}
S^{(g)}_{\ul{\mu\nu}}&=-\Pi_{\ul{\mu\nu}}\,,\nonumber\\
S^{(\Phi)}_{\ul{i\,\mu\nu}}&=-\gamma_2\,\Phi_{\ul{i\,\mu\nu}}
+\tfrac{1}{2}\,N^{\ul{\alpha}}\,N^{\ul{\beta}}\,\Phi_{\ul{i\,\alpha\beta}}\,\Pi_{\ul{\mu\nu}}\nonumber\\
&\quad+\gamN^{\ul{jk}}\,N^{\ul{\alpha}}\,\Phi_{\ul{i\,j\alpha}}\,\Phi_{\ul{k\,\mu\nu}}\,,\nonumber\\
S^{(\Pi)}_{\ul{\mu\nu}}&=2\,g^{\ul{\alpha\beta}}\,\big(\gamN^{\ul{ij}}\,
\Phi_{\ul{i\,\alpha\mu}}\,\Phi_{\ul{j\,\beta\nu}}-\Pi_{\ul{\alpha\mu}}\,\Pi_{\ul{\beta\nu}}
-g^{\ul{\delta\gamma}}\Gamma_{\ul{\mu\alpha\delta}}\Gamma_{\ul{\nu\beta\gamma}}\big)\nonumber\\
&\quad-2\,\big(\nabla_{(\ul{\mu}}H_{\ul{\nu})}+\gamma_3\,\Gamma^{\ul{\alpha}}{}_{\ul{\mu\nu}}C_{\ul{\alpha}}
-\tfrac{1}{2}\gamma_4\,g_{\ul{\mu\nu}}\Gamma^{\ul{\alpha}}C_{\ul{\alpha}}\big)\nonumber\\
&\quad-\tfrac{1}{2}N^{\ul{\alpha}}N^{\ul{\beta}}\Pi_{\ul{\alpha\beta}}\Pi_{\ul{\mu\nu}}
- N^{\ul{\alpha}}\,\gamN^{\ul{ij}}\,\Pi_{\ul{\alpha i}}\,\Phi_{\ul{j\,\mu\nu}}\nonumber\\
&\quad+\gamma_0\,\big[2\,\delta^{\ul{\alpha}}{}_{(\ul{\mu}}N_{\ul{\nu})}
-g_{\ul{\mu\nu}}\,N^{\ul{\alpha}}\big]C_{\ul{\alpha}}\,,\label{eqn:ghg11_sources}
\end{align}
where we also write,
\begin{align}
\Gamma_{\ul{\alpha\mu\nu}}\equiv
\gamN^{\ul{i}}{}_{(\ul{\mu}|}\Phi_{\ul{i}\,|\ul{\nu})\ul{\alpha}}
-\tfrac{1}{2}\gamN^{\ul{i}}{}_{\ul{\alpha}}\Phi_{\ul{i\,\mu\nu}}+N_{(\ul{\mu}}\Pi_{\ul{\nu})\ul{\alpha}}
-\tfrac{1}{2}N_{\ul{\alpha}}\Pi_{\ul{\mu\nu}}\,.\nonumber
\end{align}
The equations are subject to the GHG constraints,
\begin{align}
C_{\ul{\mu}}&=\Gamma_{\ul{\mu}}+H_{\ul{\mu}}\equiv g^{\ul{\alpha\beta}}\Gamma_{\ul{\mu\alpha\beta}}
+H_{\ul{\mu}}=0\,,
\end{align}
and reduction constraints, 
\begin{align}
C_{\ul{i\,\mu\nu}}&=\p_{\ul{i}}g_{\ul{\mu\nu}}-\Phi_{\ul{i\,\mu\nu}}=0\,,\label{eqn:ghg11_reduction}
\end{align}
which can be viewed as the definition of~$\Phi_{\ul{i\,\mu\nu}}$ from the first order reduction
of the original second order GHG equations. The {\it gauge source functions}~$H_{\ul{\mu}}$ 
can be taken as arbitrary functions of the coordinates, or else of the metric 
components~$g_{\ul{\mu\nu}}$. In the latter case it is understood that the reduction 
constraints~\eqref{eqn:ghg11_reduction} are to be used to replace the resulting derivatives of the 
metric in~\eqref{eqn:ghg11_sources}. There are further constraints, which are essentially the 
Hamiltonian and momentum constraints of~$3+1$ GR, given by,
\begin{align}
\HN&=\gamN^{\ul{ij}}\,\gamN^{\ul{kl}}\left(\p_{\ul{k}}\Phi_{\ul{ijl}}-\p_{\ul{k}}\Phi_{\ul{lij}}\right)+S^{(H)}\,,\nonumber\\
\MN_{\ul{i}}&=\gamN^{\ul{jk}}\left(\p_{[\ul{j}}\Pi_{\ul{i}]\ul{k}}+\tfrac{1}{2}d_{\ul{j}}\Phi_{\ul{ki}N}
-\tfrac{1}{2}d_{\ul{i}}\Phi_{\ul{jk}N}\right)+S^{(M)}_{\ul{i}}\,,
\end{align}
with non-principal parts,
\begin{align}
S^{(H)}&=\gamN^{\ul{ij}}\,\gamN^{\ul{kl}}\left(\Gamma^{\ul{\mu}}{}_{\ul{jk}}\Gamma_{\ul{\mu il}}
-\Gamma^{\ul{\mu}}{}_{\ul{ij}}\Gamma_{\ul{\mu kl}}\right)\,,\nonumber\\
S^{(M)}_{\ul{i}}&=\gamN^{\ul{jk}}\left(-\tfrac{1}{2}\Pi_{\ul{j}[\ul{i}}\Phi_{\ul{k}]NN}
+\tfrac{1}{2}\gamN^{\ul{lm}}\Phi_{\ul{mkj}}\Phi_{\ul{il}N}\right.\nonumber\\
&\quad\quad\quad\quad\left.+2\Gamma_{\ul{\mu}N[\ul{i}}\Gamma^{\ul{\mu}}{}_{\ul{k}]\ul{j}}\right)\, .
\end{align}
Here the~$N$ indices denote contraction with the normal vector~$N^a$, and the notation~$d$ for the derivative
means that any such contraction is not commuted through the partial derivative. See~\cite{LinSchKid05} 
or~\cite{HilWeyBru15} for more details. Notice that, in the notation of~\cite{LinSchKid05}, we have fixed the 
formulation parameter~$\gamma_1=0$, which is not the standard choice, but is convenient in what follows. The PDE 
system is symmetric hyperbolic, and has characteristic variables, 
\begin{align}
u^{\hat{0}}_{\ul{\mu\nu}}&=g_{\ul{\mu\nu}}\,,\nonumber\\
u^{\hat{\pm}}_{\ul{\mu\nu}}&=\Pi_{\ul{\mu\nu}}\mp S^{\ul{i}}\,\Phi_{\ul{i\,\mu\nu}}
-\gamma_2\,g_{\ul{\mu\nu}}\,,\nonumber\\
u^{\hat{B}}_{\ul{j\,\mu\nu}}&=\qN^{\ul{i}}{}_{\ul{j}}\,\Phi_{\ul{i\,\mu\nu}}\,,\label{eqn:CV_upper}
\end{align} 
with speeds,
\begin{align}
v^{\hat{0}}=B^S\,,\quad v^{\hat{\pm}}=B^S\pm A\,,\quad v^{\hat{B}}=B^S\,,
\end{align}
respectively, where~$S^a$ is an arbitrary vector of unit magnitude, spatial with respect to~$N^a$, 
and we have defined the projection operator~$\qN^a{}_b=\gamN^a{}_b-S^aS_b$. Here and elsewhere an 
index~$S$ denotes contraction with~$S^a$.

%%%%%%%%%%%%%%%%%%%%%%%%%%%%%%%%%%%%%%%%%%%%%%%%%%%%%%%%%%%%
\paragraph*{Shell-coordinates:} Let us now make a change of spatial 
coordinates~$X^{\ul{\mu}}\to X^{\ul{\mu}'}=(T,X^{\ul{i}'})$, and use the associated coordinate 
basis vectors to represent the spatial index in the reduction 
variable~$\Phi_{\ul{i'\mu\nu}}\equiv (\phi^{\textrm{Sh}})^{\ul{i}}{}_{\ul{i}'}\Phi_{\ul{i\,\mu\nu}}$. 
The relevant parts of the Jacobians are named,
\begin{align}
\label{eq:abbrevs_trafos_cart_shells}
(\phi^{\textrm{Sh}})^{\ul{i}}{}_{\ul{i}'}=\p_{\ul{i}'}X^{\ul{i}}\,,\qquad\quad\quad 
(\Phi^{\textrm{Sh}})^{\ul{i}'}{}_{\ul{i}}=\p_{\ul{i}}X^{\ul{i}'}\,.
\end{align}
We will refer to the~$X^{\ul{i}'}$ coordinates as `shell-coordinates',
bearing in mind the particular coordinates used in the~\texttt{bamps} code that
we employ for our numerical experiments~\cite{HilWeyBru15}.
Intuitively one may think of the shell coordinates as being 
spherical polar, although in practice we make another choice so as to avoid coordinate singularities.
In fact the specific form of the shell-coordinates is irrelevant.
We are concerned only with the asymptotic radial behavior of the 
transformation, that is we understand as shell-coordinates all spherical-like, i.e.\ containing a radial and 
two `angle', coordinates $X^{\ul{i}'}=(R,\vartheta^{A'})=(R,\theta,\phi)$, with indices~$A'$ labeling the angular 
coordinates, which are related to the Cartesian coordinates like,
\begin{align}
X^{\ul{i}}=R\,\Theta^{\ul{i}}(\theta,\phi)\,,
\end{align}
with the scalar functions~$\Theta^{\ul{i}}$ such that~$\sum_{\ul{i}=1}^3(\Theta^{\ul{i}})^2=1$. 
Here and throughout we use upper case indices~$A,B,C$ to denote the angular components. Thus we can write,
\begin{align}
R^2&=\sum_{\ul{i}=1}^3(X^{\ul{i}})^2\,,\qquad\vartheta^{A'}=\vartheta^{A'}(\Theta^{\ul{i}})\,,
\end{align}
and so we have the inverse Jacobian,
\begin{align}
\label{eqn:Jacobian_ShellCoordinates}
(\Phi^{\textrm{Sh}})^{\ul{i}'}{}_{\ul{i}}=
\left(\begin{array}{cc}
\p_{\ul{i}}R\,\, & R^{-1}(\delta^{\ul{j}}{}_{\ul{i}}-\Theta^{\ul{j}}\p_{\ul{i}}R)
\,\p_{(\Theta^{\ul{j}})}\vartheta^{A'}
\end{array}\right)\,.
\end{align}
The hope here is that, when combined with the transformation from the shell to 
hyperboloidal coordinates, the various~$O(R)$ and~$O(R^{-1})$ terms cancel in the composite 
Jacobians to give a regular transformation in the limit to null-infinity. The prior change to the
shell-coordinate basis for the reduction variable~$\Phi_{\ul{i'\mu\nu}}$ is needed so that this 
works out. We note in passing that using this change of variables might give a slight 
improvement in accuracy at large radii, even on spatial slices that terminate at spatial 
infinity, because their use allows us to avoid the numerical computation of products
of potentially badly conditioned matrices like~$(\Phi^{\textrm{Sh}})^{\ul{j}'}{}_{\ul{i}}$ 
and~$(\phi^{\textrm{Sh}})^{\ul{j}}{}_{\ul{j}'}$ when computing derivatives. 

%%%%%%%%%%%%%%%%%%%%%%%%%%%%%%%%%%%%%%%%%%%%%%%%%%%%%%%%%%%%
\paragraph*{The first order GHG formulation in shell-coordinates:}
Under the transformation~\eqref{eq:abbrevs_trafos_cart_shells} the field 
equations~\eqref{eqn:GHG_11} take the form,
\begin{align}
\p_Tg_{\ul{\mu\nu}}&=B^{\ul{i}'}\p_{\ul{i}'}g_{\ul{\mu\nu}}+A\,S^{(g)}_{\ul{\mu\nu}}\,,\nonumber\\
\p_T\Phi_{\ul{i'\mu\nu}}&=B^{\ul{j}'}\p_{\ul{j}'}\Phi_{\ul{i'\mu\nu}}-A\,\p_{\ul{i}'}\Pi_{\ul{\mu\nu}}
+\gamma_2\,A\,\p_{\ul{i}'}g_{\ul{\mu\nu}}\nonumber\\
&\quad+A\,\,S^{(\Phi)}_{\ul{i'\mu\nu}}\,,\nonumber\\
\p_T\Pi_{\ul{\mu\nu}}&=B^{\ul{i}'}\p_{\ul{i}'}\Pi_{\ul{\mu\nu}}-A\,\gamN^{\ul{i'j}'}
\,\p_{\ul{i}'}\Phi_{\ul{j'\mu\nu}}+A\,S^{(\Pi)'}_{\ul{\mu\nu}}\,.\label{eqn:GHG_shell}
\end{align}
The {\it modified} source terms are, 
\begin{align} 
\label{eqn:GHG_sources_shell}
S^{(\Phi)}_{\ul{i'\mu\nu}}&=(\phi^{\textrm{Sh}})^{\ul{i}}{}_{\ul{i}'}S^{(\Phi)}_{\ul{i\,\mu\nu}}
-  A^{-1}\Phi_{\ul{i\,\mu\nu}} B^{\ul{j}}\p_{\ul{j}}(\phi^{\textrm{Sh}})^{\ul{i}}{}_{\ul{i}'}
\,,\nonumber\\
S^{(\Pi)'}_{\ul{\mu\nu}}&=S^{(\Pi)}_{\ul{\mu\nu}}
-\Phi_{\ul{j'\mu\nu}}\gamN^{\ul{ij}}\p_{\ul{i}}(\Phi^{\textrm{Sh}})^{\ul{j}'}{}_{\ul{j}}\,.
\end{align}
Note that when evaluating the source terms whilst using the shell-coordinates we 
view~$\Phi_{\ul{i\,\mu\nu}}\equiv (\Phi^{\textrm{Sh}})^{\ul{i'}}{}_{\ul{i}}\Phi_{\ul{i'\mu\nu}}$.
The constraints transform in the obvious way. Naturally such a change of 
coordinates does not affect hyperbolicity, and the characteristic variables transform 
trivially.

%%%%%%%%%%%%%%%%%%%%%%%%%%%%%%%%%%%%%%%%%%%%%%%%%%%%%%%%%%%%
\subsection{Dual Foliation shells adapted GHG}
\label{subsection:DF_GHG}
%%%%%%%%%%%%%%%%%%%%%%%%%%%%%%%%%%%%%%%%%%%%%%%%%%%%%%%%%%%%

Let us now introduce an arbitrary new coordinate system~$x^\mu$.
The Jacobian mapping between the upper case shell-coordinates~$X^{\ul{\mu}'}$ and the lower 
case~$x^\mu$ coordinates,~${J^{\ul{\mu}'}{}_\mu= \frac{\partial X^{\ul{\mu}'}}{\partial x^\mu}}$, 
can be expressed as,
\begin{align}
J&=\left(\begin{array}{cc}
 A^{-1}W(\alpha-\beta^iv_i)
& \alpha\,\pi^{\ul{i}'}+\beta^i\phi^{\ul{i}'}{}_i \\
-A^{-1}Wv_i 
& \phi^{\ul{i}'}{}_i
\end{array}\right)\,.\label{eqn:J}
\end{align}
Likewise the inverse Jacobian is,
\begin{align}
J^{-1}&=\left(\begin{array}{cc}
\alpha^{-1}W(A-B^{\ul{i}'}V_{\ul{i}'}) & A\,\Pi^i+B^{\ul{i}'}\Phi^i{}_{\ul{i}'}\\
-\alpha^{-1}WV_{\ul{i}'} & \Phi^i{}_{\ul{i}'}
\end{array}\right)\,.\label{eqn:J_inv}
\end{align}
Eqs~\eqref{eqn:J} and \eqref{eqn:J_inv} define $\phi^{\ul{i}'}{}_i$ and $\Phi^i{}_{\ul{i}'}$ respectively.
It was shown in the DF paper that a first order evolution system in upper case coordinates of the form,
\begin{align}
\p_T\mathbf{u}&=\left(\,A\,\mathbf{A}^{\ul{p}'}+B^{\ul{p}'}\,\mathbf{1}
\,\right)\p_{\ul{p}'}\mathbf{u}+A\,\mathbf{S}\,,\label{eqn:1st_Upper}
\end{align}
where~$\mathbf{u}$ is the state vector,~$\mathbf{A}^{\ul{p}'}$ are arbitrary matrices called the `principal matrices',
and $\mathbf{S}$ contains the sources, can be rewritten in the lower case coordinate system as,
\begin{align}
\big(\mathbf{1}+\mathbf{A}^{\ul{V}}\big)\p_t\mathbf{u}&=
\alpha\,W^{-1}\left(\mathbf{A}^{\ul{p}'}\,(\varphi^{-1})^p{}_{\ul{p}'}-
\big(\mathbf{1}+\mathbf{A}^{\ul{V}}\big)\Pi^p\right)\p_p\mathbf{u}\nonumber\\
&\quad+\alpha\,W^{-1}\,\mathbf{S}\,.\label{eqn:Change_Coord}
\end{align}
Here we have used the projected 
Jacobian~${ \varphi^{\ul{i}'}{}_{i}=\gamN^{\ul{i}'}{}_{\ul{\mu}'}J^{\ul{\mu}'}{}_{i} }$, see Eq~\eqref{eq:proj_jacobians},
and the convention~$\mathbf{A}^{\ul{V}} \equiv \mathbf{A}^{\ul{i}'}V_{\ul{i}'}$. Our first 
order GHG system is by construction of the form~\eqref{eqn:1st_Upper}.
We can read off the principal matrices,
\begin{align}
\mathbf{A}^{\ul{p}'}&=
\left(
\begin{array}{ccc}
0 & 0 & 0 \\
\gamma_2\,\delta^{\ul{p}'}{}_{\ul{i}'} & 0 & -\delta^{\ul{p}'}{}_{\ul{i}'}\\
0 & -\gamN^{\ul{p}'\ul{j}'} & 0
\end{array}
\right)\,.
\end{align}
Inversion of the coefficient~$(\mathbf{1}+\mathbf{A}^{\ul{V}})$ is possible whenever the Lorentz factor~$W$ is bounded
and yields,
\begin{align}
\left(\mathbf{1}+\mathbf{A}^{\ul{V}}\right)^{-1}&=
\left(
\begin{array}{ccc}
1 & 0 & 0 \\
-\gamma_2\,W^2V_{\ul{i}'} & \gBN^{\ul{j}'}{}_{\ul{i}'} & W^2V_{\ul{i}'}\\
-\gamma_2(W^2-1) & W^2V^{\ul{j}'} & W^2
\end{array}
\right)\,,
\end{align}
with~$\gBN^{\ul{j}'}{}_{\ul{i}'}=\gamN^{\ul{j}'}{}_{\ul{i}'}+W^2V^{\ul{j}'}{}V_{\ul{i}'}$,
as defined by Eq~\eqref{eq:boost_metric_explicit_formulas}.
Inserting this into Eq~\eqref{eqn:Change_Coord}, we obtain the DF GHG equations of motion,
\begin{align}
\p_tg_{\ul{\mu\nu}}&=(\beta^p-\alpha\,v^p)\p_pg_{\ul{\mu\nu}}+\alpha\,W^{-1}s^{(g)}_{\ul{\mu\nu}}
\,,\nonumber\\
d_t\Phi_{i\,\ul{\mu\nu}}&=\left(\beta^p\delta^j{}_i
-\alpha v^p\delta^j{}_i+\alpha W^2v_i(\mathbbmss{g}^{-1})^{pj}\right)d_p\Phi_{j\,\ul{\mu\nu}}
\nonumber\\
&\quad
+\alpha\,W^{-1}\mathbbmss{g}^p{}_i\left(\gamma_2\,\p_pg_{\ul{\mu\nu}}-\p_p\Pi_{\ul{\mu\nu}}
\right)
+\alpha\,W^{-1}s^{(\Phi)}_{i\,\ul{\mu\nu}}\,,\nonumber\\
\p_t\Pi_{\ul{\mu\nu}}&=
\beta^p\p_p\Pi_{\ul{\mu\nu}}-\gamma_2\,\alpha v^p\p_pg_{\ul{\mu\nu}}
-\alpha\,W(\mathbbmss{g}^{-1})^{pi}d_p\Phi_{i\,\ul{\mu\nu}}\nonumber\\
&\quad+\alpha\,W^{-1}s^{(\Pi)}_{\ul{\mu\nu}}\,,\label{eqn:GHG_DF}
\end{align}
where for compactness we used again the boost metric and the notation,
\begin{align}
 \label{eq:notation_contraction_projJac}
 d_\mu\Phi_{i\,\ul{\mu\nu}} \equiv \varphi^{\ul{i}'}{}_i\p_\mu\Phi_{\ul{i'\mu\nu}}\,,
\end{align}
to abbreviate contraction with the projected Jacobian, with the Jacobian outside the derivative.
This notation is used to avoid objects with indices in both tensor bases in the principal part.
In our implementation we simply multiply by the inverse ${(\varphi^{-1})^{i}}_{\ul{i}'}$ so that 
we have a standard evolution system.  The source terms are,
\begin{align}
\label{eqn:sources_DFGHG}
s^{(g)}_{\ul{\mu\nu}}&=S^{(g)}_{\ul{\mu\nu}}\,,\nonumber\\
s^{(\Phi)}_{\ul{i'\mu\nu}}&=S^{(\Phi)}_{\ul{i'\mu\nu}}+W^2V_{\ul{i}'}
\left(V^{\ul{j}'}S^{(\Phi)}_{\ul{j'\mu\nu}}
+S^{(\Pi)'}_{\ul{\mu\nu}}-\gamma_2S^{(g)}_{\ul{\mu\nu}}\right)\,,\nonumber\\
s^{(\Pi)}_{\ul{\mu\nu}}&=\gamma_2S^{(g)}_{\ul{\mu\nu}}+W^2\left(
V^{\ul{i}'}S^{(\Phi)}_{\ul{i'\mu\nu}}+S^{(\Pi)'}_{\ul{\mu\nu}}-\gamma_2S^{(g)}_{\ul{\mu\nu}}
\right)\,.
\end{align}
The harmonic constraints can be written as before, but the reduction constraints 
become more complicated,
\begin{align}
C_{\ul{i'\mu\nu}}&=(\varphi^{-1})^i{}_{\ul{i}'}\p_ig_{\ul{\mu\nu}}+V_{\ul{i}'}\,\Pi_{\ul{\mu\nu}}
-\Phi_{\ul{i'\mu\nu}}\,.
\end{align}
The Hamiltonian constraint becomes,
\begin{align}
\label{eq:H_constraint_DF}
\HN &= 
(\mathbbmss{g}^{-1})^{ij}(\mathbbmss{g}^{-1})^{kl}(d_k\Phi_{ijl}-d_k\Phi_{lij})\nonumber\\
&\quad-(\mathbbmss{g}^{-1})^{ij}(d_N\Phi_{ijV}-d_N\Phi_{Vij})+S^{(H)'}\,,
\end{align}
and similarly we obtain for the momentum constraint,
\begin{align}
\label{eq:M_constraint_DF}
&\MN_i= (\mathbbmss{g}^{-1})^{jk}\Big(d_{[j}\Pi_{i]k}+\tfrac{1}{2}d_j\Phi_{kiN}-\tfrac{1}{2}d_i\Phi_{jkN}
\Big)+S^{(M)'}_i\nonumber\\
&+W(\mathbbmss{g}^{-1})^{jk}\Big(d_N\Pi_{k[i}v_{j]}+\tfrac{1}{2}d_N\Phi_{kiN}v_j-\tfrac{1}{2}d_N\Phi_{jkN}v_i\Big)\,.
\end{align}
Generalizing our earlier convention, we use~$d$ here to denote lower case partial derivatives of the GHG 
variables where any multiplication by the projected Jacobian or contraction with~$N^a$ or~$V^a$ remains outside 
the derivative; for example~$\MN_i=\varphi^{\ul{i}'}{}_i\MN_{\ul{i}'}$ and, 
\begin{align}
d_i\Pi_{jk}
&=\varphi^{\ul{j}'}{}_j\varphi^{\ul{k}'}{}_k d_i\Pi_{\ul{j'k'}}\nonumber\\
&=\varphi^{\ul{j}'}{}_j\varphi^{\ul{k}'}{}_k(\phi^{\textrm{Sh}})^{\ul{j}}{}_{\ul{j}'}
(\phi^{\textrm{Sh}})^{\ul{k}}{}_{\ul{k}'}\p_i\Pi_{\ul{jk}}\,.
\end{align}
Note that in Eqs~\eqref{eq:H_constraint_DF},~\eqref{eq:M_constraint_DF}
the terms involving~${\p_N=N^{\mu}\p_{\mu}}$ are to be substituted from the equations of motion~\eqref{eqn:GHG_DF} 
using~$N^a=W(n^a+v^a)$ as defined in Eq~\eqref{eqn:N_ndecomp}. We do not explicitly substitute here
because the first order GHG variables are not naturally~$3+1$ decomposed so the resulting equations 
become artificially complicated. For a more geometric formulation of the constraints see the DF paper. 
The sources in Eqs~\eqref{eq:H_constraint_DF},~\eqref{eq:M_constraint_DF}
are modified to take care of the change to shell-coordinates,
\begin{align}
S^{(H)'}&=S^{(H)}+\gamN^{\ul{ij}}\gamN^{\ul{kl}}\,\Phi_{\ul{i'jl}}
\p_{\ul{k}}(\Phi^{\textrm{Sh}})^{\ul{i}'}{}_{\ul{i}}\nonumber\\
&\quad-\gamN^{\ul{ij}}\gamN^{\ul{kl}}\,\Phi_{\ul{l'ij}}
\p_{\ul{k}}(\Phi^{\textrm{Sh}})^{\ul{l}'}{}_{\ul{l}}\,,\nonumber\\
S^{(M)'}_{\ul{i}}&=S^{(M)}+\tfrac{1}{2}\gamN^{\ul{jk}}\,\Phi_{\ul{k'i}N}\p_{\ul{j}}(\Phi^{\textrm{Sh}})^{\ul{k}'}{}_{\ul{k}}
\nonumber\\
&\quad
-\tfrac{1}{2}\gamN^{\ul{jk}}\,\Phi_{\ul{j'k}N}\p_{\ul{i}}(\Phi^{\textrm{Sh}})^{\ul{j}'}{}_{\ul{j}}\,.
\end{align}
Finally we have the hypersurface constraints,
\begin{align}
D_{[i}(A^{-1}Wv_{j]})=0\,,\qquad D_{[i}\phi^{\ul{i}}{}_{j]}=0\,.\label{eqn:Hyp_Cons}
\end{align}
We assume for now that the change of coordinates is given by some a priori known functions so that without further
discussion symmetric hyperbolicity is unaffected by the transformation. The characteristic variables of
the transformed system can be written,
\begin{align}
\label{eq:char_vars_hypGHG}
u^{\hat{0}}_{\ul{\mu\nu}}&=g_{\ul{\mu\nu}}\,,\nonumber\\
u^{\hat{\pm}}_{\ul{\mu\nu}}&=\Pi_{\ul{\mu\nu}}\mp
\frac{W}{\sqrt{1+(v^{\mathbbmss{s}})^2}}(\mathbbmss{g}^{-1})^{ij}\mathbbmss{s}_j
\Phi_{i\,\ul{\mu\nu}}-\gamma_2g_{\ul{\mu\nu}}\,,\nonumber\\
u^{\hat{B}}_{j\,\ul{\mu\nu}}&=\perpdsq^i{}_j \; \Phi_{i\,\ul{\mu\nu}}+W\;\perpdsq^i{}_j \, v_{i}
\,\Pi_{\ul{\mu\nu}}\,,
\end{align}
with speeds,
\begin{align}
\beta^{\mathbbmss{s}}-\alpha\,v^{\mathbbmss{s}}\,,\quad\beta^{\mathbbmss{s}}\pm
\alpha\sqrt{1+(v^{\mathbbmss{s}})^2}\,,
\quad\beta^{\mathbbmss{s}}-\alpha\,v^{\mathbbmss{s}}\,,
\end{align}
respectively. Geometrically these are of course the same variables that we had 
before in Eq~\eqref{eqn:CV_upper}. Here, the symbol $\mathbbmss{s}$ denotes an arbitrary unit vector, 
spatial with respect to~$n^a$, that is normalized against the boost metric, i.e.\ in 
general~$\mathbbmss{s}^i \mathbbmss{s}_i \neq 1$, while~$(\mathbbmss{g}^{-1})^{ij}\mathbbmss{s}_i\mathbbmss{s}_j=1$.
We label the projection operator related to $\mathbbmss{s}$ by,
\begin{align}
\perpdsq^i{}_j& :=\delta^i{}_j-(\mathbbmss{g}^{-1})^{ik}\mathbbmss{s}_k\mathbbmss{s}_j\,. 
\label{eq:2+1projector_dollars}
\end{align}
This definition allows us to write the characteristic variables~\eqref{eq:char_vars_hypGHG} in a natural way, 
avoiding complicated coupling between the transverse and longitudinal blocks of the principal symbol 
as would occur if we were to use a projection operator defined like~\eqref{eq:2+1projector_dollars}
but with~$\gamma^{ik}$ instead of $(\mathbbmss{g}^{-1})^{ik}$. Notice the drawback of this strategy 
is that the representation of the lightspeeds,~$\beta^{\mathbbmss{s}}\pm\alpha\sqrt{1+(v^{\mathbbmss{s}})^2}$,
is slightly more complicated than usual. Furthermore note that the notation~$\perpdsq^i{}_j$ is used because
the 2-metric induced from~$\mathbbmss{g}_{ij}$ by a $2+1$-split against~$\mathbbmss{s}_i$ is not the same object,
$\mathbbmss{q}_{ij} = \mathbbmss{g}_{ij} - \mathbbmss{s}_i \mathbbmss{s}_j \neq \gamma_{ik} \perpdsq^k{}_j$.

%%%%%%%%%%%%%%%%%%%%%%%%%%%%%%%%%%%%%%%%%%%%%%%%%%%%%%%%%%%%%%%%%%%%%%%%%%%%%%%%%%%%%%%
\section{Hyperboloidal coordinates }
\label{Section:Hyperboloidal_Coordinates}
%%%%%%%%%%%%%%%%%%%%%%%%%%%%%%%%%%%%%%%%%%%%%%%%%%%%%%%%%%%%%%%%%%%%%%%%%%%%%%%%%%%%%%%

When modeling asymptotically flat spacetimes, the GW signal at infinity 
with respect to the outgoing characteristic direction, i.e.\ `future null-infinity', serves 
as an idealization for what we expect to measure on Earth from a distant astrophysical source. 
Usually numerical simulations that compute GWs restrict the domain to a finite region and 
thus entail extrapolation errors when evaluating the signal to null-infinity. Also, physical 
errors are caused by imperfect boundary conditions. Thus for high accuracy, and as a matter of 
principle, it is desirable to include null-infinity in the computational domain. One possible 
procedure is the usage of the `hyperboloidal compactification' technique, which means 
i)~to boost the time function such that its level sets are spacelike everywhere but asymptote 
towards null-infinity, and ii)~to compress the infinite spatial domain into a finite coordinate 
interval.

In linear blackhole perturbation theory the technique allows highly accurate computations
of quasi-normal modes and tail decay 
rates~\cite{ZenNunHus08,Zen10,RacTot11,Jas12,ZenKhaBur14,HarBerBru13} and consequently has 
become the standard way of computing the GW-signal from extreme-mass-ratio 
configurations~\cite{ZenKha11,BerNagZen11,BerNagZen10,BarBuoHug12,TarBuoKha14,HarBerNag14,HarLukBer15}.
The use of such hyperboloidal coordinates with the Einstein equations is however more problematic 
because the compactification results in terms that are singular at null-infinity, and require 
careful attention. Building on the earlier work of Penrose~\cite{Pen63,Pen65a}, 
Friedrich~\cite{Fri81a,Fri86} was able to regularize the system by working with a 
conformally related metric and an expanded set of equations. These equations have been treated 
numerically by, for example, Frauendiener~\cite{Fra98,Fra98a} and~H\"ubner~\cite{Hub99}. See 
the forthcoming monograph by Valiente-Kroon~\cite{Val16} for a comprehensive introduction. An 
alternative approach, suggested by Zengino\u{g}lu~\cite{Zen08}, is to work with a conformal 
metric, with the conformal factor either satisfying particular equations or else being simply a given function, but to treat the remaining formally singular terms without performing a full-blown 
regularization. Much progress has been made in this direction, particularly assuming axial 
or spherical symmetry~\cite{CalGunHil05,ZenHus06,MonRin08,Rin10,VanHusHil14}. Nonetheless, solving the 
Einstein equations in full 3d on hyperboloidal slices remains an unsolved problem. 

In this section we will analyze the applicability of the hyperboloidal layer coordinates for
usage with our general DF GHG system~\eqref{eqn:GHG_DF}. We will first review the construction of 
layer coordinates and then compute the explicit coefficients of~\eqref{eqn:GHG_DF} for this choice.
Studying the asymptotics, it turns out that the hyperboloidal layer coordinates do not give 
sharp control of certain quantities in a dynamical spacetime. Consequently, we devise a generalized 
setup, in which the hyperboloidal slices are `waggled' in order to guarantee that the 
outgoing radial coordinate lightspeed is controlled.

%%%%%%%%%%%%%%%%%%%%%%%%%%%%%%%%%%%%%%%%%%%%%%%%%%%%%%%%%%%%
\subsection{Hyperboloidal Layers}\label{subsection:layers}
%%%%%%%%%%%%%%%%%%%%%%%%%%%%%%%%%%%%%%%%%%%%%%%%%%%%%%%%%%%%

%%%%%%%%%%%%%%%%%%%%%%%%%%%%%%%%%%%%%%%%%%%%%%%%%%%%%%%%%%%%
\begin{figure}[t]
  \begin{center}
    \includegraphics[width=0.45\textwidth]{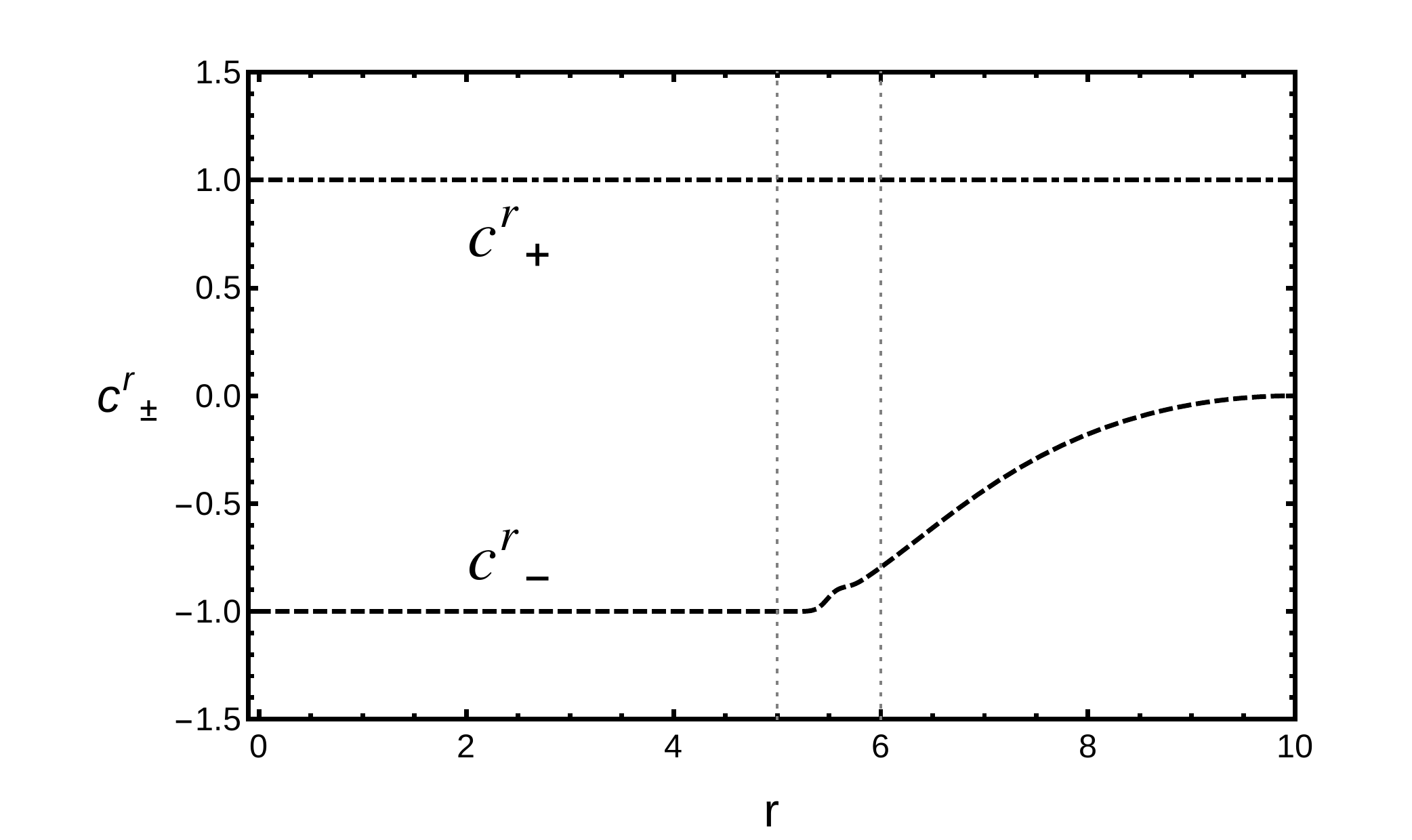}
    \caption{\label{fig:Cr_MK} Ingoing and outgoing radial 
    coordinate lightspeeds corresponding to a hyperboloidal 
    compactification of the Minkowksi spacetime. In this example
    the hyperboloidal layer begins at~$r=5$, with 
    a transition zone spanning the interval from $r=5$ to $r=6$,
    and~$r=10$ corresponding to null-infinity;
    cf.~Eqs~\eqref{eqn:hyp_layer}-\eqref{eqn:hyp_layer_transitionfun}
    with $R_i=5$, $S=10$, $\Delta r=1$. 
    The outgoing radial lightspeed is constant, 
    the idea being that outgoing waves then propagate out 
    without distortion. The ingoing radial lightspeed displays
    the three different regions, the interior, the smooth transition, 
    and the hyperboloidal region.
    }
  \end{center}
\end{figure}
%%%%%%%%%%%%%%%%%%%%%%%%%%%%%%%%%%%%%%%%%%%%%%%%%%%%%%%%%%%%

%%%%%%%%%%%%%%%%%%%%%%%%%%%%%%%%%%%%%%%%%%%%%%%%%%%%%%%%%%%%
\paragraph*{Hyperboloidal basics:} Let us specify the above ideas of hyperboloidal 
coordinates. Following~\cite{Zen10} but using slightly different notation,
we write the transformation
from spherical coordinates $(T,R)$ to hyperboloidal coordinates $(t,r)$
in terms of the `height-function' $H(R)$ and the `compress-function'
$\Omega(r)$ like,
\begin{align}
 \label{eqn:hyp_trafo}
  T = T(t,r) = t + H(R) \, , \quad R = R(r) = \Omega(r)^{-1}r \,.
\end{align}
The angular coordinates remain unchanged~$\vartheta^A=\vartheta^{A'}$. The 
term height-function alludes to the fact that~$H$ literally governs the offset
of the hyperboloidal $t$-slices from the $T$-slices in a conformal diagram. The 
compress function squeezes the infinite spatial domain to a finite interval, 
with~$\Omega(r)=0$ determining the coordinate value of~$r$ that corresponds 
to~$R=\infty$. Certain weak requirements have to be imposed on~$H$ and~$\Omega$, 
see~\cite{Zen10} for details. Intuitively the effect of raising the slices is 
to increase the outgoing coordinate lightspeed. On the other hand a 
radial compactification alone forces both the ingoing and outgoing lightspeeds 
to vanish near the compactification boundary. The basic idea of using hyperboloidal compactification 
is to choose both functions in tandem so that one obtains regular outgoing lightspeeds, 
preferably going to some desired limiting value at null-infinity. A consequence of this 
is that the incoming lightspeeds vanish at null-infinity. This means that incoming 
pulses of radiation from a neighborhood of null-infinity will be distorted as they 
propagate. By assumption there should be no such incoming pulses, but rather only 
low-frequency features. See~\cite{CalGunHil05} for further discussion of how the 
compress and height functions are connected. Choosing a suitable hyperboloidal 
compactification depends upon the data we are interested in treating. 

%%%%%%%%%%%%%%%%%%%%%%%%%%%%%%%%%%%%%%%%%%%%%%%%%%%%%%%%%%%%
\paragraph*{Hyperboloidal compactification of Minkowski:} Following the ideas 
of~\cite{Mon00}, the most natural height function in the Minkowski spacetime would be 
given by~${H = \sqrt{\kappa^2 + R^2}}$ with $\kappa=\rm{const.}$. A simple choice for 
the compress function is~${\Omega=(1-r^2)/(2\kappa)}$, which places null-infinity at the 
coordinate value $r=1$. These choices result in radial coordinate lightspeeds 
of~${c^r_{\pm}=\pm(\pm1+r)^2/(2\kappa)}$. In order to have outgoing waves remain undistorted through the 
whole hyperboloidal slice, e.g.~$c^r_+=+1$, one has to modify the above simple choices~\cite{Zen10}. 
Starting from a natural choice for the compress-function,
\begin{align}
\Omega = 1 - \frac{r^2}{S^2} \, ,
\label{eqn:spec_comp}
\end{align}
where~$S$ defines the coordinate value of null-infinity, demanding unit outgoing radial 
coordinate lightspeed gives~$H'=1-R'^{-1}$. Here and elsewhere~$H'\equiv \frac{dH}{dR}$ 
and~$R'\equiv\frac{dR}{dr}$. The resulting height-function is given by,
\begin{align}
H = \frac{ 2 R^2 + S^2 - \sqrt{ 4 R^2 S^2 +S^4} } {2 R} \,, 
\label{eqn:spec_MK_height}
\end{align}
but interestingly this explicit form is actually not needed. The coefficients
of the DF evolution equations can be built solely from the knowledge of~$H'$ and~$R'$,
see quantities computed in Sec~\eqref{subsection:Hyp_Jac}. Finally note that the highly symmetric 
Minkowski background allows us to make choices for the height-function which are only~$R$-dependent, like 
Eq~\eqref{eqn:spec_MK_height}, while generally a suitable choice for the height-function may be spacetime 
dependent.

%%%%%%%%%%%%%%%%%%%%%%%%%%%%%%%%%%%%%%%%%%%%%%%%%%%%%%%%%%%%
\paragraph*{Hyperboloidal layers:} The transformation to hyperboloidal coordinates 
starts from spherical coordinates, which in our code are only available in the outer region 
of the computational domain. This prompts us to employ the hyperboloidal compactification
only outside of a sphere of radius~$R_{i}$, which is called the `interface'. This procedure, 
devised by Zengino\u{g}lu in~\cite{Zen10}, see also~\cite{BerNagZen11}, is called the `hyperboloidal
layer'-technique. Starting from the general height-function approach, Eqs~\eqref{eqn:hyp_trafo}, the
layer setup is achieved by modifying the compactification to,
\begin{align}
 \label{eqn:hyp_layer}
 R-R_i = \frac{r-R_i}{\Omega}\,,
\end{align}
with a suitable modification of~$\Omega$. When using Eq~\eqref{eqn:spec_comp} for~$\Omega$,
the following modification is needed,
\begin{align}
 \label{eqn:hyp_layer_compressfunction}
 \Omega(r) = 1 - \left( \frac{r-R_i}{S-R_i} \right)^2 \tilde{\chi}(r)\,,
\end{align}
where~$\tilde{\chi}(r)$ is a smooth transition function on the interval $(R_i,R_i+\Delta r)$
between the values~$0$ and~$1$, i.e.\ $ \tilde{\chi}(R_i) = 0$ and $ \tilde{\chi}(R_i+\Delta r) = 1$.
In numerical experiments, see~Sec~\ref{Section:Numerical_Experiments}, one might 
take~\eqref{eqn:hyp_layer_compressfunction} with the transition function,
\begin{align}
\label{eqn:hyp_layer_transitionfun}
  \begin{cases}
   0\,,                                                       & r \leq R_i\,, \\
   \frac{1}{2} \left[ 1 + \tanh\left(\frac{1}{R_i-r} - \frac{1}{r-(R_i+\Delta r)} \right) \right] \,,
   & r\in (R_i,R_i+\Delta)\,, \\
   1\,,                 & R_i+\Delta r \leq r\leq S\,,
  \end{cases}
\end{align}
where~$\Delta r$ is usually chosen to be small compared to~${S-R_i}$. Most often~\cite{Zen10} the Heaviside step 
function, i.e.\ the~$\Delta r \to 0$ limit of Eq~\eqref{eqn:hyp_layer_transitionfun}, is chosen for~$\tilde{\chi}(r)$. 
Since we wish to use a pseudospectral method it seems advantageous to make the transition function smooth. Unfortunately 
it turns out that resolving the transition function may require very high resolution, so we postpone testing 
this setup for future work and also use the Heaviside function in our numerical experiments. This is not as bad as it
initially sounds because we align the jump with grid boundaries so that the numerical derivatives never see the
discontinuity. In summary, the transformation~\eqref{eqn:hyp_layer} then means that~$R=r$ identically before the interface, 
and~$R=R(r)$ is compactified afterwards. As before, the height-function is simply set according to $H'=1-R'^{-1}$, which 
gives~$H=0$ for~$R<R_i$, as expected, when fixing the integration constant as wanted. To understand the structure of a 
computational domain that uses the hyperboloidal layer, we can consider the radial coordinate light speeds associated to 
the above transformations on a Minkowski background. The radial lightspeeds read~$c^r_{\pm}=\pm(R'(1 \mp H'))^{-1}$ which means 
for our choice of the height function $c^r_+=1$ and $c^r_-=(1-2 R')^{-1}$. In Fig.~\ref{fig:Cr_MK} the speeds are plotted using 
the hyperboloidal layer built upon~\eqref{eqn:hyp_layer}-\eqref{eqn:hyp_layer_transitionfun}.

%%%%%%%%%%%%%%%%%%%%%%%%%%%%%%%%%%%%%%%%%%%%%%%%%%%%%%%%%%%%
\subsection{The hyperboloidal DF approach with height-function controlled slicings}
\label{subsection:Hyp_Jac}
%%%%%%%%%%%%%%%%%%%%%%%%%%%%%%%%%%%%%%%%%%%%%%%%%%%%%%%%%%%%

Now we want to compute the quantities appearing in the DF GHG system~\eqref{eqn:GHG_DF}
for the specific choice of hyperboloidal layer coordinates.
Roughly speaking, assuming knowledge of the variables $A,B^{\ul{i}'}, \gamN_{\ul{i'j}'}$
throughout the (initial) hyperboloidal slice, we need to compute the 
lower case variables~$\alpha, \beta^i, \gamma_{ij}$, the projected 
Jacobians~$\varphi^{\ul{i}'}{}_{i}, (\varphi^{-1})^{i}{}_{\ul{i}'}$, and the boost 
vectors $v_i,V_{\ul{i}'}$. Note that $W$ and~$\mathbbmss{g}_{ij}$ can be computed from these as well.
Thus we need the Jacobians for the transformation from the shells-adapted GHG coordinates $X^{\ul{\mu}'}=(T,R,\vartheta^{A'})$
to hyperboloidal coordinates $x^\mu=(t,r,\vartheta^A)$, Eq~\eqref{eqn:hyp_trafo}.
In this case the Jacobian reads,
\begin{align}
\label{eqn:J_HypLay}
 J_{\rm{hyp}} &= 
 \left(\begin{array}{cccc}
  1     & 0  & 0 & 0 \\
  H' R' & R' & 0 & 0 \\
  0     & 0  & 1 & 0 \\
  0     & 0  & 0 & 1 \\
       \end{array}\right)\, , 
\end{align}
with obvious entries in the inverse. These Jacobians will
have to be compared with Eqs~\eqref{eqn:J} and~\eqref{eqn:J_inv}
to deduce the desired DF quantities.

%%%%%%%%%%%%%%%%%%%%%%%%%%%%%%%%%%%%%%%%%%%%%%%%%%%%%%%%%%%%
\paragraph*{$2+1$ decomposition:}
Before going into the calculation, we introduce some additional notation for 
$2+1$-decompositions of our spatial slices, which will allow for 
shorter relations between the upper case variables and the desired lower case quantities.
To perform the $2+1$-splits, we first define the outward pointing unit vector normal
to surfaces of constant~$R$, namely,
\begin{align}
\label{eq:def_Ngamma_metric_2+1normal}
S_{\ul{i}'} := L {}^\textrm{\tiny{(N)}}\!D_{\ul{i}'}R .
\end{align}
We denote the radially~$2+1$ decomposed form of the upper case metric $\gamN_{\ul{i'j}'}$
with respect to the normal~$S^a$ by,
\begin{align}
\gamN_{\ul{i}'\ul{j}'}&=
\left(\begin{array}{cc}
L^2 + \slN_{A'} \slN^{A'} & \slN_{A'}\\
\slN_{B'} &\qN_{A'B'}
\end{array}
\right) \, ,
\end{align}
with~$L$ the length scalar,~$\slN^{A'}$ the slip vector, and $\qN_{A'B'}$ the induced~$2$-metric.
Analogously, we introduce,
\begin{align}
\label{eq:def_gamma_metric_2+1normal}
  s_i := l D_i r \; ,
\end{align}
to decompose the lower case metric $\gamma_{ij}$, and denote the associated length scalar, slip vector and
induced $2$-metric by~$\{l,b^A,q_{AB}\}$.
We also decompose the boost metric $\mathbbmss{g}_{ij}$ along the vector,
\begin{align}
 \label{eq:def_boost_metric_2+1normal}
 \mathbbmss{s}_i := \mathbbmss{l} \mathbbmss{D}_i r \; ,
\end{align}
which is also normal to $r$-surfaces but defined to be unit against the boost metric,
$ (\mathbbmss{g}^{-1})^{ij} \mathbbmss{s}_i \mathbbmss{s}_j = 1$.
We label the associated length scalar, slip-vector and
induced $2$-metric~$\{\mathbbmss{l},\mathbbmss{b}^A, \mathbbmss{q}_{AB} \}$,
with $\mathbbmss{q}_{ij} := \mathbbmss{g}_{ij} - \mathbbmss{s}_i \mathbbmss{s}_j$.
Note that here and in what follows the symbols~$S_{\ul{i}'}, s_i,\mathbbmss{s}_i$ denote very particular spatial 
unit normal vectors while in the context of the PDE-analysis in Sec~\ref{Section:DF_GHG}, e.g.\ in 
Eq~\eqref{eqn:CV_upper}, they were used to denote arbitrary spatial unit vectors.
We avoid introducing distinct notation because 
the definitions~\eqref{eq:def_Ngamma_metric_2+1normal},~\eqref{eq:def_gamma_metric_2+1normal}
and~\eqref{eq:def_boost_metric_2+1normal} still comply with the demand to be spatial unit vectors,
where we just remove the arbitrariness in their choice for the majority of the paper.
See also Tab~\ref{tab:2p1_unit_vectors} for a summary of
spatial unit vectors used in the course of this paper and their mutual relations.

%%%%%%%%%%%%%%%%%%%%%%%%%%%%%%%%%%%%%%%%%%%%%%%%%%%%%%%%%%%%
\paragraph*{Coordinate lightspeeds:}
In the upper case shell coordinates, $X^{\ul{\mu}'}$, we consider the coordinate lightspeeds along null geodesics 
with tangent vector proportional to~$N^a+S^a$. We refer to these as the coordinate lightspeeds along~$S_{\ul{i}'}$, 
and adopt a similar terminology when working in lower case coordinates. We find that the upper case coordinate lightspeeds 
along~$S_{\ul{i}'}$ can be written as,
\begin{align}
C^R_{\pm}&=-B^R\pm A\,L^{-1}\,, \nonumber \\
C^{A'}_{\pm}&=-B^{A'}\mp\slN^{A'}\,A\,L^{-1}\,.
\end{align}
Interestingly, many relations will turn out to be most naturally written in terms of these 
coordinate lightspeeds. We are mainly interested in situations when the~$S^a$ vector 
is close to radial, i.e.\ $\slN^{A'} \approx 0$. Given that $B^{A'}$ vanishes asymptotically, this also
means that the angular components~$C^{A'}_{\pm}$ are assumed to be small. This precludes the use of corotating
coordinates for binary spacetimes in the asymptotic region, but the assumption could be revisited in future
work. Based on these assumptions, we refer to~$C^R_{\pm}$ as the incoming and outgoing `radial' lightspeeds
respectively, although this is not strictly true, as just explained.

%%%%%%%%%%%%%%%%%%%%%%%%%%%%%%%%%%%%%%%%%%%%%%%%%%%%%%%%%%%%
\paragraph*{Upper case boost vector:} Now we can go about actually calculating the desired quantities.
For the upper case boost covector $V_{\ul{i}'}$ we 
can use the $\partial t/\partial x^i$ entries of $J_{\rm{hyp}}^{-1}$ to see
that the angular components vanish. Exploiting also the~$\partial t/\partial T$ entry, see~\eqref{eqn:J_inv}, we 
can specify the radial component and so obtain,
\begin{align}
V_R&=\frac{AH'}{1+B^RH'}\,,\qquad V_{A'}=0\,.
\end{align}
By assumption we know the upper case spatial metric,
so this is sufficient to obtain the full 
upper case boost vector $V^R=V_R/L^2$ and $V^{A'}=-\slN^{A'} \, V_R/L^2$.

%%%%%%%%%%%%%%%%%%%%%%%%%%%%%%%%%%%%%%%%%%%%%%%%%%%%%%%%%%%%
\paragraph*{Lower case metric and boost vectors:}
Similarly we can use and combine the other equations
that result from comparing the Jacobians.
We obtain the lower case lapse and shift,
\begin{align}
\alpha&=\frac{A}{\big(1-C_+^R\,H'\big)^{\tfrac{1}{2}}\big(1-C_-^R\,H'\big)^{\tfrac{1}{2}}}\,,\nonumber\\
\beta^r&=\frac{B^{R}+C_+^RC_-^R\,H'}{R'\,\big(1-C_+^R\,H'\big)\big(1-C_-^R\,H'\big)}\,,\nonumber\\
\beta^A&=\frac{2B^{A'}+(C_-^RC_+^{A'}+C_+^RC_-^{A'})\,H'}{2\,\big(1-C_+^R\,H'\big)\big(1-C_-^R\,H'\big)}\,.
\label{eqn:gauge_hyp}
\end{align}
The metric components are,
\begin{align}
l&=L\,R'\,\big(1-C_+^R\,H'\big)^{\tfrac{1}{2}}\big(1-C_-^R\,H'\big)^{\tfrac{1}{2}}\,,\nonumber\\
b^A&=R'\,\bN^{A'}+R'\,H'\,(B^{A'}+B^R\,\bN^{A'})\,,\nonumber\\
q^{AB}&=\qN^{A'B'}\,.
\label{eqn:metric_hyp}
\end{align}
The lower case boost covector is given by, 
\begin{align} 
  Wv_r &=-AH'R'\,, \nonumber \\
  Wv_A &=0\,,
\end{align}
where the Lorentz factor is,
$$ 
W=\frac{2-(C_+^R+C_-^R)H'}
{2\,\big(1-C_+^R\,H'\big)^{\tfrac{1}{2}}\big(1-C_-^R\,H'\big)^{\tfrac{1}{2}}}\,.
$$
For convenience, we also state the contravariant components,
\begin{align}
Wv^r&=-\frac{H'\big(C_+^R-C_-^R\big)^2}{4\,A\,R'\big(1-C_+^R\,H'\big)\big(1-C_-^R\,H'\big)}
\,,\nonumber\\
Wv^A&=
-\frac{H'\,C_+^{A'}}{2\,L\,\big(1-C_+^R\,H'\big)}
+\frac{H'\,C_-^{A'}}{2\,L
\,\big(1-C_-^R\,H'\big)}
\,.\nonumber\\
\end{align}
Thus we can compute the coordinate lightspeeds in the direction of~$s_i$
and find,
\begin{align}
c_\pm^r&=-\beta^r\pm \alpha\,l^{-1}=\frac{C_\pm^R}{R'\,\big(1-C_\pm^R\,H'\big)}\,,\nonumber\\
c_\pm^A&=-\beta^A\mp b^A\alpha\,l^{-1}=\frac{C_\pm^{A'}}{1-C_\pm^R\,H'}\,.\label{eqn:cpm_hyp}
\end{align}
It is remarkable that the results of the coordinate
transformations can be so cleanly expressed in 
terms of the upper case lightspeeds, and $R'$ and $H'$.

%%%%%%%%%%%%%%%%%%%%%%%%%%%%%%%%%%%%%%%%%%%%%%%%%%%%%%%%%%%%
\paragraph*{Projected Jacobians:}
We also require the projected Jacobians.
These can be computed from the matrices $\phi^{\ul{i}'}{}_i, \Phi^i{}_{\ul{i}'}$,
i.e.\ the spatial parts of the Jacobians~\eqref{eqn:J} and~\eqref{eqn:J_inv},
and the boost covectors. We find,
\begin{align}
\varphi^{R}{}_{r}&=R'\,(1+H'\,B^R)\,,\quad 
(\varphi^{-1})^{r}{}_{R}=\frac{1}{R'\big(1+B^R\,H'\big)}\, ,\nonumber\\
\varphi^{A'}{}_{r}&=R'\,H'\,B^{A'}\,,\qquad \quad
(\varphi^{-1})^{A}{}_{R}=\frac{-H'B^{A'}}{1+B^R\,H'}\,,\nonumber\\
 \varphi^{A'}{}_{A}&=\delta^{A'}{}_{A}\,,\qquad \qquad \quad
 (\varphi^{-1})^{A}{}_{A'}=\delta^{A}{}_{A'}\,,
\end{align}
and the remaining components vanish.

%%%%%%%%%%%%%%%%%%%%%%%%%%%%%%%%%%%%%%%%%%%%%%%%%%%%%%%%%%%%
\paragraph*{Boost metric:}
Finally the boost metric can be computed directly from~$\gamma_{ij},W$ and~$v_i$,
or by inverting Eq~(48) of the DF paper. From this we obtain,
\begin{align}
\label{eq:boost_metric_DFrelations}
\mathbbmss{l}&=L\,R'\,\big(1+B^R\,H'\big)\,,\nonumber\\
\mathbbmss{b}^A&=b^A
\,,\nonumber\\
(\mathbbmss{q}^{-1})^{AB}&=q^{AB}\,.
\end{align}
The inverse~$\mathbbmss{q}^{-1}$ is easily checked to be,
\begin{align}
(\mathbbmss{q}^{-1})^{ij}&:=(\mathbbmss{g}^{-1})^{ij}
-(\mathbbmss{g}^{-1})^{ik}(\mathbbmss{g}^{-1})^{jl}\mathbbmss{s}_k\mathbbmss{s}_l\, .
\end{align}
Note that in Eq~\eqref{eq:boost_metric_DFrelations} we really have to write $\mathbbmss{q}^{-1}$ because
$(\mathbbmss{q}^{-1})^{ij} \neq \mathbbmss{q}^{ij}$, 
where $ \mathbbmss{q}^{ij} \equiv \gamma^{ik} \gamma^{jl} \mathbbmss{q}_{kl} $
refers to the $2$-metric induced by the $2+1$-split along $\mathbbmss{s}_i$.
Recall that $\mathbbmss{s}_i$ is the unit normal vector to surfaces of constant $r$
but normalized so that it has unit magnitude with respect to the boost metric,
see Eq~\eqref{eq:def_boost_metric_2+1normal}. 

%%%%%%%%%%%%%%%%%%%%%%%%%%%%%%%%%%%%%%%%%%%%%%%%%%%%%%%%%%%%
\paragraph*{Summary:}
With the collected expressions of this subsection we can build the full evolution equations~\eqref{eqn:GHG_DF}. 
In fact, our asymptotic analysis in the next section~\ref{subsection:asymptotics} indicates that sharper control 
of the radially outgoing coordinate lightspeeds may be desirable. But in our first toy model numerical tests we will 
use the height-function approach. 

%%%%%%%%%%%%%%%%%%%%%%%%%%%%%%%%%%%%%%%%%%%%%%%%%%%%%%%%%%%%
\subsection{Asymptotics of the DF GHG system with hyperboloidal layer coordinates}\label{subsection:asymptotics}
%%%%%%%%%%%%%%%%%%%%%%%%%%%%%%%%%%%%%%%%%%%%%%%%%%%%%%%%%%%%

In this section we want to consider the asymptotic behavior of the lower case quantities appearing in the
DF GHG system~\eqref{eqn:GHG_DF}. In the first instance we are concerned with making the weakest possible 
reasonable assumptions on the initial data and on the height-function transformation under which
the resulting equations are regular.

%%%%%%%%%%%%%%%%%%%%%%%%%%%%%%%%%%%%%%%%%%%%%%%%%%%%%%%%%%%%
\paragraph*{Flatness assumptions:} Choosing suitable assumptions for the metric is the trickiest part of the present
work. Ideally we want to make the class of spacetimes under consideration large enough so that every
physically interesting possibility is contained. But we have to be careful in imposing sufficiently strong assumptions
that the scheme we are deriving has sufficiently good behavior near null-infinity so that the equations are both 
analytically and numerically tractable. Our working definition of asymptotic flatness is that, in a neighborhood of 
null-infinity there exist (nonunique) preferred asymptotically Cartesian coordinates~$\tilde{X}^{\tilde{\mu}}$ and that 
the metric components in this coordinate basis tend to those of the Minkowski metric in a global inertial frame at some 
slow rate. We can build shell coordinates from the spatial~$\tilde{X}^{\tilde{i}}$ in the standard way as in 
Sec~\ref{subsection:Shells}. We assume that the associated outgoing radial lightspeed is {\it exactly} unity, and that 
partial derivatives of the Cartesian components with respect to both the time derivative and the associated shell 
coordinate basis vectors maintain the same weak fall-off. The GHG coordinates~$X^{\ul\mu}$ do not match exactly the 
preferred coordinates~$\tilde{X}^{\tilde{\mu}}$, but we assume that they are close to the preferred coordinates in the 
following precise sense. We assume that in a neighborhood of a point on null-infinity we have flat space plus an error 
term,
\begin{align}
g_{\ul{\mu\nu}}=\eta_{\ul{\mu\nu}}+o(R^{-{\epsilon}})\,,\label{eqn:flat_ass1}
\end{align}
where~$\epsilon>0$, $\eta_{\ul{\mu\nu}}$ is the Minkowski metric in global inertial coordinates. We also assume that
first derivatives of these components satisfy,
\begin{align}
\p_{\ul{i}'}g_{\ul{\mu\nu}}=o(R^{-{\epsilon}})\,,\quad\p_Tg_{\ul{\mu\nu}}=o(R^{-{\epsilon}})\,.\label{eqn:flat_ass2}
\end{align}
so that as in the preferred coordinate basis no fall-off is lost upon taking derivatives. We transcribe these
assumptions to the first order reduction variables in the obvious way. We could place more sophisticated assumptions
here by distinguishing between derivatives along radially outgoing and incoming null curves. Here we expect that our
assumptions can be weakened so that the angular derivatives~$\p_{A'}g_{\ul{\mu\nu}}=o(R^{1-\epsilon})$, without breaking the
asymptotic analysis that follows. Nevertheless we prefer to make the stronger assumption because otherwise we can not
guarantee regularity of the evolved variable~$\Phi_{A'\ul{\mu\nu}}$. Finally we assume that derivatives of the outgoing
radial coordinate lightspeed fall-off slightly better than other first derivatives,
\begin{align}
\p_TC_+^R=\p_{\ul{i}'}C_+^R=o(R^{-1-{\delta}})\,,\label{eqn:flat_ass3}
\end{align}
for some~$\delta>0$. In other words we assume that the GHG coordinates tend to the preferred coordinates at large
radius. Note that it is desirable to relax this assumption, but we have been unable to do so, see 
Eq~\eqref{eq:change_of_lightspeed} below for the general expression. Theoretically, data with values of~$\epsilon$ satisfying,
\begin{align}
 1/2\leq \epsilon < 1 \,,
\end{align}
are expected to allow well-defined evolution~\cite{Bie09}, and optimally we would like to be able to handle that
whole range. Therefore we work primarily under this assumption. In many physical examples however we expect to
have~$\epsilon\sim1$, or rather the error term should be~$O(R^{-1})$. We will occasionally refer to this stronger
fall-off as a `stricter' assumption. A natural way to arrive at `required' fall-off conditions would be to insist
that the Trautmann-Bondi mass is well-defined, perhaps in so doing using the Hamiltonian and momentum constraints. Since
we are interested in performing free-evolution numerically, we do not do so. Before actually deducing the asymptotics of
the relevant upper case quantities, note that we are working in GHG-shell coordinates so that we first have to translate
our assumptions~\eqref{eqn:flat_ass1}-\eqref{eqn:flat_ass3} on the asymptotics to the transformed 
components~$g_{\ul{\mu'\nu'}}$. We find,
\begin{align}
 A          =&  \, 1 + o(R^{-{\epsilon}}) \;,  \nonumber \\
 B^R        =&  \,o(R^{-{\epsilon}})   \;,\nonumber \\
 B^{A'}     =&  \,o(R^{-{\epsilon}-1}) \;,\nonumber \\ 
 L          =& \,1 + o(R^{-{\epsilon}}) \;,\nonumber \\
 \slN^{A'}  =& \, o(R^{-{\epsilon}-1}) \;, \nonumber \\
 \qN^{A'B'} =& \, \eta^{A'B'} + o(R^{-{\epsilon}-2})
 \; .  
\end{align}
The coordinate lightspeeds therefore go like,
\begin{align}
C_\pm^R&=\pm1+o(R^{-{\epsilon}}) \; , \nonumber \\
C_\pm^{A'}&=o(R^{-{\epsilon}-1}) \; .
\end{align}
More generally one might like to replace the GHG formulation with some other choice. For that the assumptions made above
will still be required, but the following analysis of the asymptotics would have to be repeated.
It follows from these assumptions that a generic component of the Christoffel symbol is of order,
\begin{align}
\Gamma^{\ul{\kappa}}{}_{\ul{\mu\nu}}&=o(R^{-\epsilon})\,.
\end{align}
To discuss the regularity of the height-function hyperboloidal DF GHG system we now need to establish the asymptotics of
the transformations to the hyperboloidal coordinates.

%%%%%%%%%%%%%%%%%%%%%%%%%%%%%%%%%%%%%%%%%%%%%%%%%%%%%%%%%%%%
\paragraph*{Assumptions on the transformations using standard hyperboloidal coordinates:} 
As a first intuitive choice we will discuss
the radial transformation defined by Eq~\eqref{eqn:spec_comp}.
This choice is known to work well in perturbation theory.
Let us consider what the radial transformation~\eqref{eqn:spec_comp}
means at large radii,
\begin{align}
\frac{R'}{R^2}&=
\frac{2}{S^2}+\frac{1}{SR}+o(R^{-1})\,.  \label{eqn:compacti_asy_n2}
\end{align}
Without prescribing $H$ explicitly, our assumption on its asymptotics
are motivated by demanding $c^r_+ \rightarrow +1$.
Inspecting Eq~\eqref{eqn:cpm_hyp} this means to demand that 
in the neighborhood of null-infinity $H$ is such that,
\begin{align}
1-C_+^RH'&=\frac{1}{2}\frac{S^2}{R^2}+o(R^{-2})\,.\label{eqn:height_asy_n2}
\end{align}
Note that Eq~\eqref{eqn:height_asy_n2} implies asymptotically
$H' \rightarrow+1$, as expected. Looking at the relations~\eqref{eqn:gauge_hyp},\eqref{eqn:metric_hyp},
we can then find the asymptotics of the lower case quantities.
Starting with the angular components of the shift vector and the coordinate lightspeeds,
\begin{align}
\beta^A&=o(R^{1-\epsilon})\,,\qquad c^A_{\pm}=o(R^{1-\epsilon})\,,\label{eqn:height_gauge_n2}
\end{align}
we immediately encounter a very unpleasant feature.
To avoid the boundary rotating artificially and, more seriously, to avoid divergent
coordinate lightspeeds in a numerical implementation,
we would need to have~$\epsilon\geq1$ here; but this precludes practically all data of physical
interest. Note however that the stricter~$O(R^{-1})$ flatness assumption mentioned above would lead to
unproblematic angular shift components and coordinate light speeds. To cover the whole range of~$\epsilon$,
it might be possible to allow more freedom in the shell coordinates in order to gain a factor~$R^{-1}$ in
the respective entries of the Jacobians. But we want to use the shell coordinates, and so have to find another
solution. As it turns out, we can simply allow ourselves more freedom in the compactification.

%%%%%%%%%%%%%%%%%%%%%%%%%%%%%%%%%%%%%%%%%%%%%%%%%%%%%%%%%%%%
\paragraph*{Assumptions on the transformations using Calabrese \textit{et al.}'s hyperboloidal coordinates:} 
Instead of sticking to the choice~\eqref{eqn:spec_comp},
Calabrese \textit{et al.} suggested a class of radial transformations like Eq~(A3) in~\cite{CalGunHil05},
which satisfy asymptotically~$R'\sim R^n$, or more specifically,
\begin{align}
  \frac{R'}{R^{n}} =  \frac{ 2}{ S^{n} \, (n-1)} + O(R^{-1})\,,\label{eqn:compacti_asy_n}
\end{align}
valid for the real parameter~$1<n$. As discussed in~\cite{CalGunHil05},~$0<n\leq 2$ needs to hold 
to have any chance of numerical stability, and we need~$1<n$ to draw infinity to a finite coordinate.
Henceforth we take this generalized form of compactification and assume,
\begin{align}
 1<n \leq 2 \; .
\end{align}
Note that Eq~\eqref{eqn:compacti_asy_n2} corresponds to the 
case~$n=2$. Now, our requirement that~$c^r_+ \rightarrow +1$ implies,
\begin{align}
1-C_+^RH'&= \frac{S^n (n-1)}{ 2 \,  R^{n} }+o(R^{-n})\,.\label{eqn:height_asy_n}
\end{align}
As discussed later on, in dynamical spacetimes it may be desirable to obtain sharper control on the slicing.
We thus generalize the pure height-function approach~\eqref{eqn:hyp_trafo} in Sec~\ref{subsection:LS_control}.
But for the moment let us assume to take the height-function approach with~\eqref{eqn:compacti_asy_n}
and~\eqref{eqn:height_asy_n} satisfied. We will next investigate the entailed asymptotics of the
relevant quantities, and it will turn out that they can be made well-behaved for suitable choices of~$n$.

%%%%%%%%%%%%%%%%%%%%%%%%%%%%%%%%%%%%%%%%%%%%%%%%%%%%%%%%%%%%
\paragraph*{Asymptotics of the lower case quantities:}
We start with the asymptotics of the lower case lapse and shift,
\begin{align}
\alpha   &\simeq \frac{R^{\frac{n}{2}}}{S^{\frac{n}{2}}\sqrt{n-1}} \, 
\big(1+o(R^{-\epsilon})\big)\,,\nonumber\\
\beta^r  &\simeq-\frac{1}{2} \left( 1+o(R^{-\epsilon}) \right) \,,\nonumber\\
\beta^A  &=o(R^{-\epsilon-1+n})\,,\label{eqn:height_gauge_working}
\end{align}
and the spatial metric,
\begin{align}
l      &\simeq \frac{2R^{\frac{n}{2}}}{S^{\frac{n}{2}}\sqrt{n-1}}   \big(1+o(R^{-\epsilon})\big)\,,\nonumber\\
b^A    &= o(R^{-\epsilon-1+n})\,,\nonumber\\
q^{AB} &=\eta^{AB}+o(R^{-\epsilon-2})\,.\label{eqn:height_met_working}
\end{align}
In this and the following paragraph the symbol~$\simeq$ denotes equality up to a factor of one plus
a function that vanishes as~$R\to\infty$. It follows that with~$n=2$ the lower case spatial metric takes
the form~$\Omega^2\tilde{\gamma}_{\mu\nu}$, with~$\Omega=O(R)$ and~$\tilde{\gamma}_{\mu\nu}$ regular,
consistent with what we would expect from the conformal approach.
As an interesting aside, from the spatial metric we can infer
the asymptotics of the lower case time vector~$t^a=\alpha\,n^a+\beta^a$.
Contrary to naive intuition, $t^a$ is not null at~$r=S$. Instead, splitting off the part of the shift 
transverse to $s^a$ and dividing by the lapse gives the vector~$n^a+\alpha^{-1}\beta^s s^a$, which satisfies,
\begin{align}
g_{ab}(n^a+\alpha^{-1}\beta^s s^a)(n^b+\alpha^{-1}\beta^s s^b)&=o(R^{-\epsilon})\,,
\label{eqn:working_def_nullinf}
\end{align}
and so {\it is} null at~$r=S$. We can take this as a working definition of a slice being hyperboloidal, 
also sometimes referred to as asymptotically null (here in the sense that the slice terminates at null-infinity).
Back to the study of the relevant quantities for our evolution system, the coordinate
lightspeeds~\eqref{eqn:cpm_hyp} are,
\begin{align}
c^r_+&\simeq1+o(R^{-\epsilon})\,,\qquad c^r_-=O(R^{-n})\,,\nonumber\\
c^A_{\pm}&=o(R^{-\epsilon-1+n})\,.
\end{align}
This means that, if we want to control~$c^A_{\pm}$ and keep outgoing waves well-resolved,
we need to take~$n$ such that,
\begin{align}
 \label{eqn:demand_n_smaller_eps+1}
 n\leq 1 + \epsilon  < 2  \; .
\end{align} 
Note that here the control of~$c_+^r$ is very weak, and may thus provide a breaking point
in applications. The components of the boost vectors go like,
\begin{align}
\label{eq:asymptotics_boostvectors}
V_R    &=1+o(R^{-\epsilon})\,,\nonumber\\
V_{A'} &=o(R^{-\epsilon})\,,\nonumber\\
V^R    &=1+o(R^{-\epsilon}) \,,\nonumber\\
V^{A'} &= o(R^{-\epsilon-1}) \,,\nonumber\\
W      &\simeq \frac{R^{\frac{n}{2}}}{S^{\frac{n}{2}}\sqrt{n-1}} \, \big(1+o(R^{-\epsilon})\big)\,,\nonumber\\
Wv_A   &= o(R^{ - \epsilon} ) \,,\nonumber\\
Wv_r   &= - \frac{2 R^n}{ S^n(n-1) } \left(1+o(R^{-\epsilon})\right)\,,\nonumber\\
Wv^r   &\simeq-\frac{1}{2} \left( 1 +o(R^{-\epsilon}) \right) \,,\nonumber\\
Wv^A   &= o(R^{n-\epsilon-1} ) \,.
\end{align}
In fact, we could be even stricter in the angular components of the boost vectors here,
writing~$V_{A'}=0$ and~$Wv_A =0$. The reason for describing these as~$o(R^{-\epsilon})$
is to make the asymptotic properties shown in Eqs~\eqref{eq:asymptotics_boostvectors} directly
transferable to our slice-waggling setup, presented below in Sec~\ref{subsection:LS_control}. The boost 
metric~$\mathbbmss{g}_{ij}$ follows directly from the above so we omit the explicit expressions. It will 
become useful below to consider the quantity~$\Pi^i=W v^i - \alpha^{-1} W \beta^i=-A^{-1}B^{\ul{i}'}\Phi^i{}_{\ul{i}'}$, 
see~Eq~\eqref{eq:upper_case_Pi_abbrev}, with asymptotics,
\begin{align}
 \Pi^r=o(R^{-n-\epsilon}) \;, \qquad \Pi^A=o(R^{-\epsilon-1}) \; .
\end{align}
For convenience, we also state the asymptotics of the inverse projected Jacobians,
\begin{align}
(\varphi^{-1})^{r}{}_{R}      &= \frac{1}{ R'} (1+o(R^{-\epsilon})) ,\nonumber\\
(\varphi^{-1})^{r}{}_{A'} &= o(R^{-n-\epsilon})  ,\nonumber\\
(\varphi^{-1})^{A}{}_{R}      &= o(R^{-\epsilon-1}) ,\nonumber\\
(\varphi^{-1})^{A}{}_{A'} &= {\delta^A}_{A'} + o(R^{-\epsilon-1})   \,,
\end{align} 
where again we have rounded up the strict results~$(\varphi^{-1})^{r}{}_{A'}=0$
and~$(\varphi^{-1})^{A}{}_{A'} = \delta^A{}_{A'}$
for the sake of sustaining validity with our later choice of coordinates. We are now ready to evaluate
the asymptotic behavior in the DF GHG equations of motion~\eqref{eqn:GHG_DF}
under our flatness assumptions~\eqref{eqn:flat_ass1}-\eqref{eqn:flat_ass3} and the
transformation assumptions~\eqref{eqn:compacti_asy_n}-\eqref{eqn:height_asy_n}.

%%%%%%%%%%%%%%%%%%%%%%%%%%%%%%%%%%%%%%%%%%%%%%%%%%%%%%%%%%%%
\paragraph*{Regularity of the principal part:} 
Let us recite the evolution equations~\eqref{eqn:GHG_DF} for convenience and
start with analyzing the coefficients appearing in the principal part. 
The first evolution equation is,
\begin{align}
\p_tg_{\ul{\mu\nu}}&=(\beta^p-\alpha\,v^p)\p_pg_{\ul{\mu\nu}}+\alpha\,W^{-1}s_{\ul{\mu\nu}}^{(g)}\,,
\end{align}
which is regular in the principal part since,
\begin{align}
 {\beta^r-\alpha\,v^r=o(R^{-n-\epsilon})}  \; , \quad {\beta^A-\alpha\,v^A=o(R^{-\epsilon-1})} \; .
\end{align}
To evaluate the regularity in the evolution equations of the reduction variables~$\Phi_{\ul{i'\mu\nu}}$,
we must abandon the concise notation `$d_\mu$', used in Eq~\eqref{eqn:GHG_DF}, and take care of
the additional~$\varphi$ and~$\varphi^{-1}$ factors. After a few simplifications of the resulting
coefficients, the evolution of the radial component~$\Phi_{\ul{R\mu\nu}}$ can be written as,
\begin{align}
& \p_t\Phi_{R\ul{\mu\nu}}=\,
     \alpha\,W^{-1} \, {\gBN^{\ul{p}'}}_{R} (\varphi^{-1})^p{}_{\ul{p}'}            
      \left(
       \gamma_2  \p_p g_{\mu\nu} - \p_p \Pi_{\mu\nu} 
      \right)  \nonumber \\
  & \quad  + \alpha\,W^{-1}
      \left( 
       -W^2 V_R \gamN^{\ul{p}'\ul{j}'} (\varphi^{-1})^p{}_{\ul{p}'}
       - \Pi^p \delta^{\ul{j}'}_R 
      \right)
      \p_p \Phi_{\ul{j'\mu\nu}}  \nonumber \\
  & \quad + \alpha\,W^{-1} s^{(\Phi)}_{R\ul{\mu\nu}} \; .
\label{eq:recitation_hypDF_EvoEq_Phi_R}
\end{align}
To analyze the asymptotics of the coefficients,
first note that,
\begin{align}
 \alpha\,W^{-1}=1+o(R^{-\epsilon}) \; .
\end{align}
Looking at the radial component, we find, 
\begin{align}
 \label{eqn:asymps_coeffs_DFGHG_PhiR_1stline}
 {\gBN^{\ul{p}'}}_{R} (\varphi^{-1})^r{}_{\ul{p}'}
 &\simeq \frac{1}{2} (1+o(R^{-\epsilon})) \, , \nonumber \\
 {\gBN^{\ul{p}'}}_{R} (\varphi^{-1})^A{}_{\ul{p}'}
 &= o(R^{-\epsilon-1+n}) \, ,
\end{align}
where we have used the combined knowledge about asymptotics gathered in the preceding discussion.
Thus the first line of Eq~\eqref{eq:recitation_hypDF_EvoEq_Phi_R} is unproblematic. For the second line
we have to inspect the combinations,
\begin{align}
\label{eqn:asymps_coeffs_DFGHG_PhiR_2ndline}
 -W^2 V_R \gamN^{\ul{p}'R} (\varphi^{-1})^r{}_{\ul{p}'}
 & \simeq -\frac{1}{2} (1+o(R^{-\epsilon})) \, ,  \\
 -W^2 V_R \gamN^{\ul{p}'A'} (\varphi^{-1})^r{}_{\ul{p}'}
 & = o(R^{-\epsilon-1}) \, , \nonumber \\
 -W^2 V_R \gamN^{\ul{p}'R} (\varphi^{-1})^A{}_{\ul{p}'}
 & = o(R^{-\epsilon-1+n}) \, , \nonumber \\
 -W^2 V_R \gamN^{\ul{p}'A'} (\varphi^{-1})^A{}_{\ul{p}'}
 & = O(R^{n-2}) (1+o(R^{-\epsilon})) \; \nonumber ,  
\end{align}
which are manifestly regular. As the~$\Pi^i$ vanish asymptotically, it follows that all coefficients in the
principal part of the radial component of Eq~\eqref{eq:recitation_hypDF_EvoEq_Phi_R} take a regular limit.
The evolution of the angular component can be written in the same form,
\begin{align}
\label{eq:recitation_hypDF_EvoEq_Phi_A}
& \p_t\Phi_{A'\ul{\mu\nu}}=\,
     \alpha\,W^{-1} \, {\gBN^{\ul{p}'}}_{A'} (\varphi^{-1})^p{}_{\ul{p}'}            
      \left(
       \gamma_2  \p_p g_{\mu\nu} - \p_p \Pi_{\mu\nu} 
      \right)  \nonumber \\
  & \quad  + \alpha\,W^{-1}
      \left( 
       -W^2 V_{A'} \gamN^{\ul{p}'\ul{j}'} (\varphi^{-1})^p{}_{\ul{p}'}
       - \Pi^p \delta^{\ul{j}'}_{A'} 
      \right)
      \p_p \Phi_{\ul{j'\mu\nu}}  \nonumber \\
  & \quad + \alpha\,W^{-1} s^{(\Phi)}_{A'\ul{\mu\nu}} \; .
\end{align}
The asymptotics can be seen to be regular analogously to the procedure for the radial component,
i.e.\ by inspecting again the terms of Eqs~\eqref{eqn:asymps_coeffs_DFGHG_PhiR_1stline} and~\eqref{eqn:asymps_coeffs_DFGHG_PhiR_2ndline}
with the downstairs~$R$-components replaced by~$A'$-components everywhere. We omit the explicit repetition
since obviously~$V_{A'}$ is even better behaved asymptotically than~$V_{R}$, and consequently
${\gBN^{\ul{p}'}}_{A'}$ is better behaved than ${\gBN^{\ul{p}'}}_{R}$. Finally we have,
\begin{align}
\p_t\Pi_{\ul{\mu\nu}}=&
    -\alpha \,W \, \gamN^{\ul{p}'\ul{j}'} (\varphi^{-1})^p{}_{\ul{p}'} \p_p \Phi_{\ul{j'\mu\nu}}
    + \beta^p \p_p \Pi_{\ul{\mu\nu}}   \nonumber \\
 &  - \alpha \, \gamma_2 \, v^p \p_p g_{\ul{\mu\nu}}    
    + \alpha \,W^{-1} s^{(\Pi)}_{\ul{\mu\nu}} \,,
\end{align}
where all coefficients have been checked already, taking
account of $\alpha \sim W$ and relations~\eqref{eqn:asymps_coeffs_DFGHG_PhiR_2ndline}.
We therefore conclude that the principal part of the DF
GHG equations of motion~\eqref{eqn:GHG_DF} is regular near and at null-infinity. 
That said, because the terms involved often take the form of ratios of functions blowing up at the same rate, it is clear that care is needed in any numerical implementation. Note also that although it may seem convenient to discard 
the~$\gamma_2$ constraint damping term altogether, numerical experiments show it to be an important 
ingredient to the method~\cite{LinSchKid05}. 

%%%%%%%%%%%%%%%%%%%%%%%%%%%%%%%%%%%%%%%%%%%%%%%%%%%%%%%%%%%%
\paragraph*{Regularity of the source terms:} With the principal part checked,
we are now concerned with the source terms of Eq~\eqref{eqn:GHG_DF}. We assume that,
\begin{align}
\gamma_0=o(R^{-\epsilon})=\gamma_2\,,\qquad H_{\ul{\mu}}=o(R^{-2\epsilon})=\p_{\ul{\nu}} H_{\ul{\nu}}\,,
\end{align}
whereas~$\gamma_3$ and~$\gamma_4$ may be of order unity. For convenience of the reader, we re-quote the
explicit source terms as given already in Eqs~\eqref{eqn:sources_DFGHG},
\begin{align}
\label{eqn:sources_DFGHG_re-quoted}
s^{(g)}_{\ul{\mu\nu}}&=S^{(g)}_{\ul{\mu\nu}}\,,\nonumber\\
s^{(\Phi)}_{\ul{i'\mu\nu}}&=S^{(\Phi)}_{\ul{i'\mu\nu}}+W^2V_{\ul{i}'}
\left(V^{\ul{j}'}S^{(\Phi)}_{\ul{j'\mu\nu}}
+S^{(\Pi)'}_{\ul{\mu\nu}}-\gamma_2S^{(g)}_{\ul{\mu\nu}}\right)\,,\nonumber\\
s^{(\Pi)}_{\ul{\mu\nu}}&=\gamma_2S^{(g)}_{\ul{\mu\nu}}+W^2\left(
V^{\ul{i}'}S^{(\Phi)}_{\ul{i'\mu\nu}}+S^{(\Pi)'}_{\ul{\mu\nu}}-\gamma_2S^{(g)}_{\ul{\mu\nu}}
\right)\,.
\end{align}
In the DF GHG evolution system, Eqs~\eqref{eqn:GHG_DF}, these source terms
are always accompanied with an asymptotic `one', $\alpha W^{-1}\sim 1$,
so the task really is to make sure that the source terms~\eqref{eqn:sources_DFGHG_re-quoted} are regular
at null-infinity. Let us start with the simplest part, the~$g_{\ul{\mu\nu}}$-equation.
Looking back at Eqs~\eqref{eqn:ghg11_sources}, we see that
\begin{align}
 S^{(g)}_{\ul{\mu\nu}}=-\Pi_{\ul{\mu\nu}}\,.
\end{align}
Our assumption~${\Pi_{\ul{\mu\nu}}=o(R^{-\epsilon})}$, made in Eqs~\eqref{eqn:flat_ass1}-\eqref{eqn:flat_ass3},
already guarantees that this source term is regular at null-infinity.
Turning to~$s^{(\Phi)}_{\ul{i'\mu\nu}}$ and~$s^{(\Pi)}_{\ul{\mu\nu}}$ in Eqs~\eqref{eqn:sources_DFGHG_re-quoted}, 
we see first that the `naked' terms~$S^{(\Phi)}_{\ul{i'\mu\nu}}$ and~$\gamma_2S^{(g)}_{\ul{\mu\nu}}$
have to be checked for regularity, but both fall off by our basic assumptions on the GHG variables. 
Furthermore, we see the terms,
\begin{align}
 W ^2 V^{\ul{i}'}S^{(\Phi)}_{\ul{i'\mu\nu}} \; , W^2 S^{(\Pi)'}_{\ul{\mu\nu}}, \; W^2 \gamma_2\,S^{(g)}_{\ul{\mu\nu}} \;.
\end{align}
Here we have a dangerous~$W^2\sim O(R^n)$ factor, so we have to make sure that within these combinations
the growth of the~$W^2$ prefactor can be compensated. That the 
combination~$W^2 \gamma_2\,S^{(g)}_{\ul{\mu\nu}}$ falls off fast enough follows directly from our assumptions.
We thus need to check only the other two.
We start by consulting equations~\eqref{eqn:GHG_sources_shell},
which give the primed source terms and are re-quoted here to present the argument,
\begin{align} 
\label{eqn:GHG_sources_shell_cpy}
 V^{\ul{i}'} S^{(\Phi)}_{\ul{i'\mu\nu}}&= V^{\ul{i}'} (\phi^{\textrm{Sh}})^{\ul{i}}{}_{\ul{i}'}S^{(\Phi)}_{\ul{i\,\mu\nu}}
- V^{\ul{i}'}  A^{-1}\Phi_{\ul{i\,\mu\nu}} B^{\ul{j}}\p_{\ul{j}}(\phi^{\textrm{Sh}})^{\ul{i}}{}_{\ul{i}'}
\,,\nonumber\\
 S^{(\Pi)'}_{\ul{\mu\nu}}&=S^{(\Pi)}_{\ul{\mu\nu}}
-\Phi_{\ul{j'\mu\nu}}\gamN^{\ul{ij}}\p_{\ul{i}}(\Phi^{\textrm{Sh}})^{\ul{j}'}{}_{\ul{j}}\,.
\end{align}
We see that~$S^{(\Phi)}_{\ul{i'\mu\nu}}$ and~$S^{(\Pi)'}_{\ul{\mu\nu}}$ 
are built from $S^{(\Phi)}_{\ul{i\,\mu\nu}}$, $S^{(\Pi)}_{\ul{\mu\nu}}$,
and terms involving the Jacobians between Cartesian and
shell coordinates~$\phi^{\textrm{Sh}}$,$\Phi^{\textrm{Sh}}$.
Therefore, we need to analyze the asymptotics of these Jacobians,
which is most easily done assuming standard spherical polar
coordinates and which then can be transferred to the more general shell
coordinates~Eq~\eqref{eqn:Jacobian_ShellCoordinates}. One finds that 
\textit{at most},
\begin{align}
 & {(\phi^{\textrm{Sh}})^{\ul{i}}}_{R}\sim O(1) \; , \quad 
   {(\phi^{\textrm{Sh}})^{\ul{i}}}_{A'}\sim O(R) \; , \quad 
 \p_{\ul{j}}\phi^{\textrm{Sh}} \sim O(1) \; , \nonumber \\
 &\Phi^{\textrm{Sh}} \sim O(1) \; , \quad 
 \p_{\ul{j}}\Phi^{\textrm{Sh}} \sim O(R^{-1})  \; ,
 \label{eqn:worst_falloffs_ShellJacobians}
\end{align}
where this schematic way of writing has to be understood as considering the `worst' components,
where worst refers to the fall-off-rates.
Thus special care is needed whenever ${(\phi^{\textrm{Sh}})^{\ul{i}}}_{A'}$
is involved because we have to compensate an additional order $O(R)$.
So, using Eqs~\eqref{eqn:worst_falloffs_ShellJacobians},
the source terms~\eqref{eqn:GHG_sources_shell_cpy} can be schematically written as,
\begin{align}
  \label{eqn:GHG_sources_shell_cpy_scheme}
  V^{R} S^{(\Phi)}_{R\ul{\mu\nu}} & \sim 
       (1 + o(R^{-\epsilon})) \left( O(S^{(\Phi)}_{\ul{i\mu\nu}}) +  O(\Phi_{\ul{i\,\mu\nu}}) \; o(R^{-\epsilon}) \right) \; ,\nonumber \\    
  V^{A'} S^{(\Phi)}_{A'\ul{\mu\nu}} & \sim 
       o(R^{-1-\epsilon}) \left( O(R) \, O(S^{(\Phi)}_{\ul{i\mu\nu}})  +  O(\Phi_{\ul{i\,\mu\nu}}) \; o(R^{-\epsilon}) \right)\; ,\nonumber \\      
  S^{(\Pi)'}_{\ul{\mu\nu}} & \sim O(S^{(\Pi)}_{\ul{\mu\nu}} ) +  O(\Phi_{\ul{j'\mu\nu}} ) O(R^{-1})  \; ,
\end{align}
where we have also assumed the worst components of the upper-case spatial metric,
$\gamN^{\ul{ij}} \sim O(1) + o(R^{-\epsilon})$ and $B^{\ul{j}} \sim O(R^{-\epsilon})$.
We have disentangled the composition of~$V^{\ul{i}'}S^{(\Phi)}_{\ul{i'\mu\nu}}$ and~$S^{(\Pi)'}_{\ul{\mu\nu}}$ up
to the point that we need the original source terms, given in Eqs~\eqref{eqn:ghg11_sources}.
It remains to take account of the asymptotics of these original source terms. 
We want to convince the reader that both original source terms can be made,
\begin{align}
 S^{(\Phi)}_{\ul{i\,\mu\nu}} = o(R^{-2 \epsilon}) = S^{(\Pi)}_{\ul{\mu\nu}} .
  \label{eq:asymp_origSources}
\end{align}
This would be enough to cancel the~$O(R^n)$ growth of the $W^2$ factor if we tighten the
demand on~$n$ made in Eq~\eqref{eqn:demand_n_smaller_eps+1} to, 
\begin{align}
  1 < n < 2 \epsilon \; ,
\end{align}
which is always possible for $\epsilon > \frac{1}{2}$.
Glancing back at Eqs~\eqref{eqn:ghg11_sources}, we see that both terms in Eqs~\eqref{eq:asymp_origSources} are
composed primarily of quadratic terms like~$\Pi^2,\Phi^2,\Phi \, \Pi$ and~$\Gamma^2$ plus a few
non-quadratic terms. The fall-off of the quadratic terms is given by our assumptions on the
fall-offs of the fields, Eqs~\eqref{eqn:flat_ass1}. Concretely, we have assumed that $\Phi_{\ul{i\,\mu\nu}}
\sim o(R^{-\epsilon})$ as well as~$ \Pi_{\ul{\mu\nu}}\sim o(R^{-\epsilon})$, so the quadratic terms are at
least~$o(R^{-2\epsilon})$. That the remaining terms are also~$o(R^{-2\epsilon})$ follows from our assumptions
on~$\gamma_0,\gamma_2$ and~$H_{\ul{\mu}}$. 
Finally we have convinced ourselves that the terms in Eqs~\eqref{eqn:sources_DFGHG_re-quoted} are regular.

%%%%%%%%%%%%%%%%%%%%%%%%%%%%%%%%%%%%%%%%%%%%%%%%%%%%%%%%%%%%
\paragraph*{DF formalism vs.\ standard conformal compactification:}
It would be interesting to fully understand the relationship between 
the hyperboloidal DF approach and the standard conformal compactification approach. Here 
we discuss this point only superficially. In the conformal approach one defines a regular 
metric conformally related to the physical metric according to,
\begin{align}
\tilde{g}_{\mu\nu}&=\Omega^2g_{\mu\nu}=\Omega^2J^{\ul{\mu}'}{}_\mu J^{\ul{\nu}'}{}_\nu g_{\ul{\mu'\nu'}}\,,
\end{align}
with~$\Omega=O(R^{-1})$ near null-infinity and a compactification with~${n=2}$.
One can view the conformal factor very roughly as being the magnitude of the inverse 
Jacobian squared. In fact, in the conformal approach the choice $n=2$ is mandatory to 
make things work out. To understand that this is the case, let us consider Minkwoski 
spacetime in standard spherical coordinates. The metric in the hyperboloidal coordinate 
basis then reads,
\begin{align}
 \eta_{\mu\nu} &= 
  \left( \begin{array}{cccc}
   -1  & -H' R' & 0 & 0 \\
  -H' R' & R'^2 (1-H'^2) & 0 & 0  \\
  0 & 0 & R^2 & 0  \\
  0 & 0 & 0 & R^2 \sin^2\theta
  \end{array} \right)   \nonumber \\  
  & \sim 
  \left( \begin{array}{cccc}
   -1  & O(R^n) & 0 & 0 \\
  O(R^n) & \, \; \quad O(R^n) \quad \; \,   & 0 & 0  \\
  0 & 0 & R^2 & 0  \\
  0 & 0 & 0 & R^2 \sin^2\theta
  \end{array} \right)  \; .
\end{align}
In the conformal approach $\tilde{g}_{\mu\nu}=\Omega^2 \eta_{\mu\nu}$ is supposed to be 
regular so we definitely need $\Omega \sim O(R^{-n/2})$ to compensate the~$O(R^{n})$ 
components. For $n>2$ the conformal metric would
asymptotically become singular because~$O(R^2) \Omega^2 \rightarrow 0$, and 
for $n<2$ it would become infinite because $O(R^2) \Omega^2 \rightarrow \infty$.
So we see that the conformal metric is regular only for $n=2$. The argument holds 
analogously for more general spacetimes. We conclude that the conformal approach 
restricts the freedom of the compactification. In our hyperboloidal DF formalism we 
hope to make use of this additional freedom, so to treat initial data with weaker 
fall-off requirements. It thus seems plausible that fixing~$n=2$ in the DF approach 
brings about some conceptual similarity with the conformal approach. For example, in 
both cases one would then need to assume initial data with~$O(R^{-1})$ fall-off of the 
metric's Cartesian components. Recall that when using~$n=2$ in our DF approach we 
needed to replace our assumption~$o(R^{-\epsilon})$ on the fall-off of the metric to the 
stricter condition~$O(R^{-1})$ for regularity of the evolution system, and similarly 
for~$\Pi_{\ul{\mu\nu}}$. Another difference between the two approaches concerns the 
numerical benefits of explicitly regular limits in contrast to implicitly regular 
limits of the form~$O(R)/O(R)$. For example, looking at first derivatives of the 
conformal metric,
\begin{align}
\p_i\tilde{g}_{\mu\nu}&=\Omega^2J^{\ul{\mu}'}{}_\mu J^{\ul{\nu}'}{}_\nu J^{\ul{\alpha}'}{}_i\p_{\ul{\alpha}'}g_{\ul{\mu'\nu'}}
+ 2 \, J^{\ul{\alpha}'}{}_i(\p_{\ul{\alpha}'}\ln\Omega)\,\tilde{g}_{\mu\nu}
\nonumber\\
&\quad+2 \, \Omega^2g_{\ul{\mu'\nu'}}J^{\ul{\alpha}'}{}_i\big(\p_{\ul{\alpha}'}J^{\ul{\mu}'}{}_{(\mu} \big) J^{\ul{\nu}'}{}_{\nu)}\,,
\end{align}
we see that we pick up terms which are formally singular
but which should take a regular limit under our asymptotics assumptions. Such formally singular terms would presumably 
appear in the explicit GHG evolution equations for the conformal metric, which would then need to be carefully processed 
using L`H\^opital's rule, as was done in~\cite{MonRin08}, before numerical implementation. As a matter of fact, when 
using~$n=2$ in our wave equation experiments, without regularizing the variables, we encounter the same problem and find 
formally singular terms in the sources. We wish to avoid computing and using these limits, and therefore advocate using 
the `generalized harmonic basis', for example, for the representation of our variables combined with a hyperboloidal 
compactifition with~$n<2$. Another justification argument for using~$n<2$ is that we may then deal with slower fall-off 
than~$O(R^{-1})$.

%%%%%%%%%%%%%%%%%%%%%%%%%%%%%%%%%%%%%%%%%%%%%%%%%%%%%%%%%%%%
\paragraph*{Heuristic comparison of our flatness assumptions
with the conformal definition of asymptotic flatness:} 
From the textbook~\cite{Wal84} definition of 
asymptotic flatness at null-infinity it follows that there are coordinates~$X^{\ul{\mu}}=(T,X,Y,Z)$, with associated tensor 
basis, in which the metric takes the form,
\begin{align}
g_{\ul{\mu\nu}}&=\eta_{\ul{\mu\nu}}+O(V^{-1})\,,
\end{align}
with~$V \sim T+R$ and~$R$ formed from~$X,Y$ and~$Z$ as usual. Note that~$V$ is not required to be a null coordinate at finite 
value. Defining an asymptotically hyperboloidal time coordinate~$t=T-H(R)$ so that~$H(R)\sim R$ for large~$R$, then on a slice 
of constant~$t$, we have~$O(V^{-1})\sim O(R^{-1})$, and so the fall-off we obtain from the standard definition, which requires 
a suitable conformal compactification, is compatible with what we find necessary in the extreme case~$n=2$. We therefore 
expect that our~$o(R^{-\epsilon})$ definition of asymptotic flatness at null-infinity is truly weaker than the conformal 
definition, although to rigorously show this a more careful consideration may be required. Despite the difficulty, it is desirable 
to show explicitly that given initial data satisfying the~$o(R^{-\epsilon})$ requirements, these asymptotics are propagated, at 
least locally in time, in the development of the data. We leave this to future work. 

%%%%%%%%%%%%%%%%%%%%%%%%%%%%%%%%%%%%%%%%%%%%%%%%%%%%%%%%%%%%
\paragraph*{Preservation of assumption~\eqref{eqn:compacti_asy_n}:} The previous results on the asymptotics of the lower-case variables all 
rest on condition~\eqref{eqn:compacti_asy_n}. If this condition were violated the fragile construction would fail catastrophically; 
the lower case lapse and shift explode and any numerical approximation approach is doomed. Therefore we at least need to see whether 
or not our flatness assumptions imply that~\eqref{eqn:compacti_asy_n} is preserved. Suppose that the condition is fulfilled at 
some~$t=\bar{t}$. Then we can write,
\begin{align}
H'&=\left[1-\frac{S^n(n-1)}{2R^n}+o(R^{-n})\right]\frac{1}{\bar{C}^R_+}\,,
\end{align}
where~$\bar{C}^R_+$ denotes the outgoing lightspeed~$C^R_+,$ resricted to the~$t=\bar{t}$ slice. Plugging this 
into~\eqref{eqn:compacti_asy_n} at some later time~$t$ gives,
\begin{align}
1-C^R_+H'&=\frac{S^n(n-1)}{2R^n}+o(R^{-n})+\frac{\Delta^R_+}{\bar{C}^R_+}\left[1+O(R^{-n})\right]\,,
\end{align}
with~$\Delta^R_+=C^R_+-\bar{C}^R_+$. Evidently the last term on the right hand side can be absorbed by the second 
if~$\Delta^R_+=o(R^{-n})$. Assuming existence of the solution satisfying our flatness assumptions we have,
\begin{align}
\label{eq:change_of_lightspeed}
\Delta^R_+&=\int_{\bar{t}}^t\textrm{d}t'(\p_{t'}C^R_+)=\int_{\bar{t}}^t\textrm{d}t'(\p_TC^R_+)\nonumber\\
&\sim(t-\bar{t})\,o(R^{-1-\delta})=o(R^{-1-\delta})\,,
\end{align} 
where we integrate along curves of fixed~$X^{\ul{i}'}$, or equivalently fixed~$x^i$, and must explicitly use our assumption that 
derivatives of the coordinate lightspeed~$C^R_+$ fall-off faster than generic first derivatives~\eqref{eqn:flat_ass3}. Thus we find 
that choosing~$1<n\leq 1+\delta$ is sufficient to preserve~\eqref{eqn:compacti_asy_n}.

%%%%%%%%%%%%%%%%%%%%%%%%%%%%%%%%%%%%%%%%%%%%%%%%%%%%%%%%%%%%
\subsection{The hyperboloidal DF approach with dynamical lightspeed control}
\label{subsection:LS_control}
%%%%%%%%%%%%%%%%%%%%%%%%%%%%%%%%%%%%%%%%%%%%%%%%%%%%%%%%%%%%

%============================================================
% TABLE I: different n-spatial and N-spatial, unit normal vectors for 2+1 splits
%============================================================
\begin{table*}[t]
\caption{ 
Summary of the different spatial unit normal vectors used in the course of this work
for the various 2+1 splits against $R$, $r$ and $u$ respectively. Here $u$ refers to an outgoing 
null coordinate with $u=t-r$ as used in Sec~\ref{subsection:LS_control} to construct
`waggled' hyperboloidal slices with a dynamical lightspeed control.
}
\centering
  \begin{tabular}[t]{ l | l | l | l | l }
  \hline\hline
    &  {Definition}  &  {Normalization}  & Splits & Note  \\ 
  \hline\hline
    \multirow{2}{*}{N-spatial} &
      $S_{\ul{i}'} := L {}^\textrm{\tiny{(N)}}\!D_{\ul{i}'} R$      &  $L^{-2} := \gamN^{RR}$ &  $\gamN^{\ul{i'j'}}$   &\\
    & $S^+_{\ul{i}'} := -L_+ {}^\textrm{\tiny{(N)}}\!D_{\ul{i}'} u$  &  $L_+^{-2} := \gamN^{uu}$ & $\gamN^{\ul{i'j'}}$ &
    $S^+_{\ul{i}'}= E_+^{-1} V_{\ul{i}'}^+$     \\    
    \hline\hline
    \multirow{2}{*}{n-spatial} &
      $ s_i := l \;  D_i r = -l \;  D_i u $  & $l^{-2} := \gamma^{rr} = \gamma^{uu}$   &  $\gamma^{ij}$ &  \\
    & $ \mathbbmss{s}_i := \mathbbmss{l}\,\mathbbmss{D}_i r = -\mathbbmss{l} \mathbbmss{D}_i u $ & 
    $ \mathbbmss{l}^{-2} := ( \mathbbmss{g}^{-1} )^{rr} = ( \mathbbmss{g}^{-1} )^{uu}$ 
        &  $(\mathbbmss{g}^{-1})^{ij}$ & $\mathbbmss{s}_i =  (\mathbbmss{l}/l) s_i $  \\
  \hline\hline    
  \end{tabular} 
\label{tab:2p1_unit_vectors}
\end{table*}

%=============================================================

The shortcoming of the pure height-function approach, namely the weak control of~$c_+^r$, that we encountered in
the previous subsection can be overcome using the flexibility of the DF formalism. We will investigate here a
simple alternative, namely to `waggle' the hyperboloidal slices in such a way that the outgoing radial coordinate
light speed is fixed at will, whilst maintaining the asymptotics and conclusions of Sec~\ref{subsection:asymptotics}. 

%%%%%%%%%%%%%%%%%%%%%%%%%%%%%%%%%%%%%%%%%%%%%%%%%%%%%%%%%%%%
\paragraph*{Waggled hyperboloidal slices:}  The idea is to use an outgoing null 
coordinate~${u=u(T,R,\vartheta^{A'})}$ in the construction of
the `hyperboloidal' time-coordinate $t$.
Let us therefore relax the height-function transformation of the time coordinate
while fixing the radial compactification as before, and write,
\begin{align}
t=u+r\,,&\qquad r=r(R)\,,\label{eqn:hyp_dynamical}
\end{align}
again with the angular coordinates unchanged~${\vartheta^A=\vartheta^{A'}}$. 
We may still speak of~$t=\rm{const.}$ slices as hyperboloidal
in the sense of~\eqref{eqn:working_def_nullinf}. The transformation~$\eqref{eqn:hyp_dynamical}$
is certainly not a unique good choice, but does seem a natural way to control coordinate
lightspeeds in the hyperboloidal coordinate system.
To see this, we only need to understand
that the demand for~$u$ to be a null coordinate
means $u$ is a solution to the eikonal 
equation,
\begin{align}
 g^{ab}\nabla_au\nabla_bu =0 \; \; .
 \label{eqn:Eikonal_eqn}
\end{align}
Expanding this equation and using the coordinate light speeds
along the radial direction, $c^r_{\pm}=-\beta^r \pm \alpha/l$, the eikonal equation implies,
\begin{align}
 \frac{1}{\alpha^2} \left( 1 - c^r_+ \right) 
                    \left( 1 - c^r_- \right) = 0 \; .
 \label{eq:Eikonal_condition_coordinate_speeds}
\end{align}
Demanding that away from null-infinity~$c^r_- < 0$, 
the outgoing coordinate lightspeed
along~$s^i$ thus has to satisfy,
\begin{align}
c_+^r&=1\, .
\end{align}
Since we did not take care to control the \textit{ingoing}
radial coordinate light speed, the analogue expression for $c^r_-$ is not as 
trivial, see Eq~\eqref{eqn:Cpm_hyp_waggle}. We note though that one 
could further modify the coordinate transformation~\eqref{eqn:hyp_dynamical} 
to control both outgoing and ingoing coordinate light speeds. This would likely 
result in even more symmetric expressions throughout the next paragraphs, but the current 
procedure suffices to arrive at an asymptotically well-behaved hyperboloidal 
evolution system. We therefore leave this further generalization to future work.

%%%%%%%%%%%%%%%%%%%%%%%%%%%%%%%%%%%%%%%%%%%%%%%%%%%%%%%%%%%%
\paragraph*{The optical Jacobian:}  
In practice the waggled hyperboloidal transformation~\eqref{eqn:hyp_dynamical}
requires the construction of the null-coordinate $u$, or rather the associated 
Jacobian. In analytically known spacetimes this is no problem, e.g.\ in Minkowski spacetime, we 
can use $u=T-R$. In a dynamical context one will not be able to give algebraic relations 
\textit{a priori}. Instead, we will have to fix $u$ on the initial slice and then adapt the
transformation dynamically. In fact we exploit here that the DF formalism does not 
require~$u$ explicitly but only the Jacobians. We will therefore evolve the Jacobians of the 
transformation as independent variables such that~$u$ remains a null coordinate.
Instead of evolving~$(\partial_T u,\partial_{\ul{i}'}u)$ directly, we introduce an equivalent set 
of variables,
\begin{align}
V^+_{\ul{i}'} &:=  - \partial_{\ul{i}'}u  = \alpha^{-1}WV_{\ul{i}'} 
+ \Phi^r{}_{\ul{i}'}  \; , \nonumber \\
E_+           &:= N^{\ul{\mu}'} \partial_{\ul{\mu}'} u 
               = \alpha^{-1}W-\Pi^r\; ,
               \label{eqn:optical_Jacobian_Vars}
\end{align}
which we refer to as the `optical Jacobians'. The~$+$ marker here 
denotes that the quantities are associated with the \textit{outgoing} 
null-coordinate, see Sec~IV of the DF paper for a double null foliation,
and may occur as a super- or subscript as required to avoid clashing with indices.
The second equality in~\eqref{eqn:optical_Jacobian_Vars} already
shows how the optical Jacobians are related
to the entries of the DF~Jacobian~\eqref{eqn:J_inv},
which will be specified momentarily.
First note that we can express the eikonal Eq~\eqref{eqn:Eikonal_eqn}
in terms of the new variables,
\begin{align} 
 E_+^2 = \gamN^{\ul{i'j'}} \, V^+_{\ul{i}'}V^+_{\ul{j}'}\,,
 \label{eq:EikonalEquation_OpticalJacobian}
\end{align}
which implies that one could completely drop the variable~$E_+$
in favor of $V^+_{\ul{i}'}$ if one so wishes.
Finally, note the use of the optical Jacobians must be done carefully,
as one would expect coordinate singularities to be a significant
danger if these coordinates are employed in the strong-field region.
Therefore it will also be of interest to develop a natural
way to make the transition between lower and upper case coordinates 
start smoothly at some fixed coordinate radius. This is sketched in 
Sec~\ref{subsection:Dynamical_Layers}.

%%%%%%%%%%%%%%%%%%%%%%%%%%%%%%%%%%%%%%%%%%%%%%%%%%%%%%%%%%%%
\paragraph*{The waggled-hyperboloidal Jacobian:}
We will now repeat the steps of Sec~\ref{subsection:Hyp_Jac}
and compare the explicit Jacobians with the DF Jacobians 
to deduce all needed DF quantities.
For the transformation to waggled hyperboloidal slices, Eq~\eqref{eqn:hyp_dynamical},
we have the inverse Jacobian,
\begin{align}
\label{eqn:J_waghyp}
 J^{-1}_{\rm{waghyp}} &= 
 \left(\begin{array}{cccc}
  \frac{ \p u }{ \p T}           & 0       & 0 & 0 \\
  \frac{ \p u }{ \p R} + R'^{-1} & R'^{-1} & 0 & 0 \\
  \frac{ \p u }{ \p \theta}      & 0       & 1 & 0 \\
  \frac{ \p u }{ \p \phi}        & 0       & 0 & 1 \\
       \end{array}\right)\, , 
\end{align}
with obvious inverse. By comparison with Eqs~\eqref{eqn:J} and~\eqref{eqn:J_inv},
we can again collect the prescriptions that relate the lower case quantities of the 
waggled-hyperboloidal coordinates to the given upper case ones.
As in the height-function approach, we have that the spatial part,
\begin{align}
\Phi^i{}_{\ul{i}'}&\equiv\p_{\ul{i}'}x^i\,,\label{eqn:wag_Jac1}
\end{align}
is given \textit{a priori} algebraically, since $R'$ is given.
Exploiting that the lower case spatial coordinates
are $T$-independent as before, we also have that,
\begin{align}
\Pi^i=-A^{-1}B^{\ul{i}'}\Phi^i{}_{\ul{i}'}\,,\label{eqn:wag_Jac2}
\end{align}
which can be evaluated since we know Eq~\eqref{eqn:wag_Jac1}.
Then, equations~\eqref{eqn:optical_Jacobian_Vars} 
provide,
\begin{align}
  \alpha^{-1}W & = E_+-R'^{-1}A^{-1}B^{R} \; , \nonumber \\
  -\alpha^{-1}WV_{\ul{i}'} &= -V^+_{\ul{i}'} + \Phi^r{}_{\ul{i}'} \; .
  \label{eqn:wag_Jac3}
\end{align}
We thus have at hand all DF-quantities of 
the inverse Jacobian~\eqref{eqn:J_inv} relating
the waggled-hyperboloidal coordinates with the shells-adapted GHG
coordinates.
We will now use these information to compute the
lower case quantities from the upper case quantities
and the optical Jacobians. The upper case boost vectors follow immediately,
\begin{align}
 V_R  &=  \frac{V_R^+ - R'^{-1}} { E_+ + \Pi^r }  \,,\qquad
 V_{A'} =  \frac{V_{A'}^+ } { E_+ + \Pi^r }  \,,
 \label{eqn:waghyp_Vboost}
\end{align}
and one could further use $ \Pi^r = - A^{-1} B^R R'^{-1}$.
Combining equations~\eqref{eqn:wag_Jac1},~\eqref{eqn:wag_Jac3} and
introducing the shorthands,
\begin{align}
\chi_{(\pm)}&=C^R_+ \pm C^R_-\,,\qquad 
\chi_{(\times)}=\frac{2\,C^R_+ C^R_- }{ C^R_+ - C^R_-}\,,
\end{align}
the Lorentz-factor can be written as,
\begin{align}
W = \frac{ (E_+ + \Pi^r)\,\sqrt{ A R'} }
          { 
\Big(  \chi_{(-)} V^+_S  
+ \chi_{(+)}\,E_+ 
+ \chi_{(\times)}\,(LR')^{-1}\Big)^{\frac{1}{2}}}\, \; , 
\label{eqn:waghyp_W}
\end{align}
where we have introduced~$V^+_S := V^+_{\ul{i}'} S^{\ul{i}'}$.
The lapse is given by,
\begin{align}
 \alpha = \frac{ \sqrt{A R'} }
          { 
            \Big( \chi_{(-)}  V^+_S + \chi_{(+)}\,E_+ 
+ \chi_{(\times)}\,(LR')^{-1}\Big)^{\frac{1}{2}}}\, \; .
\end{align}
The lower case length scalar is,
\begin{align}
l&=2 \frac{\Big(
\chi_{(-)}  V^+_S +\chi_{(+)}\,E_+
+\chi_{(\times)}\,(LR')^{-1}
\Big)^{\frac{1}{2}}}
{ \chi_{(-)}  V^+_S +\chi_{(+)}\,E_+}\, \sqrt{A R'}\,,
\label{eqn:waghyp_l}
\end{align}
and putting together the previous relations we find that the radial component 
of the shift is,
\begin{align}
\beta^r&=-\frac{1}{2}-\frac{\chi_{(\times)}(LR')^{-1}}{2\left( \chi_{(-)} V^+_S +\chi_{(+)}\,E_+ 
+\chi_{(\times)}\,(LR')^{-1}\right)}\,.
\end{align}
This is all we need in order to compute the coordinate light speeds according to their 
general definitions, see the first equality in equations~\eqref{eqn:cpm_hyp}.
We will use two more shorthands,
\begin{align}
\chi_{(\pm)}^{A'}&=\frac{C_+^{A'}\pm C_-^{A'}}{L(C_+^R-C_-^R)}\,,\quad 
\chi_{(\times)}^{A'}= 2\frac{C_-^RC_+^{A'}+ C_+^RC_-^{A'}}{L(C_+^R-C_-^R)^2}\,.
\end{align}
The angular components of the shift are,
\begin{align}
\beta^A&=-\frac{\left(V_+^{A'}+E_+\,\chi_{(+)}^{A'}+(LR')^{-1}\chi_{(\times)}^{A'}\right)}
{ \chi_{(-)} V^+_S + \chi_{(+)}\,E_+ + \chi_{(\times)}\,(LR')^{-1}}\,AR'\,.
\end{align}
The slip vector is,
\begin{align}
  b^A&= \frac{2 AR'}{ \chi_{(-)} V_S^++\chi_{(+)}E_+}\left( \chi_{(\times)}^{A'} (L R')^{-1} - V_+^{A'}
  - E_+\,\chi_{(+)}^{A'}\right)\nonumber\\
&\quad -\,\frac{2\,\chi_{(\times)} \chi_{(-)}}
{\big(\chi_{(-)} V_S^++\chi_{(+)}E_+\big)^2}\left(V_+^{A'}+E_+\,\chi_{(+)}^{A'}\right)\,,
\end{align}
and, defining,
\begin{align}
  \hat{\chi}_{(\pm)}^{A'}&=\, -\chi_{(-)}^{A'} \pm \frac{ V_S^+ \pm E_+  }{ \chi_{(-)} V_S^+
    + \chi_{(+)} E_+} \chi_{(-)} \chi_{(+)}^{A'} \, \nonumber \\
& \quad + \frac{\chi_{(-)} \mp \chi_{(+)} }{ \chi_{(-)} V_S^+ +\chi_{(+)} E_+} V_+^{A'}
\end{align}
the two-metric is,
\begin{align}
q^{AB}&=\qN^{A'B'}+ \hat{\chi}_{+}^{(A'} \hat{\chi}_{-}^{B')}\,.
\end{align}
The boost vector can be deduced from the definition of~$\Pi^i$,
\begin{align}
Wv^i&=\Pi^i+\alpha^{-1}W\beta^i\,.
\end{align}
Concerning the coordinate lightspeeds, we find that, as expected,~$c^r_+ = 1$. For the ingoing direction we obtain,
\begin{align}  
c^r_-&= \frac{\chi_{(\times)}(LR')^{-1}}{ \chi_{(-)} V^+_S+\chi_{(+)}\,E_+ 
+\chi_{(\times)}\,(LR')^{-1}}\, .
\label{eqn:Cpm_hyp_waggle}
\end{align}
We omit explicit expressions for the  angular lightspeeds and boost metric, but they are easily constructed from the 
foregoing results. 

%%%%%%%%%%%%%%%%%%%%%%%%%%%%%%%%%%%%%%%%%%%%%%%%%%%%%%%%%%%%
\paragraph*{Evolution subsystem for the optical Jacobian:}  
It remains to establish evolution equations for $E_+,V^+_{\ul{i}'}$,
which we deal with next. We can easily find the evolution of $V^+_{\ul{i}'}$ by
exploiting that it is defined as a gradient of the scalar $u$,
and using the hypersurface constraints, see~\eqref{eqn:Hyp_Cons},
\begin{align}
 \partial_T V^+_{\ul{i}'} = - \partial_{\ul{i}'} \partial_T u \;.
 \label{eqn:idea_evo_V+}
\end{align}
Then, we rewrite $\partial_T u$ in terms of $E_+,V^+_{\ul{i}'}$,
by virtue of the definition of $E_+$ in~\eqref{eqn:optical_Jacobian_Vars},
\begin{align}
 \partial_T u = A E_+ - B^{\ul{i}'} V^+_{\ul{i}'} \; .
\end{align}
This gives the evolution equation,
\begin{align}
   \partial_T V^+_{\ul{i}'} =
   - A {}^\textrm{\tiny{(N)}}\!D_{\ul{i}'} E_+
   - E_+ {}^\textrm{\tiny{(N)}}\!D_{\ul{i}'} A
   + {}^\textrm{\tiny{(N)}}\!D_{\ul{i}'} ( B^{\ul{j}'} V_{\ul{j}'}^+ ) \; ,
   \label{eqn:intermediate_dtV+}
\end{align}
which will be manipulated a bit more momentarily.
The missing equation for $E_+$ is found from the
condition that~$u$ is a null coordinate,
which means $u$ is a solution to the eikonal equation.
We can thus employ~\eqref{eq:EikonalEquation_OpticalJacobian}
to rewrite spatial derivatives of $E_+$,
\begin{align}
 {}^\textrm{\tiny{(N)}}\!D_{\ul{i}'} E_+  
   &= E_+^{-1} V_+^{\ul{j}'} {}^\textrm{\tiny{(N)}}\!D_{\ul{i}'} V_{\ul{j}'}^+ 
    = E_+^{-1} V_+^{\ul{j}'} {}^\textrm{\tiny{(N)}}\!D_{\ul{j}'} V_{\ul{i}'}^+ \nonumber \\
   &=:S_+^{\ul{j}'} {}^\textrm{\tiny{(N)}}\!D_{\ul{j}'} V_{\ul{i}'}^+ 
   \equiv    {}^\textrm{\tiny{(N)}}\!D_{(S_+)} V_{\ul{i}'}^+ \; ,
   \label{eqn:DiE+toDiVj}
\end{align}
Here we have introduced the spatial unit normal vector $S^+_{\ul{i}'}$
associated with a 2+1 split against $u$,
\begin{align}
 S^+_{\ul{i}'} := - L^+ {}^\textrm{\tiny{(N)}}\!D_{\ul{i}'} u = L^+ V^+_{\ul{i}'} \; . 
 \label{eqn:S+_unit_against_u}
\end{align}
Note that the normalization together with the eikonal equation implies $L^+ = E_+^{-1}$.
Coming back to the search for $\partial_T E_+$,
we evaluate $\partial_T E_+^2$ according to~\eqref{eqn:Eikonal_eqn},
use~\eqref{eqn:intermediate_dtV+} to remove the appearance of~$\partial_T V^+_{\ul{i}'}$,
and express $\partial_T \gamN^{\ul{i'j'}}$ through the extrinsic 
curvature~$\KN_{\ul{i'j'}}$ of the constant~$T$-slice.
After a few steps we arrive at the following
equations of motion for the optical Jacobians,
\begin{align}
\p_T\ln E_+&=\Lie_{(B-AS_+)}\ln E_++A\big(\KN_{S_+S_+}-\Lie_{S_+}\ln A\big)\,,\nonumber\\
\p_TV^+_{\ul{i}'}&=-A\DN_{(S_+)}V^+_{\ul{i}'}-E_+\DN_{\ul{i}'}A+\Lie_BV^+_{\ul{i}'}
\,.\label{eqn:Opt_Jac_EOM}
\end{align}
In view of the reduction constraints~\eqref{eqn:ghg11_reduction}
we can treat derivatives of $A$ and $B^{\ul{i}'}$ as source terms so that
the subsystem~\eqref{eqn:Opt_Jac_EOM} is only minimally coupled to the first order GHG system,
as hoped for.
The second equation can be further abbreviated
using 
\begin{align}
 {}^\textrm{\tiny{(N)}}\!D_{\ul{i}'} E_+ 
 = {}^\textrm{\tiny{(N)}}\!D_{\ul{i}'}  \left( V^+_{\ul{j}'} S_+^{\ul{j}'} \right)
 \; , \label{eqn:helper_to_get_dTV+_withC}
\end{align}
so that the evolution system becomes,
\begin{align}
\p_T \ln E_+&=  \left( B^{\ul{j}'} - A S_+^{\ul{j}'} \right) 
               \partial_{\ul{j}'} \ln E_+ 
             + A \, S^{(E_+)} \,,\nonumber\\
\p_T V^+_{\ul{i}'}&= \left( B^{\ul{j}'} - A S_+^{\ul{j}'} \right) 
               \partial_{\ul{j}'} V^+_{\ul{i}'} 
             + A \, S^{(V^+)}_{\ul{i}'} \; .  
\label{eqn:Opt_Jac_EOM_advectionform}
\end{align}
The resulting equations are thus simply advection 
equations, and so the characteristic variables are trivially constructed.
The source terms are given by,
\begin{align}
 S^{(E_+)}           & =  \KN_{S_+S_+} - \Lie_{S_+}\ln A \, , \nonumber \\
 S^{(V^+)}_{\ul{i}'} & =  A^{-1}  V^+_{\ul{j}'} \p_{\ul{i}'} 
                      \big( B^{\ul{j}'} - A S_+^{\ul{j}'} \big) \, .               
\end{align}
For $S^{(V^+)}_{\ul{i}'}$ we have used,
\begin{align}
  E_+ {}^\textrm{\tiny{(N)}}\!D_{\ul{i}'} A \equiv (V^+_{\ul{j}'} S_+^{\ul{j}'}) {}^\textrm{\tiny{(N)}}\!D_{\ul{i}'}A
     = V^+_{\ul{j}'} {}^\textrm{\tiny{(N)}}\!D_{\ul{i}'} \big( A S_+^{\ul{j}'} \big) \; ,
\end{align}
where the second equality follows from the fact that $ \gamN_{\ul{i'j'}} S_+^{\ul{i}'} S_+^{\ul{j}'}=1$
so that $V^+_{\ul{j}'} A  {}^\textrm{\tiny{(N)}}\!D_{\ul{i}'} S_+^{\ul{j}'} =0 $.
Thus, we have everything for the DF transformation.

%%%%%%%%%%%%%%%%%%%%%%%%%%%%%%%%%%%%%%%%%%%%%%%%%%%%%%%%%%%%
\paragraph*{Dual foliation transformation of the optical Jacobian equations of motion:} 
Before transforming the subsystem~\eqref{eqn:Opt_Jac_EOM_advectionform},
recall that, as observed above, the optical Jacobians are coupled only through 
source terms to the rest of the equations of motion.
Therefore, the form of the DF GHG equations~\eqref{eqn:GHG_DF}
is unaffected by their presence,
apart from the various coefficients taking different values.
Obviously equations~\eqref{eqn:Opt_Jac_EOM_advectionform} are of the form~\eqref{eqn:1st_Upper}.
So let us apply the DF recipe~\eqref{eqn:Change_Coord}.
We read off the principal matrices and contract with $V_{\ul{i}'}$,
\begin{align}
   \mathbf{A}^{\ul{p}'} &= -S_+^{\ul{p}'} \mathbf{1} \; , \nonumber \\
  \Rightarrow 
  (\mathbf{1}+\mathbf{A}^{\ul{V}}) &= (1-S_+^{\ul{j}'} V_{\ul{j}'}) \mathbf{1} \nonumber \\
                                   &= (1+ W v_j(\varphi^{-1})^j{}_{\ul{i}'}\,S_+^{\ul{i}'} )  \mathbf{1} \; .
  \label{eqn:inverse_factor_DF_opticalJac_subsystem}
\end{align}
Computing the optical Jacobian equations of motion in the lower case coordinates, it becomes helpful to note that the
spatial unit normal vector $s_i$ of~\eqref{eq:def_gamma_metric_2+1normal}
with upper indices can be written as,
\begin{align}
s^i &=\frac{ (\varphi^{-1})^i{}_{\ul{i}'}\,S_+^{\ul{i}'} }{W(1+Wv_j (\varphi^{-1})^j{}_{\ul{i}'}\,S_+^{\ul{i}'})}+v^i\,,
\label{eq:unitvector_ploppingout_opticalJacobianEOMs}
\end{align}
It is then straightforward to calculate that the equations of motion
remain advection-like,
\begin{align}
\p_t\ln E_+ &=(\beta^p-\alpha\,s^p) \p_p \ln E_+       +\alpha\,W^{-1}\, s^{(E_+)}\,,\nonumber\\
\p_tV^+_{\ul{i}'}&=(\beta^p-\alpha\,s^p) \p_p V^+_{\ul{i}'} +\alpha\,W^{-1}s^{(V_+)}_{\ul{i}'}\,.
\label{eqn:DF_Opt_Jac_EOM}
\end{align}
The source terms follow immediately,
\begin{align}
s^{(E_+)}        &=\big(1+Wv_j\, (\varphi^{-1})^j{}_{\ul{j}'}\,S_+^{\ul{j}'} \big)^{-1}S^{(E_+)}\,,\nonumber\\
s^{(V_+)}_{\ul{i}'}&=\big(1+Wv_j\, (\varphi^{-1})^j{}_{\ul{j}'}\,S_+^{\ul{j}'} \big)^{-1} S^{(V^+)}_{\ul{i}'}\,,
\end{align}
which, as noted above, must be evaluated in terms of the GHG reduction variables without taking derivatives.
This can be achieved by rewriting the term involving the gradient of~$S_+^{\ul{i}'}$. Since these source terms 
are linear in the generalized harmonic connection coefficients and do not come with any parameters we can tune,
we already see that these terms will be the trickiest for maintaining regularity. 

%%%%%%%%%%%%%%%%%%%%%%%%%%%%%%%%%%%%%%%%%%%%%%%%%%%%%%%%%%%%
\paragraph*{Asymptotics of the lower case quantities:} Finally, to convince ourselves that 
the waggled-hyperboloidal coordinate transformations lead to regular equations at future 
null infinity, we can again check the asymptotic behavior of the just computed lower case 
quantities. We therefore assume,
\begin{align}
  E_+ &= 1 + o(R^{-\epsilon}) \,, \quad
  V^+_{R} = 1 + o(R^{-\epsilon}) \, , \nonumber \\
  V^+_{A'} &= o(R^{-\epsilon}) \, ,
 \label{eqn:optical_Jacs_ass}
\end{align}
on the new evolution variables $E_+, V^+_{\ul{i}'}$, complementing the assumptions made in 
Sec~\ref{subsection:asymptotics} on the GHG variables. This is consistent with what we expect to have in a
blackhole spacetime. For example, for the Schwarzschild blackhole we can explicitly construct the outgoing
null coordinate~${u=T-R^*}$, where~${R^*=R+2 M \ln(R/2M-1)}$ is the tortoise coordinate.
Then we can compute~${E_+=1+O(R^{-1})}$,~${V^+_{R}=1+O(R^{-1})}$, and~${V^+_{A'}=0}$,
which shows that our restrictions~\eqref{eqn:optical_Jacs_ass}
are even weaker than needed for this special case.
Now, employing the same assumptions for the radial
compactification as before, i.e. Eq~\eqref{eqn:compacti_asy_n},
we are ready to compute the asymptotics of the lower
case quantities by inspection of 
equations~\eqref{eqn:waghyp_Vboost}-\eqref{eqn:Cpm_hyp_waggle}. Remarkably it turns out 
that the results are {\it unchanged} with respect to the pure height-function 
approach, as presented in Sec~\ref{subsection:asymptotics}. 

%%%%%%%%%%%%%%%%%%%%%%%%%%%%%%%%%%%%%%%%%%%%%%%%%%%%%%%%%%%%
\paragraph*{Asymptotics of the optical Jacobian system:} We must additionally consider the limiting behavior 
of the new part of the evolution system, Eq~\eqref{eqn:DF_Opt_Jac_EOM}. In the principal part we have to 
consider,
\begin{align}
\beta^r-\alpha\,s^r&=-c_+^r=-1\,,\nonumber\\
\beta^A-\alpha\,s^A&=-c_+^A=o(R^{-\epsilon-1+n})\,,
\end{align}
which, by construction, are regular. For the source terms we have to look more carefully. First note that from 
our basic flatness assumptions it follows that shell-coordinate derivatives of the metric represented in the 
shell-coordinate basis satisfy,
\begin{align}
\p_{\ul{i}'}L&=o(R^{-\epsilon})\,,\nonumber\\
\p_{\ul{i}'}\bN_{A'}&=o(R^{1-\epsilon})\,,\nonumber\\
\p_{\ul{i}'}\qN_{A'B'}&=o(R^{2-\epsilon})\,.
\end{align}
Moreover the difference~$\delta^{\ul{i}'}=S_+^{\ul{i}'}-S^{\ul{i}'}$ between the two unit upper case spatial normal 
vectors satisfies,
\begin{align}
\delta^{\ul{i}'}&=o(R^{-1-2\epsilon})=\delta_R\,,\qquad \delta_{A'}=o(R^{-\epsilon})\,.
\end{align}
Consider therefore the source term~$S^{(V^+)}_{\ul{i}'}$, which we can manipulate as follows,
\begin{align}
&AS^{(V^+)}_{\ul{i}'}=V^+_{\ul{j}'}\p_{\ul{i}'}B^{\ul{j}'}-E_+ \p_{\ul{i}'}A 
+ AE_+\Gamma^{S_+}{}_{S_+\ul{i}'} \nonumber\\
&\,\,=-\frac{L}{L_+}\p_{\ul{i}'}C^R_++\frac{1}{L_+}\delta_{\ul{j}'}\p_{\ul{i}'}B^{\ul{j}'}
+\frac{A}{L_+}\,\delta^{\ul{j}'}\,S^{\ul{k}'}\,\p_{\ul{i}'}\gamN_{\ul{j'k'}}\nonumber\\
&\,\,\quad+\frac{A}{L_+}\delta^{\ul{j}'}\delta^{\ul{k}'}\Gamma_{\ul{j'k'i}'}\nonumber\\
&\,\,=o(R^{-1-\delta})\,,
\end{align}
where in this equation~$\Gamma^{\ul{k}'}{}_{\ul{i'j'}}$ denotes the upper case spatial Christoffel symbol in 
shell-coordinates. As in the height-function approach we needed to use the fast fall-off of derivatives 
of~$C_+^R$ and ultimately require that~$1<n<1+\delta$. Notice here that if we assume fall-off 
like~$\p_{\ul{i}'}C_+^R=O(R^{-2})$ instead of~$\p_{\ul{i}'}C_+^R=o(R^{-1-\delta})$ in the harmonic basis and insist 
on compactifying with~$n=2$, we will have to compute the regular limits using L`H\^opital's rule. This again 
reinforces the view that we should choose~$1<n<2$. Since~$ E_+^2=\gamN^{\ul{i'j'}}\,V^+_{\ul{i}'}V^+_{\ul{j}'}$ we 
need not discuss the final source term~$S^{(E_+)}$, although we expect it can be treated with similar arguments 
to those above by noting the specific form of~$\p_TC_+^R$. It is remarkable that the same flatness assumptions 
allowing the use of the height-function method work for the optical-Jacobians. It is not clear how to weaken 
the requirement on~$C_+^R$. Whatever the alternative approach it seems that source terms with weaker decay 
will have to be carried through the transformation to the hyperboloidal slices at some point in the setup and 
this will break the asymptotics.

%%%%%%%%%%%%%%%%%%%%%%%%%%%%%%%%%%%%%%%%%%%%%%%%%%%%%%%%%%%%
\subsection{The Dynamical Transition Layer}\label{subsection:Dynamical_Layers}
%%%%%%%%%%%%%%%%%%%%%%%%%%%%%%%%%%%%%%%%%%%%%%%%%%%%%%%%%%%%

We have now presented two candidate methods for the use of hyperboloidal slices using a radial 
compactification in combination with the DF formalism. The only open issue for the optical Jacobians
is how to transition from the harmonic coordinates~$X^{\ul{\mu}}$, or the associated shell 
coordinates~$X^{\ul{\mu}'}$, in the central region to the hyperboloidal coordinates~$x^\mu$ in the exterior. 
The aim of this section is thus to generalize the `hyperboloidal layers' transition approach outlined in 
Sec~\ref{subsection:layers} for use with the optical Jacobians of Sec~\ref{subsection:LS_control}.

%%%%%%%%%%%%%%%%%%%%%%%%%%%%%%%%%%%%%%%%%%%%%%%%%%%%%%%%%%%%
\paragraph*{Generalizing the optical Jacobians:} It is actually not difficult to perform the 
transition from harmonic to hyperboloidal coordinates. Let us reconsider the derivation 
that led to the optical Jacobian equations of motion~\eqref{eqn:Opt_Jac_EOM_advectionform},
which started essentially from the requirement that~$u$ be a solution to the eikonal 
equation. Now we replace this condition with,
\begin{align}
g^{uu}&=\chi\,, \label{eqn:transition_layer_condition}
\end{align}
where~$\chi$ is some scalar function to be specified momentarily.
First we introduce~$U:=T-R$, bearing in mind that for any asymptotically
Minkowskian spacetime~$U$ is a first approximation to a null-coordinate.
We can roughly think of~$g^{uu}$ as proportional to~$g^{UU}$, i.e.,
\begin{align}
 \chi = g^{UU} \tilde{\chi} \;.  \label{eqn:transition_cutoff_function}
\end{align}
where $\tilde{\chi}=\tilde{\chi}(r)$ is some suitably chosen transition function in~$r$.
This way~$u$ will be a null-coordinate in the region where~$\tilde{\chi}=0$,
and $u=U$ in the region where $\tilde{\chi}=1$. The intermediate region, 
where $\tilde{\chi}\neq0$ and $\tilde{\chi}\neq1$, is called the \textit{transition-layer},
and serves to smoothly connect the hyperboloidal region with the standard GHG interior.
Defining~$E_+,V^+_{\ul{i}'}$ as 
before in Eq~\eqref{eqn:optical_Jacobian_Vars} and following exactly the same steps as after 
Eq~\eqref{eqn:Opt_Jac_EOM} we obtain,
\begin{align}
\p_T \ln E_+&=  \left( B^{\ul{j}'} - A S_+^{\ul{j}'} \right) 
               \partial_{\ul{j}'} \ln E_+ 
             + A \, S^{(E_+)} \,,\nonumber\\
\p_T V^+_{\ul{i}'}&= \left( B^{\ul{j}'} - A S_+^{\ul{j}'} \right) 
               \partial_{\ul{j}'} V^+_{\ul{i}'} 
             + A \, S^{(V^+)}_{\ul{i}'} \,,  
\label{eqn:Gen_Opt_Jac_EOM_advectionform}
\end{align}
which differs from Eq~\eqref{eqn:Opt_Jac_EOM_advectionform} only
through the slightly modified sources,
\begin{align}
 S^{(E_+)}           & =   \KN_{S_+S_+} - \Lie_{S_+}\ln A -\tfrac{1}{2}E_+^{-2}\p_N\chi \, , \nonumber \\
 S^{(V^+)}_{\ul{i}'} & =  A^{-1}  V^+_{\ul{j}'} \p_{\ul{i}'} 
                      \big( B^{\ul{j}'} - A S_+^{\ul{j}'} \big)+\tfrac{1}{2}E_+^{-1}\DN_{\ul{i}'}\chi \, .               
\end{align}
Although derivatives of~$\chi$ appear, we can use the GHG reduction constraints~\eqref{eqn:ghg11_reduction} 
to rewrite them in terms of the reduction variables, and thus remove any potentially dangerous coupling in 
the principal part. These equations thus transform in the obvious way, as in~\eqref{eqn:DF_Opt_Jac_EOM}, when 
moving to~$x^\mu$ coordinates. We propose to perform the transition from the time coordinate~$T$ to~$t$ in a 
region with~$R=r$, before starting the compactification in the region where~$u$ is a true optical function. 
This is the generalization of the hyperboloidal layers approach to a slice-waggling setup. We expect that in 
practice coming up with a good choice for the transition and compactification will require some experimentation 
so we will not give the construction explicitly here. 

%%%%%%%%%%%%%%%%%%%%%%%%%%%%%%%%%%%%%%%%%%%%%%%%%%%%%%%%%%%%
\subsection{Regularization}\label{subsection:Regularization}
%%%%%%%%%%%%%%%%%%%%%%%%%%%%%%%%%%%%%%%%%%%%%%%%%%%%%%%%%%%%

%%%%%%%%%%%%%%%%%%%%%%%%%%%%%%%%%%%%%%%%%%%%%%%%%%%%%%%%%%%%
\paragraph*{Renormalized variables:} We now start to investigate to what extent the DF GHG evolution
equations can be regularized {\it explicitly}. With the assumed fall-off rates in mind we define the variables,
\begin{align}
\label{eq:rescaled_fields_GHG}
\tilde{h}_{\ul{\mu\nu}}&=R^\delta\,(g_{\ul{\mu\nu}}-\eta_{\ul{\mu\nu}})\,,\nonumber\\
\quad\tilde{\Phi}_{\ul{i'\mu\nu}}&=R^\delta\,\Phi_{\ul{i'\mu\nu}}
+\delta\,R^{-1-\delta}\,\delta^R{}_{\ul{i}'}h_{\ul{\mu\nu}}\,,\nonumber\\
\tilde{\Pi}_{\ul{\mu\nu}}&=R^\delta\,\Pi_{\ul{\mu\nu}}\,,
\end{align}
with~$\epsilon>\delta\geq0$ some constant. In the special case that the error terms in our asymptotics
assumptions can be taken as~$O(R^{-1})$ we may wish to set~$\delta=1$, but this is only possible if the nonlinear
terms have sufficient decay, which we consider in a few paragraphs. It would not be sensible to use
these variables in the strong-field region, but to avoid doing so a modification of the layers approach can be taken.
The crux of the regularization strategy, which is also described for the wave equation in Sec~\ref{subsection:WE},
is to pull as much decay as can be expected out of the evolved variables to try and obtain an evolution system with
coefficients as close to unity as possible. 

%%%%%%%%%%%%%%%%%%%%%%%%%%%%%%%%%%%%%%%%%%%%%%%%%%%%%%%%%%%%
\paragraph*{An alternative first order GHG system:} In shell-coordinates we choose a first order reduction of the GHG
system according to,
\begin{align}
\p_T\tilde{h}_{\ul{\mu\nu}}&=B^{\ul{i}'}\p_{\ul{i}'}\tilde{h}_{\ul{\mu\nu}}+A\,S^{(\tilde{h})}_{\ul{\mu\nu}}\,,\nonumber\\
\p_T\tilde{\Phi}_{\ul{i'\mu\nu}}&=B^{\ul{j}'}\p_{\ul{j}'}\tilde{\Phi}_{\ul{i'\mu\nu}}-A\,\p_{\ul{i}'}\tilde{\Pi}_{\ul{\mu\nu}}
+\gamma_2\,A\,\p_{\ul{i}'}\tilde{h}_{\ul{\mu\nu}}\nonumber\\
&\quad+A\,\,S^{(\tilde{\Phi})}_{\ul{i'\mu\nu}}\,,\nonumber\\
\p_T\tilde{\Pi}_{\ul{\mu\nu}}&=B^{\ul{i}'}\p_{\ul{i}'}\tilde{\Pi}_{\ul{\mu\nu}}-A\,\gamN^{\ul{i'}\ul{j'}}
\,\p_{\ul{i}'}\tilde{\Phi}_{\ul{j'\mu\nu}}+A\,S^{(\tilde{\Pi})}_{\ul{\mu\nu}}\,,\label{eqn:GHG_Reg}
\end{align}
with source terms,
\begin{align}
&S^{(\tilde{h})}_{\ul{\mu\nu}}=-\tilde{\Pi}_{\ul{\mu\nu}}+\delta\,R^{-1}N^R\,\tilde{h}_{\ul{\mu\nu}}\,,\nonumber\\
&S^{(\tilde{\Phi})}_{\ul{i'\mu\nu}}=-\gamma_2\,\tilde{\Phi}_{\ul{i'\mu\nu}}
+R^{-\delta}\big(\tfrac{1}{2}\,\hat{\Phi}_{\ul{i'}NN}\,\tilde{\Pi}_{\ul{\mu\nu}}+\gamN^{\ul{jk}}
\,\hat{\Phi}_{\ul{i'j}N}\,\hat{\Phi}_{\ul{k\,\mu\nu}}\big)\nonumber\\
&\,\,+\hat{\Phi}_{\ul{k'\mu\nu}}N^{\ul{j}}
\left(\p_{\ul{i}'}(\Phi^{\textrm{Sh}})^{\ul{k}'}{}_{\ul{j}}
-\delta\,R^{-1}\gamN^{\ul{k}'}{}_{\ul{i}'}(\Phi^{\textrm{Sh}})^{R}{}_{\ul{j}}\right)\,,
\end{align}
and finally the more complicated,
\begin{align}
&R^\delta S^{(\tilde{\Pi})}_{\ul{\mu\nu}}=
 -2\,\left(\tilde{\nabla}_{(\ul{\mu}}H_{\ul{\nu})}+\gamma_3\,\tilde{\Gamma}^{\ul{\alpha}}{}_{\ul{\mu\nu}}\tilde{C}_{\ul{\alpha}}
 -\tfrac{1}{2}\gamma_4\,g_{\ul{\mu\nu}}\tilde{\Gamma}^{\ul{\alpha}}\tilde{C}_{\ul{\alpha}}\right) \nonumber\\
&\,\,\,+2\,g^{\ul{\alpha\beta}}\,\left(\gamN^{\ul{ij}}\,
 \hat{\Phi}_{\ul{i\,\alpha\mu}}\,\hat{\Phi}_{\ul{j\,\beta\nu}}-\tilde{\Pi}_{\ul{\alpha\mu}}\,\tilde{\Pi}_{\ul{\beta\nu}}
-g^{\ul{\delta\gamma}}\tilde{\Gamma}_{\ul{\mu\alpha\delta}}\tilde{\Gamma}_{\ul{\nu\beta\gamma}}\right)\nonumber\\
&\,\,\,+R^{\delta-1}N^R\tilde{\Pi}_{\ul{\mu\nu}}-\tfrac{1}{2}\tilde{\Pi}_{NN}\tilde{\Pi}_{\ul{\mu\nu}}
-\gamN^{\ul{ij}}\,\tilde{\Pi}_{N\ul{i}}\,\hat{\Phi}_{\ul{j\,\mu\nu}}\nonumber\\
&\,\,\,-R^\delta\,\hat{\Phi}_{\ul{i'\mu\nu}}
\left(\gamN^{\ul{ij}}\,\,\p_{\ul{i}}(\Phi^{\textrm{Sh}})^{\ul{i}'}{}_{\ul{j}}
-(\delta+1)R^{-1}\,\gamN^{R\ul{i}'}\right)
\nonumber\\
&\,\,\,-\tfrac{\delta(\delta-1)}{R^2L^2}\tilde{h}_{\ul{\mu\nu}}
+\tilde{\gamma}_0\,\left(2\,\delta^{\ul{\alpha}}{}_{(\ul{\mu}}N_{\ul{\nu})}
-g_{\ul{\mu\nu}}\,N^{\ul{\alpha}}\right)\tilde{C}_{\ul{\alpha}}\,.\label{eqn:GHG_S_Reg}
\end{align}
The trick here is that the~$\delta$ parameter controls the coefficient of the leading order term in~$R$ near 
null-infinity. In particular the terms grouped together involving~$\p_{\ul{i}}(\Phi^{\textrm{Sh}})^{\ul{i}'}{}_{\ul{j}}$ 
cancel when~$\delta=1$. This can be seen explicitly for the wave equation in Sec~\ref{subsection:WE}. Terms appearing 
with a free-parameter can be given the desired fall-off, and since the remainder is quadratic, we gain fall-off in~$R$. 
Here we have introduced the shorthand~$\tilde{\Gamma}_{\ul{\alpha\mu\nu}}=R^\delta\,\Gamma_{\ul{\alpha\mu\nu}}$, which in terms 
of the evolved variables is, 
\begin{align}
\tilde{\Gamma}_{\ul{\alpha\mu\nu}}&=
\gamN^{\ul{i}}{}_{(\ul{\mu}|}\tilde{\Phi}_{\ul{i}\,|\ul{\nu})\ul{\alpha}}
-\tfrac{1}{2}\gamN^{\ul{i}}{}_{\ul{\alpha}}\tilde{\Phi}_{\ul{i\,\mu\nu}}+N_{(\ul{\mu}}\tilde{\Pi}_{\ul{\nu})\ul{\alpha}}
-\tfrac{1}{2}N_{\ul{\alpha}}\tilde{\Pi}_{\ul{\mu\nu}}\nonumber\\
&\,\,-\delta\,R^{-1}\gamN^R{}_{(\ul{\mu}}\tilde{h}_{\ul{\nu})\ul{\alpha}}
+\tfrac{1}{2}\,\delta\,R^{-1}\gamN^R{}_{\ul{\alpha}}\tilde{h}_{\ul{\mu\nu}}\,,
\end{align}
for the rescaled GHG Christoffel symbols and also write,
\begin{align}
\hat{\Phi}_{\ul{i'\mu\nu}}&=R^\delta\,\Phi_{\ul{i'\mu\nu}}=\tilde{\Phi}_{\ul{i'\mu\nu}}
-\delta\,R^{-1}\gamN^R{}_{\ul{i}'}\tilde{h}_{\ul{\mu\nu}}\,,\nonumber\\
\tilde{C}_{\ul{\alpha}}&=R^\delta\,\tilde{C}_{\ul{\alpha}}\,,\qquad 
\tilde{\nabla}_{\ul{\alpha}}H_{\ul{\beta}}=R^{2\delta}\,\nabla_{\ul{\alpha}}H_{\ul{\beta}}\,,\nonumber\\
\tilde{\gamma}_0&=R^\delta\,\gamma_0\,,
\end{align} 
which is justified if we assume reasonable fall-off on the gauge source functions. It is important to realize that this 
is a different system of PDEs as compared to~\eqref{eqn:GHG_11}, since we have modified the equations of motion by 
additions of the reduction constraints. The full set of constraints are easily expressed in terms of the new 
variables. 

%%%%%%%%%%%%%%%%%%%%%%%%%%%%%%%%%%%%%%%%%%%%%%%%%%%%%%%%%%%%
\paragraph*{Regularized DF GHG system:} Making now the change of independent variables to work in terms of the lower 
case coordinates, we obtain,
\begin{align}
\p_t\tilde{h}_{\ul{\mu\nu}}&=(\beta^p-\alpha\,v^p)\p_p\tilde{h}_{\ul{\mu\nu}}+\alpha\,W^{-1}s^{(\tilde{h})}_{\ul{\mu\nu}}
\,,\nonumber\\
d_t\tilde{\Phi}_{i\,\ul{\mu\nu}}&=\left(\beta^p\delta^j{}_i
-\alpha v^p\delta^j{}_i+\alpha W^2v_i(\mathbbmss{g}^{-1})^{pj}\right)d_p\tilde{\Phi}_{j\,\ul{\mu\nu}}
\nonumber\\
&\quad+\alpha\,W^{-1}\mathbbmss{g}^p{}_i\left(\gamma_2\,\p_p\tilde{h}_{\ul{\mu\nu}}-\p_p\tilde{\Pi}_{\ul{\mu\nu}}\right)
+\alpha\,W^{-1}s^{(\tilde{\Phi})}_{i\,\ul{\mu\nu}}\,,\nonumber\\
\p_t\tilde{\Pi}_{\ul{\mu\nu}}&=
\beta^p\p_p\tilde{\Pi}_{\ul{\mu\nu}}-\gamma_2\,\alpha v^p\p_p\tilde{h}_{\ul{\mu\nu}}
-\alpha\,W(\mathbbmss{g}^{-1})^{pi}d_p\tilde{\Phi}_{i\,\ul{\mu\nu}}\nonumber\\
&\quad+\alpha\,W^{-1}s^{(\tilde{\Pi})}_{\ul{\mu\nu}}\,.\label{eqn:GHG_DF_Reg}
\end{align}
Since the principal part of the new system is identical to the old one of Eq~\eqref{eqn:GHG_shell}, the system 
transforms as earlier, so the sources are once again,
\begin{align}
s^{(\tilde{h})}_{\ul{\mu\nu}}&=S^{(\tilde{h})}_{\ul{\mu\nu}}\,,\nonumber\\
s^{(\tilde{\Phi})}_{\ul{i'\mu\nu}}&=S^{(\tilde{\Phi})}_{\ul{i'\mu\nu}}+W^2V_{\ul{i}'}
\left(V^{\ul{j}'}S^{(\tilde{\Phi})}_{\ul{j'\mu\nu}}
+S^{(\tilde{\Pi})}_{\ul{\mu\nu}}-\gamma_2S^{(\tilde{h})}_{\ul{\mu\nu}}\right)\,,\nonumber\\
s^{(\tilde{\Pi})}_{\ul{\mu\nu}}&=\gamma_2S^{(\tilde{h})}_{\ul{\mu\nu}}+W^2\left(
V^{\ul{i}'}S^{(\tilde{\Phi})}_{\ul{i'\mu\nu}}+S^{(\tilde{\Pi)}}_{\ul{\mu\nu}}-\gamma_2S^{(\tilde{h})}_{\ul{\mu\nu}}
\right)\,.\label{eqn:GHG_DF_S_Reg}
\end{align}
Ultimately, up to the change of variables, we expect the continuum solutions to these equations to have the same 
asymptotic behavior as in the previous system. Although the regularized formulation still has coefficients that 
grow with~$R$, overall we have gained an order of~$R^\delta$ in various coefficients. To avoid repetition we do not give 
a full discussion of the asymptotics for this system. We expect that the improved coefficients will be helpful in the 
planned numerical implementation.

%%%%%%%%%%%%%%%%%%%%%%%%%%%%%%%%%%%%%%%%%%%%%%%%%%%%%%%%%%%%
\paragraph*{Discussion of asymptotics and the weak-null condition:}
We just examined the effect of modifying the evolution variables by multiplying them by powers of~$R$. We saw 
that multiplying up by a full-power of~$R$ is desirable since then we can remove from the equations of motion all terms relying explicitly
on fall-off of the variables at future-null infinity. What's more, this full power of~$R$ 
regularization will be helpful for the extraction of GWs. Key to the use of this renormalization is that all non-principal 
terms fall-off faster than~$O(R^{-2})$. This is because, for the hyperboloidal DF method, when using the renormalized variables we need
to multiply up the sources by terms of order~$O(R^{n+1})$, with~$1<n$. Therefore such a renormalization is not possible for non-linear wave
equations with a generic quadratic nonlinearity in first derivatives of the wave-field, but can only be naively employed for systems satisfying the 
null-condition~\cite{Sog95}. Roughly speaking this is the requirement that every term quadratic in first derivatives of the evolved 
fields contains at most one {\it bad} derivative in the~$K^a=N^a-S^a$ direction. This direction is bad because the associated derivatives 
fall-off only like~$O(R^{-1})$. The field equations of the GHG system do not satisfy the null-condition. They do however satisfy the weak-null 
condition~\cite{LinRod04} from which we naively expect, under the flatness assumptions of~\cite{LinRod04}, that only one component of the 
metric fails to fall-off like~$O(R^{-1})$ and rather goes like~$O(R^{-1}\log R)$. Such data can be dealt with under our previous 
assumptions. We would however like to exclude this logarithmic growth whenever possible. Therefore, following
for example~\cite{Tra58}, let us {\it assume} that the logarithmic growth is absent and examine what requirement is placed upon the data. For
this we strengthen our asymptotics assumptions so that in a neighborhood of null-infinity we have,
\begin{align}
g_{\ul{\mu\nu}}&=\eta_{\ul{\mu\nu}}+O(R^{-1})\,,\nonumber\\ 
\p_{\ul{\alpha}} g_{\ul{\mu\nu}}&=-\tfrac{1}{2}L_{\ul{\alpha}}\p_{K}g_{\ul{\mu\nu}}+O(R^{-2})\,,\nonumber\\
\p_{\ul{\alpha}} \p_{\ul{\beta}} g_{\ul{\mu\nu}}&=\tfrac{1}{4}L_{\ul{\alpha}}L_{\ul{\beta}}\p_{K}\p_{K}g_{\ul{\mu\nu}}+O(R^{-2})\,,
\label{eq:Asy_Stronger}
\end{align}
with~$L^a=N^a+S^a$ and~$K^a=N^a-S^a$, and~$\p_{K}g_{\ul{\mu\nu}}=O(R^{-1})=\p_{K}\p_{K}g_{\ul{\mu\nu}}$. 
For the sake of this discussion we assume that the gauge source functions~$H_{\ul{\mu}}$ and their derivatives decay sufficiently rapidly that 
they may safely be ignored. The terms quadratic in first derivatives in the GHG system then all satisfy the null-condition except those like,
\begin{align}
M(\p_Kg,\p_Kg)&=\p_Kg_{\ul{\mu\nu}}\left(\frac{1}{2}g^{\ul{\mu\alpha}}g^{\ul{\nu\beta}}
-\frac{1}{4}g^{\ul{\mu\nu}}g^{\ul{\alpha\beta}}\right)\p_Kg_{\ul{\alpha\beta}}\,.\nonumber
\end{align}
To exploit this observation in a free-evolution setup in the first order GHG system, we must add constraints in the form,
\begin{align}
C_{\ul{\mu}}C_{\ul{\mu}}-L^{\ul{\alpha}}C_{\ul{\alpha}}\p_Kg_{\ul{\mu\nu}}\,,
\end{align}
to the equation of motion for~$\Pi_{\ul{\mu\nu}}$. Defining the modified~$M_{\textrm{GHG}}(\p_Kg,\p_Kg)$ by making further adjustments using the Harmonic
constraints we arrive at,
\begin{align}
  M_{\textrm{GHG}}&=\frac{1}{4}\qN^{\ul{\alpha\beta}}d_Kg_{\ul{\alpha\beta}}d_Kg_{KL}
  \nonumber\\
  &+\p_Kg_{\ul{\mu\nu}}\left(\frac{1}{2}\qN^{\ul{\mu\alpha}}\qN^{\ul{\nu\beta}}
  -\frac{1}{4}\qN^{\ul{\mu\nu}}\qN^{\ul{\alpha\beta}}\right)\p_Kg_{\ul{\alpha\beta}}\,,
\end{align}
with the difference,
\begin{align}
  &M_{\textrm{GHG}}(\p_Kg,\p_Kg)-M(\p_Kg,\p_Kg)\nonumber\\
  &=\qN^{\ul{\alpha\beta}}C_{\ul{\alpha}}d_Kg_{K\ul{\beta}}-\tfrac{1}{4}L^{\ul{\alpha}}C_{\ul{\alpha}}d_Kg_{KK}-\tfrac{1}{4}K^{\ul{\alpha}}C_{\ul{\alpha}}d_Kg_{KL}\,,\nonumber
\end{align}
and~$\qN^{ab}$ defined as elsewhere. In fact there is still quite some freedom in adjusting these expressions.
The additional modifications should also be suitably included in the~$\Pi_{\ul{\mu\nu}}$
equation. Assuming momentarily that all constraints are satisfied we then have,
\begin{align}
R_{\ul{\mu\nu}}\sim M_{\textrm{GHG}}\,L_{\ul{\mu}}L_{\ul{\nu}}+O(R^{-3})\,,
\end{align}
with~$M_{\textrm{GHG}}=O(R^{-2})$. Therefore under the assumption~\eqref{eq:Asy_Stronger} we must 
additionally require that~$M_{\textrm{GHG}}$ vanishes faster than expected so that the spacetime is asymptotically Ricci flat to order~$O(R^{-2})$. It is not at all clear that
this condition will be propagated under time-evolution by the GHG equations of motion, or if assumption~\eqref{eq:Asy_Stronger} simply breaks down.
But the additional requirement is suspiciously like that leading to the Bondi-mass loss formula and therefore seems physically reasonable. Perhaps it
would be useful to have a model problem with this structure. This discussion will be expanded by a proper investigation alongside our presentation of
the implementation of the DF GHG system on hyperboloidal slices.

%%%%%%%%%%%%%%%%%%%%%%%%%%%%%%%%%%%%%%%%%%%%%%%%%%%%%%%%%%%%%%%%%%%%%%%%%%%%%%%%%%%%%%%
\section{Numerical Experiments}
\label{Section:Numerical_Experiments}
%%%%%%%%%%%%%%%%%%%%%%%%%%%%%%%%%%%%%%%%%%%%%%%%%%%%%%%%%%%%%%%%%%%%%%%%%%%%%%%%%%%%%%%

In this section we present our numerical experiments with the DF-formalism.
Although our analytic considerations have been concerned with the hyerboloidal DF-representation
of the GHG evolution equations, their numerical solution requires some fundamental coding efforts,
plus carefully thought out initial data procedures,
which we will undertake in a follow-up work.
Therefore, numerically we stick with the flat space scalar wave-equation,
\begin{align}
 \label{eq:flat_space_wave_eq}
 \Box \psi \equiv -\partial_T^2 \psi + \partial_{\ul{i}} \partial^{\ul{i}} \psi = 0 \; ,
\end{align}
which serves as a simple but important test-problem for the DF-approach because it can be written in a first order 
form that is very similar to the first order GHG system.
We emphasize that earlier work~\cite{Zen10} already described how solutions of Eq~\eqref{eq:flat_space_wave_eq}
can be obtained along hyperboloidal time-slices. The major novelties in our approach are that we can do away with 
the regularization normally used for the wave equation, and even on hyperboloidal slices we can treat nonlinearities 
which fall-off slower than~$R^{-3}$.

%%%%%%%%%%%%%%%%%%%%%%%%%%%%%%%%%%%%%%%%%%%%%%%%%%%%%%%%%%%%
\subsection{Wave Equation}\label{subsection:WE}
%%%%%%%%%%%%%%%%%%%%%%%%%%%%%%%%%%%%%%%%%%%%%%%%%%%%%%%%%%%%

%%%%%%%%%%%%%%%%%%%%%%%%%%%%%%%%%%%%%%%%%%%%%%%%%%%%%%%%%%%%
\paragraph*{First order form in shell-coordinates:}
We prepare the presentation of the DF-version of the wave equation
by recalling that the relevant first order reduction of Eq~\eqref{eq:flat_space_wave_eq} in global inertial 
coordinates $X^{\ul{i}}$, using the variables ${\pi := -\partial_T \psi}$ and ${\phi_{\ul{i}} := \partial_{\ul{i}} \psi}$,
is given by,
\begin{align}
\p_T\psi&=-\pi\,,\nonumber\\
\p_T\phi_{\ul{i}}&=-\p_{\ul{i}}\pi+\gamma_2\p_{\ul{i}}\psi-\gamma_2\phi_{\ul{i}}\,,\nonumber\\
\p_T\pi&=-\p^{\ul{i}}\phi_{\ul{i}}\,,
\end{align}
subject to the constraint,
\begin{align}
C_{\ul{i}}=\p_{\ul{i}}\psi-\phi_{\ul{i}}\,.
\end{align}
This corresponds to the first order GHG formulation~\eqref{eqn:GHG_11} as closely as 
possible. After employing shell-coordinates $X^{\ul{i}'}$ as introduced in Sec~\ref{subsection:Shells} 
we have instead,
\begin{align}
\label{eq:flatspace_wave_eq_shells}
\p_T\psi&=-\pi\,,\nonumber\\
\p_T\phi_{\ul{i}'}&=-\p_{\ul{i}'}\pi+\gamma_2\p_{\ul{i}'}\psi-\gamma_2\phi_{\ul{i}'}\,,
\nonumber\\
\p_T\pi&=-\p^{\ul{i}'}\phi_{\ul{i}'}+S^{(\pi)}\,,
\end{align}
where the source term is,
\begin{align}
\label{eq:Spi_source_flatspace_wave_eq_shells}
S^{(\pi)}=-\phi_{\ul{i}'}\p^{\ul{i}}(\Phi^{\textrm{Sh}})^{\ul{i}'}{}_{\ul{i}}\,,
\end{align}
and the constraint transforms in the obvious way. 

%%%%%%%%%%%%%%%%%%%%%%%%%%%%%%%%%%%%%%%%%%%%%%%%%%%%%%%%%%%%
\paragraph*{Dual-Foliation representation:} 
The system~\eqref{eq:flatspace_wave_eq_shells} complies with the DF recipe-form~\eqref{eqn:1st_Upper}
so that we can simply apply the DF-formalism~\eqref{eqn:Change_Coord}, which gives,
\begin{align}
\label{eq:DF_flatspace_wave_eq}
\p_t\psi&=-\pi\,,\nonumber\\
d_t\phi_{i}&=W^3v_i(\mathbbmss{g}^{-1})^{pj}\,d_p\phi_{j}
+\mathbbmss{g}^p{}_i\left(\gamma_2\,\p_p\psi-\p_p\pi\right)+s^{(\phi)}_i\,,\nonumber\\
\p_t\pi&=Wv^p\p_p\pi-\gamma_2 W v^p\p_p\psi-W^2(\mathbbmss{g}^{-1})^{pi}d_p\phi_{i}+s^{(\pi)}\,,
\end{align}
with sources,
\begin{align}
s^{(\phi)}_{\ul{i}'}&=-\gamma_2\,\phi_{\ul{i}'}+W^2V_{\ul{i}'}
\left(\gamma_2\,\pi-\gamma_2V^{\ul{j}'}\phi_{\ul{j}'}+S^{(\pi)}\right)
\,,\nonumber\\
s^{(\pi)}&=-\gamma_2\,\pi+W^2\left(\gamma_2\,\pi-\gamma_2V^{\ul{j}'}\phi_{\ul{j}'}+S^{(\pi)}
\right)\,,\label{eqn:DF_flatspace_wave_eq_source}
\end{align}
and constraint,
\begin{align}
  C_{\ul{i}'}=(\varphi^{-1})^i{}_{\ul{i}'}\p_i\psi+V_{\ul{i}'}\,\pi-\phi_{\ul{i'}}\,.
  \label{eqn:red_constr_DFWE}
\end{align}
Taking into account the simplifications of flat-space,
the system~\eqref{eq:DF_flatspace_wave_eq} resembles the DF-version of the
GHG system, Eq~\eqref{eqn:GHG_DF},
in accordance with our above assertion that the wave equation is comparable with the GHG evolution system.

%%%%%%%%%%%%%%%%%%%%%%%%%%%%%%%%%%%%%%%%%%%%%%%%%%%%%%%%%%%%
\paragraph*{Explicit equations for shell-coordinates with the height-function approach:} 
Let us now specify the coefficients of the evolution equations~\eqref{eq:DF_flatspace_wave_eq}
to our general shell-coordinates.
The non-zero metric components in shell-coordinates, as defined in Sec~\ref{subsection:Shells}, are,
\begin{align}
 \label{eq:Minkowski_shell_coordinates}
 \eta_{\ul{\mu}'\ul{\nu}'} &= 
  \left( \begin{array}{cccc}
   -1 & 0 & 0 & 0 \\
    0 & 1  & 0 & 0  \\
    0 & 0 & \qN_{\theta\theta} & \qN_{\theta\phi}  \\
    0 & 0 & \qN_{\phi\theta} & \qN_{\phi\phi}
  \end{array} \right) \,, 
\end{align}
where $\qN_{A'B'}$ depends on the specific choice of shell-coordinates.
Thus we always have $A=1$, $B^{\ul{i}'}=(0,0,0)$, $C_R^{\pm}=\pm 1$,
and related quantities can be trivially found.
Then, making some choice for the hyperboloidal coordinate system, 
we can write equations~\eqref{eq:DF_flatspace_wave_eq} explicitly.
We do not need to waggle the slices, as proposed in Sec~\ref{subsection:LS_control},
because for a stationary background like Minkowski 
the algebraic height-function approach~\eqref{eqn:hyp_trafo} is sufficient.
The hyperboloidal quantities can thus be expressed in terms of $H'$ and $R'$.
Additionally demanding $H'=1-R'^{-1}$, we could even discard the~$H'$ term if we so wished.
To state just a few, we have, e.g., the lapse and the Lorentz factor $\alpha^2=W^2=(1-H'^2)^{-1}$,
the shift $\beta^i=-W^2(H'/R',0,0)$, and the boost vectors $v^i=-W(H'/R',0,0)$
and $V^{\ul{i}'}=(H',0,0)$. Inserting the explicit expressions into~\eqref{eq:DF_flatspace_wave_eq},
we obtain,
\begin{align}
 \label{eq:DF_flatspace_wave_eq_shellsandheightfunction}
 \p_t\psi     &= -\pi \,,\nonumber\\
 \p_t\phi_{R} &= - \frac{  \p_r \pi } { R' (1-H'^{2})}
                 - \frac{ H' \p_r \phi_R} {R' (1 - H'^{2}) }  \nonumber\\
              &  - \frac{ H' } {1 - H'^{2} } \, \qN^{A'B'} {\delta^A}_{A'} \p_{A} \phi_{B'} 
                 + \frac{H'}{1-H'^2} S^\pi                         \nonumber\\
              &  + \frac{ \gamma_2 }{ 1-H'^2}
                    \left( \frac{  \p_r \psi } {R'}
                           + H' \pi 
                           - \phi_R
                    \right)
                 \,,\nonumber \\
 \p_t\phi_{A'}&=  - {\delta^A}_{A'}\p_A \pi + \gamma_2 \left( {\delta^A}_{A'} \p_A \psi - \phi_{A'} \right) \,,\nonumber\\
 \p_t\pi      &=    - \frac{ H'  \p_r \pi } {R' (1-H'^2)} 
                  - \frac{ \p_r \phi_R  } {R' (1-H'^2)} \;  \nonumber  \\
              &   - \frac{1}{ 1 - H'^2 } \qN^{A'B'} {\delta^A}_{A'} \p_{A} \phi_{B'} 
                  + \frac{ 1}{1-H'^2} S^\pi \nonumber \\
              &   +  \frac{ \gamma_2 \, H' } { 1-H'^2}
                     \left(  \frac{ \p_r \psi }{ R'}
                           + H' \pi
                           - \phi_R
                     \right)                          \,.       
\end{align}
Evidently, at first sight the coefficients are not all regular at null-infinity.
Implicitly, however, all of them take a finite limit.
To understand this let us replace $H'=1-R'^{-1}$,
and consider the asymptotics of, e.g., $\frac{H'}{1-H'^2}$.
One obtains $ { \frac{H'}{1-H'^2}=R' \frac{1-R'^{-1}}{2-R'^{-1}} \rightarrow R'/2 = O(R^n) }$.
Following this logic, most terms can be made explicitly regular, while only a few remain problematic.
Among the problematic ones are those multiplied by $\gamma_2$ but
let us not worry about them because $\gamma_2$ can be chosen at will.
Concerning the $R'\qN^{A'B'}$ terms, we get easily rid off the apparent failure
once specifying explicit shell-coordinates because any shell-coordinates 
satisfy~$\qN^{A'B'} \sim O(R^{-2})$.
Consequently, the only real problematic term hides in $\frac{H'}{1-H'^2} S^\pi$.
Using that for shell-coordinates we have, 
\begin{align}
 \label{eq:assumptions_shell_Jacobian}
 {\p^{\ul{i}}(\Phi^{\textrm{Sh}})^{\ul{i}'}{}_{\ul{i}}=(2/R, O(R^{-2}), O(R^{-2}))} \: ,
\end{align}
we see the problematic $ \frac{H'}{1-H'^2} \phi_R \, 2/R $ term.
Here we actually depend on the correct fall-off of the field $\phi_R$.
Indeed, in three spatial dimensions solutions to the wave equation 
fall-off like~$O(R^{-1})$, and so do the variables~$\phi_R$ and~$\pi$.
Thus, for $n<2$ there is absolutely no problem because $ \frac{H'}{1-H'^2} \phi_R \, 2/R \to 0$ at null-infinity,
which supports our view that $n<2$ compactifications are advantageous.
The $n=2$ case, instead, requires us to use L`H\^opital's rule for the not-explicitly regular coefficients at 
null-infinity, as alluded to in Sec~\ref{subsection:asymptotics}, when comparing the DF-formalism with the standard 
conformal compactification. This procedure is described in Sec~\ref{subsection:Implementation} below.
Note that, if one wishes to restrict~$\gamma_2$, which was found in our asymptotic analysis in 
to be necessary, the same argument of the correct fall-off of the fields holds for the 
problematic~$\gamma_2$ terms. In summary, we have convinced ourselves now
that implicitly all coefficients take a regular limit, even for the `conformal' case~$n=2$.
The discussion of regularity for GR in Sec~\ref{subsection:asymptotics} is not more sophisticated 
than this, except that we were careful to assume fall-off much weaker than that expected 
for physical solutions.

%%%%%%%%%%%%%%%%%%%%%%%%%%%%%%%%%%%%%%%%%%%%%%%%%%%%%%%%%%%%
\paragraph*{Regularization:} Since we know that solutions to the wave equation fall off like~$R^{-1}$, there is hope 
that the coefficients can be improved by considering suitably rescaled fields. The idea is that it would already help 
if one could turn some $O(R^2)/O(R^2)$ coefficient to $O(R)/O(R)$. We therefore try and rescale the variables by powers 
of~$R$, and, following the standard conformal approach for the wave equation~\cite{Zen10}, we define,
\begin{align}
\tilde{\psi}      &=R^\delta\,\psi \,,&\quad \tilde{\pi}&=R^\delta\,\pi\,,\nonumber\\
\tilde{\phi}_R    &=R^\delta\,\phi_R+\delta\,R^{\delta-1}\psi \,,&\quad \tilde{\phi}_{A'}&=R^\delta\,\phi_{A'}\,,\label{eqn:WE_renorm}
\end{align}
with~$\delta\geq0$ some constant, and now~$\tilde{\phi}_R$ will be a regular function with reduced decay 
in~$R$. We then take as variables~$(\tilde{\psi},\tilde{\phi}_{\ul{i'}},\tilde{\pi})$. The choice 
for~$\tilde{\phi}_R$ gives~$\tilde{\phi}_{\ul{i}'}=\p_{\ul{i}'}\tilde{\psi}$ when the reduction constraints are satisfied, 
and leads to convenient cancellations in the new equations-of-motion. Note that if we view the wave equation as a model 
for GR, the analogous normalization of~$g_{\ul{\mu\nu}}$ can not be used directly, see instead the rescaled fields
in Eq~\eqref{eq:rescaled_fields_GHG}. The evolution equations 
for~$(\tilde{\psi},\tilde{\phi}_{\ul{i}'},\tilde{\pi})$ in the upper case coordinates are easily obtained, 
exploiting Eq~\eqref{eq:Minkowski_shell_coordinates}, Eq~\eqref{eq:assumptions_shell_Jacobian}
and that~$\tilde{\phi}_R \equiv \partial_R \tilde{\psi}$.
The system reads,
\begin{align}
\label{eq:flatspace_wave_eq_shells_regfields}
\p_T\tilde{\psi}  &=-\tilde{\pi}\,,\nonumber\\
\p_T\tilde{\phi}_{\ul{i}'}
                  &= -\p_{\ul{i}'} \tilde{\pi}+\gamma_2\p_{\ul{i}'}\tilde{\psi}-\gamma_2\tilde{\phi}_{\ul{i}'}
\,,\nonumber\\
\p_T\tilde{\pi}&=-\p^{\ul{i}'} \tilde{\phi}_{\ul{i}'} + \tilde{S}^\pi\,,
\end{align}
whose structure almost coincides with the `untilded' version, Eq~\eqref{eq:flatspace_wave_eq_shells},
except for the source term,
\begin{align}
 \tilde{S}^{(\pi)} &= \frac{\delta-1}{R}\Big(2\,\tilde{\Phi}_R-\frac{\delta}{R}\tilde{\psi}\Big)
 -\tilde{\phi}_{\ul{A}'}\p^{\ul{i}}(\Phi^{\textrm{Sh}})^{\ul{A}'}{}_{\ul{i}}\,,
\end{align}
see~Eq~\eqref{eq:Spi_source_flatspace_wave_eq_shells}.
Applying the DF-formalism~\eqref{eqn:Change_Coord} to the system~\eqref{eq:flatspace_wave_eq_shells_regfields},
we obtain the new DF-system,
\begin{align} 
 \label{eq:DF_flatspace_wave_eq_shellsandheightfunction_reg}
 \p_t \tilde{\psi} 
              &= - \tilde{\pi} \,,\nonumber\\
 \p_t \tilde{\phi}_{R}
              &= - \frac{ \p_r \tilde{\pi} } { R'(1-H'^2) }
                 - \frac{ H' \p_r \tilde{\phi}_R } {R' (1-H'^2) } \nonumber \\
              &  - \frac{H'}{1-H'^2} \qN^{A'B'} {\delta^A}_{A'} \p_{A} \tilde{\phi}_{B'}  
                 + \frac{ H'} {1-H'^2} \tilde{S}^{(\pi)}   \nonumber \\
              &  + \frac{ \gamma_2 } {1-H'^2}
                     \left(
                      \frac{ \p_r \tilde{\psi} } {R'}
                     + H' \tilde{\pi}
                     - \tilde{\phi}_R
                     \right)         \,,\nonumber \\
 \p_t \tilde{\phi}_{A'}
              &= - {\delta^A}_{A'} \p_A \tilde{\pi} + \gamma_2 \left( {\delta^A}_{A'}  \p_A \tilde{\psi} - \tilde{\phi}_{A'} \right) \,,\nonumber\\
 \p_t \tilde{\pi}
              &= - \frac{ H' \p_r \tilde{\pi}}{R'(1-H'^2)} 
                 - \frac{ \p_r \tilde{\phi}_R } {R'(1- H'^2)} \nonumber \\
              &  - \frac{ 1 } {1-H'^2} \qN^{A'B'} {\delta^A}_{A'} \p_{A} \tilde{\phi}_{B'}
                 + \frac{ 1 } {1-H'^2} \tilde{S}^{(\pi)} \nonumber \\
              &  + \frac{ \gamma_2 \, H' } { 1-H'^2}
                     \left(
                      \frac{ \p_r \tilde{\psi} } {R'}
                     + H' \tilde{\pi}
                     - \tilde{\phi}_R
                     \right)                    \,.     
\end{align}
Comparing with Eq~\eqref{eq:DF_flatspace_wave_eq_shellsandheightfunction},
we see that the system's structure is identical modulo the change to tilde-variables, except that the coefficient multiplying the 
\textit{worst} term contained in~$S^{(\pi)}$, can be reduced in~$\tilde{S}^{(\pi)}$ and even set to vanish if we choose~$\delta=1$. 
Thus, demanding a strong-enough fall-off of $\gamma_2$, or even setting $\gamma_2=0$, we expect that this system performs much 
`cleaner' than Eq~\eqref{eq:DF_flatspace_wave_eq_shellsandheightfunction} in a numerical implementation. In App.~\ref{App:Mink_WE_SphPol} we present the equations of motion~\ref{eq:DF_flatspace_wave_eq_shellsandheightfunction_reg} explicitly in the special case that the shell coordinates are spherical polar.

%%%%%%%%%%%%%%%%%%%%%%%%%%%%%%%%%%%%%%%%%%%%%%%%%%%%%%%%%%%%
\subsection{Implementation}\label{subsection:Implementation}
%%%%%%%%%%%%%%%%%%%%%%%%%%%%%%%%%%%%%%%%%%%%%%%%%%%%%%%%%%%%

We have implemented the wave equation in the form~\eqref{eq:DF_flatspace_wave_eq_shellsandheightfunction_reg}
in the \texttt{bamps} pseudospectral code~\cite{HilWeyBru15,Bru11,BugDieBer15}. Full details will be discussed
when we present our implementation of the DF GHG system but here we nevertheless want to highlight
some key facts.

%%%%%%%%%%%%%%%%%%%%%%%%%%%%%%%%%%%%%%%%%%%%%%%%%%%%%%%%%%%%
\paragraph*{Technical development to the \texttt{bamps} infrastructure:}
The \texttt{bamps} code relies on a multidomain method involving communication of the solution between neighboring
patches. Previously tensor variables were stored in a global Cartesian basis so no transformation of the variables
was needed for this communication. In the present setup however we need to use the shell coordinate basis for the
representation of our reduction variables~$\phi_{\ul{i}'}$, and in the case of GHG we may even need to couple solutions
of different PDE systems across patches. The latter occurs because in the strong-field region we may want to set the
GHG formulation parameter to the standard value~$\gamma_1=-1$, whereas in the hyperboloidal layer we want~$\gamma_1=0$.
To deal with this we have setup the code so that a particular physics `project' can have different equations of
motion in different patches. The transformation of variables must be taken into account in the boundary 
communication. In fact the result across different shell boundaries in the angular direction is straightforwardly 
derived: Angular tensor components in the target shell are computed as simple linear combinations of purely angular 
tensor components of the neighboring shell. The radial components transform trivially in this case.
The result across the strong-field region to shells boundaries is constructed by applying the full 
Jacobian~\eqref{eq:abbrevs_trafos_cart_shells}.

%%%%%%%%%%%%%%%%%%%%%%%%%%%%%%%%%%%%%%%%%%%%%%%%%%%%%%%%%%%%
\paragraph*{Implementation of L`H\^opital's rule regularization for formally singular terms:}
When using~$n=2$ compactification and evolving without regularizing the variables, we apply L`H\^opital's rule on the source terms~$W^2 S^{(\pi)}$
and $W^2 V_{\ul{i}'} S^{(\pi)}$ in Eq~\eqref{eq:DF_flatspace_wave_eq} to compute their finite values at~$r=S$. After performing a characteristic decomposition
of~$\phi_R = \frac{1}{2}(u^{\hat{+}} - u^{\hat{-}})$ and using that solutions to the wave equation fall off like~$u^{\hat{+}} = O(R^{-1})$
and~$u^{\hat{-}} = O(R^{-2})$, we find,
\begin{align}
\label{eqn:nonregular_lhospital}
\lim_{r \to S} W^2 S^{(\pi)} &= - \frac{1}{2} \partial_r(\pi + \phi_R) \,, \nonumber\\
\lim_{r \to S} W^2 V_{R} S^{(\pi)} &= - \frac{1}{2} \partial_r(\pi + \phi_R) \,, \nonumber\\
\lim_{r \to S} W^2 V_{A'} S^{(\pi)} &= 0 \,.
\end{align}
This is implemented in the obvious way by adjusting the sources at, and only at null-infinity.

%%%%%%%%%%%%%%%%%%%%%%%%%%%%%%%%%%%%%%%%%%%%%%%%%%%%%%%%%%%%
\paragraph*{The cartoon method and mirror symmetries on hyperboloidal slices:} 
For axisymmetry, \texttt{bamps} supports a 2D reduction, called the cartoon method~\cite{AlcBraBru99,Pre04}. This is based on 
the vanishing Lie derivative $\mathcal{L}_\phi T = 0$ for any tensor $T$ along the $\phi^{\ul{i}} = (-y,x,0)$-direction.
A transformation of the Lie vector $\phi$ was necessary to make this method compatible with the new shell coordinate basis.
We refer to the notation of~\cite{HilWeyBru15} ($x^{\ul{i}'} = (\bar{x},\bar{y},\bar{z})$) and want to state the resulting
modified cartoon formulas for patches oriented in $x$ direction (XP) and in $z$ direction (ZP):
\begin{align} 
 \label{eq:DF_cartoon}
 \text{XP:} \quad &\partial_{\bar{y}} u(\bar{x},0,\bar{z}) = 0\,,\nonumber\\
                  &\partial_{\bar{y}} v^{\bar{y}}(\bar{x},0,\bar{z}) = 0\,, \nonumber\\
                  &\partial_{\bar{y}} v^{\bar{z}}(\bar{x},0,\bar{z}) = \bar{z} v^{\bar{y}}(\bar{x},0,\bar{z})\,,\nonumber\\
 \text{ZP:} \quad &\partial_{\bar{z}} u(\bar{x},\bar{y},0) = 0\,,\nonumber\\
                  &\partial_{\bar{z}} v^{\bar{y}}(\bar{x},\bar{y},0) = -\frac{v^{\bar{z}}(\bar{x},\bar{y},0)}{\bar{y}}\,, \nonumber\\ 
                  &\partial_{\bar{z}} v^{\bar{z}}(\bar{x},\bar{y},0) = \frac{v^{\bar{y}}(\bar{x},\bar{y},0)}{\bar{y}} \,.     
\end{align}
Similarly, \texttt{bamps} is able to handle mirror symmetries along the~$x$, $y$ and $z$ directions. The corresponding 
transformation behavior of a tensor component was only implemented for Cartesian coordinates and had to be 
carried out in the shell coordinates. This results in odd or even angular vector components, depending on shell
orientation and symmetry surface. Scalar functions and radial vector components always have even behavior.

%%%%%%%%%%%%%%%%%%%%%%%%%%%%%%%%%%%%%%%%%%%%%%%%%%%%%%%%%%%%
\paragraph*{The energy method and communication of data via the penalty method:} Our numerical method relies on the
existence of a continuum PDE energy to communicate data between grids. The basic energy on the hyperboloidal slice is
built from the density,
\begin{align}
  \varepsilon&=\Lambda\,\psi^2+2\,\gamma_2\,\psi\,\big(\pi-V^{\ul{i}'}\phi_{\ul{i}'}\big)\nonumber\\
  &\quad+\big(\pi-V^{\ul{i}'}\phi_{\ul{i}'}\big)^2
  +{}^\textrm{\tiny{(N)}}(\mathbbmss{g}^{-1})^{\ul{i'j'}}\phi_{\ul{i}'}\phi_{\ul{j}'}\,,
\end{align}
with~$\Lambda^2>\gamma_2^2$. For symmetric hyperbolicity we require that this is everywhere positive definite in the evolved
variables. But on the hyperboloidal slice we have,
\begin{align}
{}^\textrm{\tiny{(N)}}(\mathbbmss{g}^{-1})^{RR}=1-V^RV^R=O(R^{-n})\,,
\end{align}
so we lose control of the radial reduction variable~$\phi_R$ as we head towards null-infinity. Remarkably this is no problem
for our numerical work, because at this boundary we do not require boundary conditions, and the required control of~$\phi_{A'}$
is maintained. We therefore use our standard pseudospectral penalty method as described in~\cite{HilWeyBru15}. By working with
regularized variables and maybe including suitable weights in the energy estimate, we expect that the missing control can be
regained. Such improved energies may even lead to superior approximation methods. This will be investigated in detail in
future work.

%%%%%%%%%%%%%%%%%%%%%%%%%%%%%%%%%%%%%%%%%%%%%%%%%%%%%%%%%%%%
\paragraph*{Computation of regular coefficients:} Although the equations of
motion~\eqref{eq:DF_flatspace_wave_eq_shellsandheightfunction_reg} have regular coefficients on the right hand side, they are
formed from divergent quantities. Therefore care is needed in the implementation if we are to maintain accuracy and avoid
`NaNs'. Fortunately the complete equations can be built from the following regular combinations of~$R'^{-1}$ and~$H'$,
\begin{align}
R^{-1}  &= \frac{\Omega^\frac{1}{n-1}}{\tilde{r}+\Omega^\frac{1}{n-1}R_i}\,,\quad
R'^{-1} = \frac{\Omega^\frac{1}{n-1}}{\Omega+\tfrac{\tilde{r}^2}{\tilde{S}^2}\tfrac{2\tilde{\chi}+\tilde{\chi}'}{n-1}},\nonumber\\
\frac{R'}{R^2}       &= \frac{\left(\Omega+\tfrac{\tilde{r}^2}{\tilde{S}^2}\tfrac{2\tilde{\chi}+\tilde{\chi}'}{n-1}\right)\Omega^\frac{2-n}{n-1}}
{\left(\tilde{r}+\Omega^{\frac{1}{n-1}}R_i\right)^2}\,,
\end{align}
where we write,
\begin{align}
\tilde{r}=r-R_i\,,\quad \tilde{S}=S-R_i\,,\quad
\Omega=1-\tilde{\chi}\frac{\tilde{r}^2}{\tilde{S}^2}\,,
\end{align}
with~$\tilde{\chi}$ a transition function between~$0$ and~$1$. These expressions can be used whilst varying~$n$, our measure of the
rate of compactification.

%%%%%%%%%%%%%%%%%%%%%%%%%%%%%%%%%%%%%%%%%%%%%%%%%%%%%%%%%%%%
\subsection{Numerical Results}\label{subsection:Num_Results}
%%%%%%%%%%%%%%%%%%%%%%%%%%%%%%%%%%%%%%%%%%%%%%%%%%%%%%%%%%%%

%============================================================
% TABLE II: Runs Performed
%============================================================
\begin{table}[t]
  \caption{Summary of wave equation experiments. $N$ refers to the number of points
    per dimension in each grid. By default we use the regularized variables with full~$O(R)$ renormalization,
    and so indicate only when not doing so. Similarly by default we filter the Chebyschev coefficients.
    The~$p$ column refers to which exponent was used for the nonlinearity~$\psi^p$ of Eq~\eqref{eqn:cubic_wave}.
    The final column refers to the compactification parameter.
  }
\centering
  \begin{tabular}[t]{ l | l | l | l | l | l | l | l }
  \hline\hline
  ID & N & $\gamma_2$ & Filter & Cartoon & Reg. & $p$ & n \\ 
  \hline\hline
  \eqref{eqn:wave_id_3D} & $11,13,\hdots,29$ & 1 & \xmark & \xmark & \xmark &   & 2.00 \\
                         & $11,13,\hdots,29$ & 1 & angular& \xmark & \xmark &   & 2.00 \\
                         & $11,13,\hdots,29$ & 1 &        & \xmark & \xmark &   & 2.00 \\
                         & $11,13,\hdots,29$ & 1 &        & \xmark &        &   & 2.00 \\
                         & $11,13,\hdots,25$ & 1 &        & \xmark & \xmark &   & 1.50 \\
                         & $11,13,\hdots,25$ & 1 &        & \xmark & \xmark & 2 & 1.50 \\
                         & $11,13,\hdots,25$ & 1 & \xmark & \xmark &        &   & 2.00 \\
                         & $11,13,\hdots,23$ & 0 &        & \xmark &        &   & 2.00 \\
                         & $11,13,\hdots,23$ & 1 &        & \xmark &        & 3 & 2.00 \\
  \hline\hline
  \eqref{eqn:wave_id_2D} & $11,13,\hdots,23$ & 1 &        & \xmark &        &  & 2.00 \\
                         & $11,13,\hdots,23$ & 1 &        & \cmark &        &  & 2.00 \\
                         & $11,13,\hdots,23$ & 1 &        & \cmark &        &  & 1.75 \\
                         & $11,13,\hdots,23$ & 1 &        & \cmark &        &  & 1.50 \\
                         & $11,13,\hdots,23$ & 1 &        & \cmark &        &  & 1.25 \\                         
    \hline\hline
  \end{tabular} 
\label{tab:runs}
\end{table}

%=============================================================

%%%%%%%%%%%%%%%%%%%%%%%%%%%%%%%%%%%%%%%%%%%%%%%%%%%%%%%%%%%%
\begin{figure*}[t]
  \begin{center}
    \includegraphics[width=\textwidth]{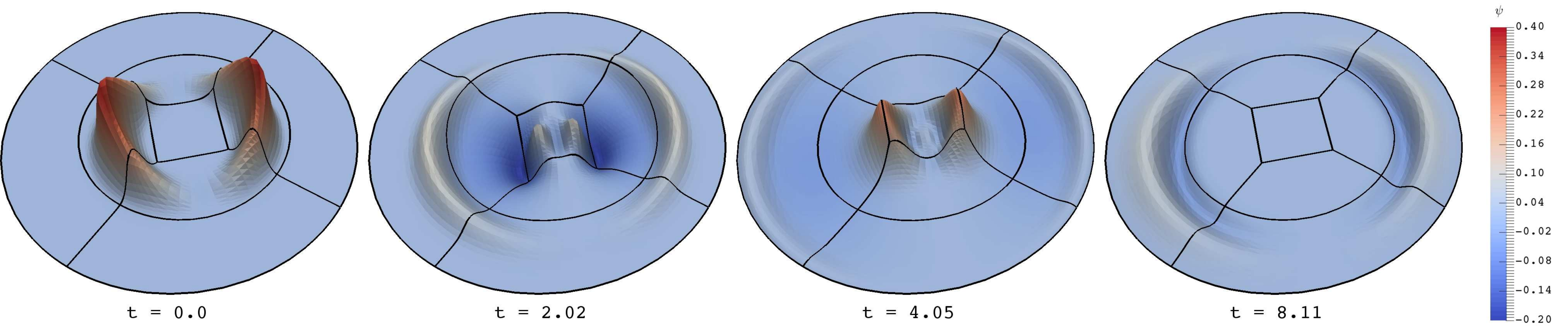}
    \caption{\label{fig:Num_DF_evo}
      The basic dynamics of the evolution of~$\psi$ in the plane~$y=0$, starting with
      initial data~\eqref{eqn:wave_id_3D}. We see that initially there is a pulse traveling outwards
      which is absorbed at null-infinity, the outer boundary of the plot. After traveling
      inwards the second pulse also propagates off of the grid. This evolution was performed
      in full 3d with~$25^3$ points per patch. The patches each consist of one grid. Their
      boundaries are marked by the thick black lines.     
      }
  \end{center}
\end{figure*}
%%%%%%%%%%%%%%%%%%%%%%%%%%%%%%%%%%%%%%%%%%%%%%%%%%%%%%%%%%%%

%%%%%%%%%%%%%%%%%%%%%%%%%%%%%%%%%%%%%%%%%%%%%%%%%%%%%%%%%%%%
\paragraph*{Initial data:} As initial data we always choose~$\pi=0$ combined with either,
\begin{align}
\label{eqn:wave_id_3D}
\psi(r, \theta, \varphi) = 
\sqrt{\frac{15}{32\pi}} e^{-\left(\frac{r-r_0}{\sigma}\right)^2} \frac{\sin^2 \theta \cos 2 \varphi}{\sqrt{1+R(r)^2}}\,,
\end{align}
or
\begin{align}
\label{eqn:wave_id_2D}
\psi(r, \theta) = \frac{e^{-r^2-z_0^2 + 2 z_0 r \cos\theta}}{\sqrt{1+R(r)^2}}\,,
\end{align}
for axisymmetry. We typically choose~$r_0=3$,~$\sigma=0.6$ and~$z_0=0.2$. From this we use reduction 
constraint~\eqref{eqn:red_constr_DFWE} to set the variable~$\phi_{\ul{i}'}$. The tests were performed on a desktop 
machine with 8~GB of memory and 4~cores. We run the code in parallel with MPI as discussed in~\cite{HilWeyBru15}. 
With our standard setup, a 3d run with~$N=23$ points per dimension computes at roughly~10 time units per hour. We always employ 
the same grid, whose setup can be understood from Fig~\ref{fig:Num_DF_evo}. We do not subdivide patches and the 
grid has~$R_i=4.5$ and~$S=7.5$. Table~\ref{tab:runs} contains a summary of the runs performed.

\begin{figure}[t]
  \begin{center}
    \includegraphics[width=0.45\textwidth]{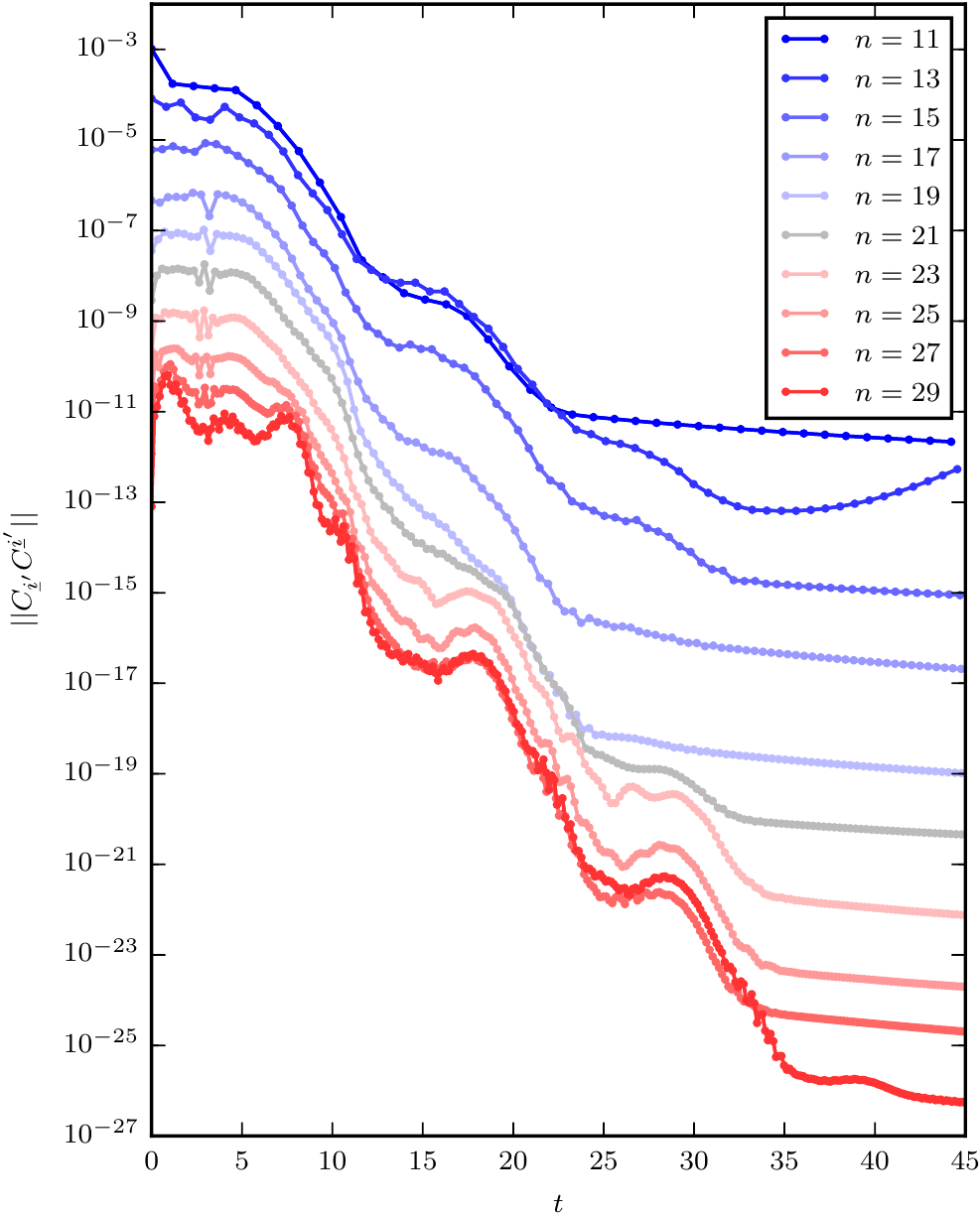}
    \caption{\label{fig:Num_Constraints}
      Convergence of the norm of the constraints~$C_{\ul{i}'}$ as resolution is increased in our wave equation 
      experiments with~$O(R)$ regularization of the evolved fields and the filter. The initial data was given 
      by~\eqref{eqn:wave_id_3D}. In fact we see slightly cleaner convergence of the constraints in these experiments 
      if we do not filter~\cite{HilWeyBru15} the Chebyschev coefficients, but for nonlinear problems we expect that the 
      filter is needed, and therefore only present results with it.        
    }
  \end{center}
\end{figure}

\begin{figure}[t]
  \begin{center}
    \includegraphics[width=0.45\textwidth]{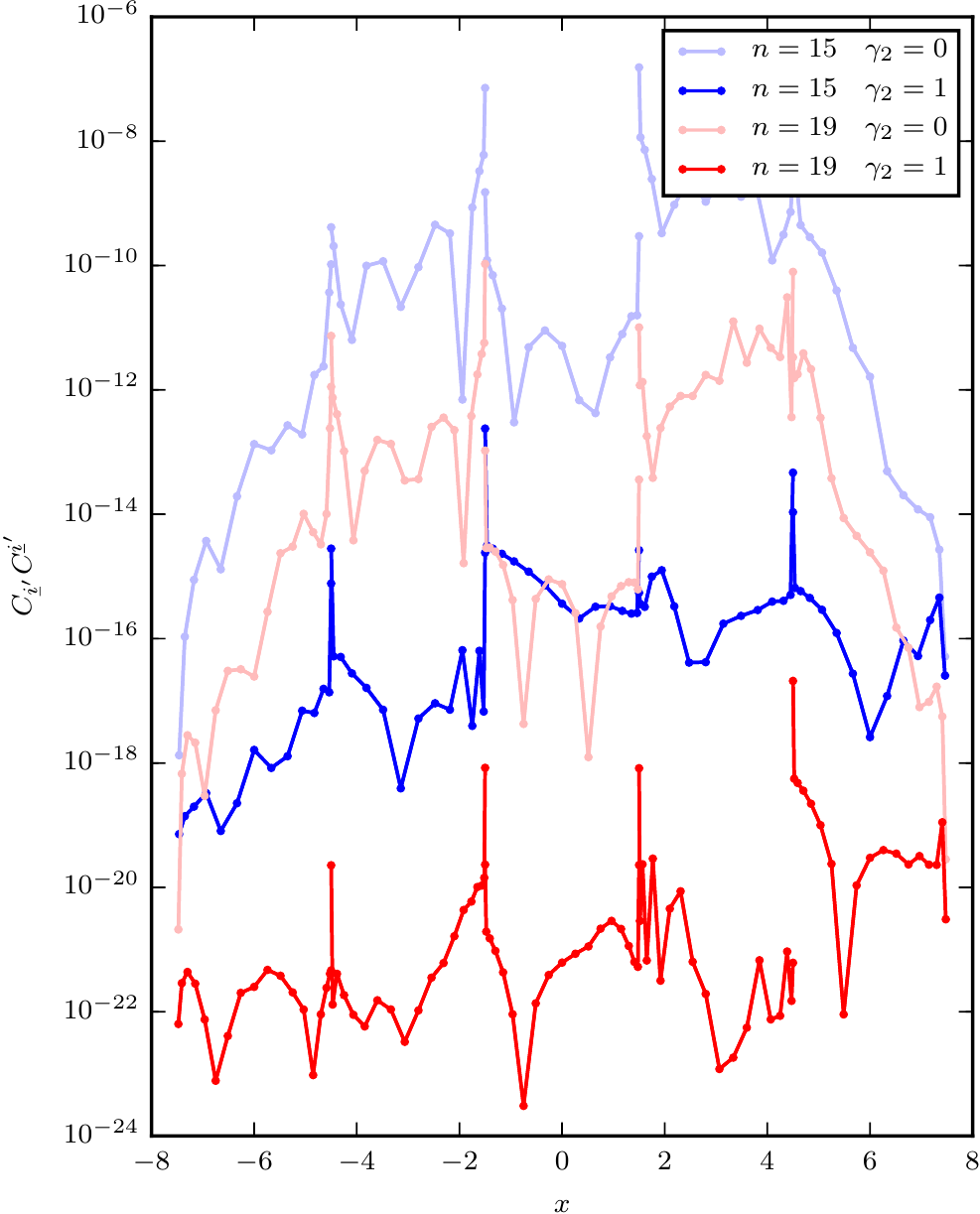}
    \caption{\label{fig:Num_Constraints_space}
      A comparison of the squared constraints~$C^{\ul{i}'}C_{\ul{i}'}$ in space at~$t=25$ in the evolution of 
      initial data~\eqref{eqn:wave_id_3D}, with and without constraint damping switched on. Note that we suppress 
      (by~$R^{-1}$) the central value of the damping parameter~$\gamma_2$ in the hyperboloidal region. Here the hyperboloidal 
      layer starts at~$R_i=4.5$ and null-infinity is at~$S=r=7.5$, and we see that the constraint damping is having 
      the desired effect.
    }
  \end{center}
\end{figure}

%%%%%%%%%%%%%%%%%%%%%%%%%%%%%%%%%%%%%%%%%%%%%%%%%%%%%%%%%%%%
\paragraph*{Basic results with~$n=2$ hyperboloidal slices:}
The dynamics of a typical evolution with initial data~\eqref{eqn:wave_id_3D} are presented in Fig~\ref{fig:Num_DF_evo}; the 
wave propagates out and leaves the domain through null-infinity almost without reflection. Using neither regularization 
of the evolved fields or filtering the basis coefficients we find that the expected fall-off of the field at the outer boundary 
is violated as~$\psi$ differs slightly from zero. This effect however converges away rapidly with resolution. Note that this
setup is particularly interesting because of the close similarity to the problem of the wave equation in the~AdS spacetime. Using 
the filter in all directions, which is expected to be necessary for nonlinear problems, we find that the method is unstable. We suspect
that this is caused by the fact that the radial filter does not respect the expected fall-off of the fields, and violates the delicate
L`H\^opital limiting procedure used at null-infinity for this setup. Adjusting the filter so that only the angular directions are treated 
cures the instability. Therefore we expect that a more carefully constructed radial filter would work. 
We now focus the discussion on the case most studied in the literature, namely~$n=2$ hyperboloidal compactification with the 
maximum~$O(R)$ renormalization of the fields. By construction now~$\psi$ vanishes at null-infinity. The outgoing wave does leave behind 
a small amount of noise which we see as~$C_{\ul{i}'}$ constraint violation. That this violation converges away rapidly with resolution is 
demonstrated in Fig~\ref{fig:Num_Constraints}. Since our asymptotics require that the constraint damping parameter~$\gamma_2$ fall-off 
as~$R$ increases, a concern may be that the constraint damping scheme is ineffective in these evolutions. In 
Fig~\ref{fig:Num_Constraints_space} we look at the constraint violation in space with and without this damping. It turns out that the 
constraint damping is still effective, possibly because the definition of the constraint~$C_{\ul{i}'}$ combined with the hyperboloidal 
coordinates conspires to suppress the appearance of violations. We performed a set of runs with initial data~\eqref{eqn:wave_id_2D} both 
with and without the cartoon symmetry reduction. We see good agreement between the two sets and comparable convergence of errors. Within 
our range of resolutions the cartoon runs ran between~$10$ and~$23$ times faster than the 3d tests, with larger speedups at higher 
resolution.

%%%%%%%%%%%%%%%%%%%%%%%%%%%%%%%%%%%%%%%%%%%%%%%%%%%%%%%%%%%%
\paragraph*{Results with $n<2$ hyperboloidal slices:} Although it is most straightforward to compare~$n=2$ results with
the literature, in our analysis of the GHG system we found that it will be more convenient to choose initial data with
compactification parameter~$n<2$, especially since then L`\^opital's rule is not required. Therefore we performed tests also with this setup. 
Starting with~$n=3/2$ we ran the code using the filter in all directions without regularizing the variables. Here we find that the instability 
that we saw previously with~$n=2$ is no longer present, and the results converge nicely. We then moved on to use the regularized variables. 
Since we saw good agreement before with and without symmetry reduction here we ran only faster cartoon tests. Qualitatively we find the same 
dynamical behavior as before, as expected. Again constraint violation converges away rapidly with resolution.

%%%%%%%%%%%%%%%%%%%%%%%%%%%%%%%%%%%%%%%%%%%%%%%%%%%%%%%%%%%%
%%%%%%%%%%%%%%%%%%%%%%%%%%%%%%%%%%%%%%%%%%%%%%%%%%%%%%%%%%%%

%%%%%%%%%%%%%%%%%%%%%%%%%%%%%%%%%%%%%%%%%%%%%%%%%%%%%%%%%%%%
%%%%%%%%%%%%%%%%%%%%%%%%%%%%%%%%%%%%%%%%%%%%%%%%%%%%%%%%%%%%

%%%%%%%%%%%%%%%%%%%%%%%%%%%%%%%%%%%%%%%%%%%%%%%%%%%%%%%%%%%%
\paragraph*{Wave equation with nonlinear sources:}
As a final test we return to~$n=2$ slices and consider the field equation,
\begin{align}
\nabla^a\nabla_a\psi-\psi^p=0\,,\label{eqn:cubic_wave}
\end{align}
in flat space, for which numerical results with a hyperboloidal compactification were previously presented with~$p=3$
in~\cite{BizZen08}. It is trivial to add the additional term to the equations of motion. Starting with the regularized 
variables and~$p=3$, we took initial data~\eqref{eqn:wave_id_3D} and found again qualitatively similar results and convergence 
of the method. Doubling the size of the initial pulse also makes little difference to the outcome, indicating that the data is 
small enough that the evolution is essentially linear. To test our method against a much more aggressive nonlinearity we 
chose~$p=2$, and evolved with~$n=3/2$, with the filter and without regularization of the evolved variables. We saw in the 
previous paragraph that this setup works without the nonlinear term; with it we see that for short times the method converges
nicely as desired, before the solution rapidly explodes. We plan to study the blow-up of solutions to nonlinear equations in 
detail in the future.

%%%%%%%%%%%%%%%%%%%%%%%%%%%%%%%%%%%%%%%%%%%%%%%%%%%%%%%%%%%%%%%%%%%%%%%%%%%%%%%%%%%%%%%
\section{Conclusion}
\label{Section:Conclusion}
%%%%%%%%%%%%%%%%%%%%%%%%%%%%%%%%%%%%%%%%%%%%%%%%%%%%%%%%%%%%%%%%%%%%%%%%%%%%%%%%%%%%%%%

Continuing our research programme on the evolution of extreme spacetimes with the {\texttt bamps}
code we turned our attention to the problem of null-infinity, for which we want an explicit 
numerical treatment. Once this is in place we hope to be able to treat the threshold of blackhole 
formation comprehensively. Our approach is to employ the DF formalism~\cite{Hil15} 
together with the generalized harmonic formulation of GR to evolve hyperboloidal initial data. 
The simplest strategy for this is to use a hyperboloidal layer~\cite{Zen10}, built upon a
height-function change to the time coordinate. But the DF formalism gives us the freedom to
{\it effectively} choose the height-function dynamically which affords sharp control of the
outgoing radial coordinate lightspeed in the hyperboloidal coordinates. The use of optical
Jacobians, or rather solutions to the eikonal equation, was key to our construction.

As a first step towards implementation of the method we tested the wave equation in flat-space using 
the DF approach. This leads to a related, but distinct set of equations as compared with 
the standard approach to the wave equation~\cite{Zen10} with the conformal treatment. The wave equation 
is a particularly good toy model for the proposed method because as pulses propagate out under GHG 
formulation, the field equations resemble more and more closely those of the flat-space wave equation. 
The experiments were a success. Firstly the reduction constraints converge rapidly as resolution is 
increased. Secondly additional nonlinear terms did not cause the method to break down. Finally the 
generalized slices of~\cite{CalGunHil05}, which give us greater deal of flexibility in the asymptotic behavior 
near null-infinity, gave comparable results to the standard choice. Although these results are no 
guarantee that the full approach will work for GR, they are very encouraging. 

Natural questions concern the generality and applicability of the proposed method. $i).$ Are suitable 
initial data readily available using standard methods? $ii).$ How closely is the construction tied 
to the generalized harmonic formulation?  $iii).$ Could the method be used for second order in space 
formulations of GR?  $iv).$ Are there other possibilities for the treatment of null-infinity within the 
same code infrastructure? To answer~$i).$ as far as initial data is concerned, hyperboloidal data 
are not available in exactly the form we would require. But there is a wealth of experience in creating 
hyperboloidal data with the conformal approach, see for example~\cite{Fra98b,BucPfeBar09,SchPanAns13,SchPanAns13a},
so it is expected that, although a research project by itself, their construction will not be a show-stopper.
$ii).$ We expect that the use of other first order formulations of GR will be straightforward with the approach 
provided that the matrix~$(\mathbf{1}+\mathbf{A}^{\ul{V}})$ remains regular for all subluminal~$V_{\ul{i}'}$. Another
possible complication is that if there are superluminal speeds in the system boundary conditions will be needed at
null-infinity, undoing one of the major advantages of the hyperboloidal compactification. $iii).$ The use of
second order in space formulations will require more care than the approach presented here, because derivatives
of the hyperboloidal Jacobians will appear in the metric equations of motion, so that the asymptotics near
null-infinity will have to be carefully reconsidered. We suggest that the answer to~$iv).$ is no. An obvious
alternative would be to use the DF approach for Cauchy-Characteristic-Matching. This could be done
by solving the first order GHG system in the weak-field region on characteristic-slices and then communicating 
data in the standard way through the boundary of the Cauchy region. This would have the advantage over 
current attempts that the same PDE system would be solved over the whole domain, but just using 
different coordinates in different regions. This should make the matching conditions more 
straightforward.

It is hoped that in the near future dynamical spacetimes will come under robust control with 
the proposed method. In that case our ongoing study of the critical collapse of gravitational 
waves will tie directly to null-infinity, and should allow significant progress in the numerical 
investigation of the weak cosmic censorship conjecture.

%%%%%%%%%%%%%%%%%%%%%%%%%%%%%%%%%%%%%%%%%%%%%%%%%%%%%%%%%%%%%%%%%%%%%%%%%%%%%%%%%%%%%%%
\begin{acknowledgments}
%%%%%%%%%%%%%%%%%%%%%%%%%%%%%%%%%%%%%%%%%%%%%%%%%%%%%%%%%%%%%%%%%%%%%%%%%%%%%%%%%%%%%%%

It is a pleasure to thank Marcus Ansorg, Sebastiano Bernuzzi, Sascha Husa,  
Rodrigo Panosso-Macedo, Sebastian M\"ockel, Juan-Antonio Valiente-Kroon, David Schinkel and 
Alex Va\~n\'o-Vi\~nuales, for helpful discussions. We are especially grateful to Juan-Antonio 
Valiente-Kroon for making his forthcoming monograph available to us before publication. This 
work was supported in part by the Graduierten-Akademie Jena. We acknowledge the use of the 
LRZ machine SuperMUC.

%%%%%%%%%%%%%%%%%%%%%%%%%%%%%%%%%%%%%%%%%%%%%%%%%%%%%%%%%%%%%%%%%%%%%%%%%%%%%%%%%%%%%%%
\end{acknowledgments}
%%%%%%%%%%%%%%%%%%%%%%%%%%%%%%%%%%%%%%%%%%%%%%%%%%%%%%%%%%%%%%%%%%%%%%%%%%%%%%%%%%%%%%%

%% %%%%%%%%%%%%%%%%%%%%%%%%%%%%%%%%%%%%%%%%%%%%%%%%%%%%%%%%%%%%%%%%%%%%%%%%%%%%%%%%%%%%%%%
\appendix
%% %%%%%%%%%%%%%%%%%%%%%%%%%%%%%%%%%%%%%%%%%%%%%%%%%%%%%%%%%%%%%%%%%%%%%%%%%%%%%%%%%%%%%%%

%%%%%%%%%%%%%%%%%%%%%%%%%%%%%%%%%%%%%%%%%%%%%%%%%%%%%%%%%%%%%%%%%%%%%%%%%%%%%%%%%%%%%%%%%
\section{Flat Space DF Wave Equation in spherical-polar coordinates}\label{App:Mink_WE_SphPol}
%%%%%%%%%%%%%%%%%%%%%%%%%%%%%%%%%%%%%%%%%%%%%%%%%%%%%%%%%%%%%%%%%%%%%%%%%%%%%%%%%%%%%%%%%

In this appendix we want to complement Sec~\ref{subsection:WE}'s analytic discussion
on the DF wave equation in flat space by choosing as shell-coordinates
the well-known spherical-polar coordinates,
coordinates
\begin{align}
  X &= R \sin\theta \cos\phi \nonumber \\
  Y &= R \sin\theta \sin\phi \nonumber \\
  Z &= R \cos\theta \,,
\end{align}
which fulfill all conditions demanded of `shell'-coordinates as defined in Sec~\ref{subsection:Shells}.
For these coordinates we obtain ${\qN_{A'B'}=\rm{diag}(R^2,R^2 \sin^2\theta)}$, and
${\p^{\ul{i}}(\Phi^{\textrm{Sh}})^{\ul{i}'}{}_{\ul{i}}=(2/R, \cot(\theta)/R^2,0)}$.
Inserting the specific expressions into Eq~\eqref{eq:DF_flatspace_wave_eq_shellsandheightfunction} yields,
\begin{align}
\p_t\phi_R&=-\frac{H'}{1-H'^2}\left(\frac{1}{R'}\p_r\phi_R+\frac{2}{R}\phi_R
-\gamma_2\,\pi+\gamma_2\,H'\phi_R\right)\nonumber\\
&\quad-\frac{H'}{1-H'^2}\left(
\frac{\p_\theta(\sin\theta\,\phi_\theta)}{R^2\sin\theta}
+\frac{\p_\phi\phi_\phi}{R^2\sin^2\theta}\right)-\gamma_2\,\phi_R\nonumber\\
&\quad+\frac{\gamma_2\,\p_r\psi-\p_r\pi}{R'(1-H'^2)}\,,
\label{eqn:shell_wave_PhiR_SphPol}
\end{align}
for the radial reduction variable, 
\begin{align}
\p_t\phi_{A'}&=\gamma_2\,(\p_A\psi-\phi_A)-\p_A\pi\,,
\end{align}
for the angular reduction variables, and,
\begin{align}
&\p_t\pi=-\frac{H'\,\p_r\pi}{R'\big(1-H'^2\big)}
-\frac{1}{1-H'^2}\left(\frac{1}{R'}\p_r\phi_R+\frac{2}{R}\phi_R\right)
\nonumber\\
&+\frac{\gamma_2\,H'}{\big(1-H'^2\big)}\left(\frac{1}{R'}\,\p_r\psi
-\phi_R+H'\,\pi\right)\,,
\nonumber\\
&-\frac{\p_\theta(\sin\theta\,\phi_\theta)}{R^2(1-H'^2)\sin\theta}
-\frac{\p_\phi\phi_\phi}{R^2(1-H'^2)\sin^2\theta}\,,
\label{eqn:shell_wave_Pi_SphPol}
\end{align}
for the wave equation proper. Evidently the spherical polar coordinate singularity would
cause problems, which is why we avoid these coordinates. Spontaneously 
here one would be concerned about the~$O(R)$ and~$O(R^2)$ coefficients multiplying the 
source terms. But again solutions to the wave equation 
fall-off like~$O(R^{-1})$, and the combination~$\phi_R-\pi$ falls off one order 
faster, so choosing~$\gamma_2=O(R^{-1})$ is sufficient for regularity of these terms as 
we approach null-infinity. Inserting the explicit expressions into the~$O(R)$ regularized 
version, see Eq~\eqref{eqn:WE_renorm} with~$\delta=1$, gives,
\begin{align}
\p_t\tilde{\psi}&=-\tilde{\pi}\,,\nonumber\\
\p_t\tilde{\phi}_{A'}&=\gamma_2\,(\p_A\tilde{\psi}-\tilde{\phi}_A)-\p_A\tilde{\pi}\,.
\end{align}
For the radial reduction variable we have
\begin{align}
\p_t\tilde{\phi}_R&=-\frac{H'}{1-H'^2}\left(\frac{1}{R'}\p_r\tilde{\phi}_R
-\gamma_2\,\tilde{\pi}+\gamma_2\,H'\tilde{\phi}_R\right)\nonumber\\
&\quad-\frac{H'}{1-H'^2}\left(
\frac{\p_\theta(\sin\theta\,\tilde{\phi}_\theta)}{R^2\sin\theta}
+\frac{\p_\phi\tilde{\phi}_\phi}{R^2\sin^2\theta}\right)-\gamma_2\,\tilde{\phi}_R\nonumber\\
&\quad+ \frac{\gamma_2\,\p_r\tilde{\psi}-\p_r\tilde{\pi}}{R'(1-H'^2)}\,,
\end{align}
Finally the wave equation becomes,
\begin{align}
\p_t\tilde{\pi}&=-\frac{H'\,\p_r\tilde{\pi}}{R'\big(1-H'^2\big)}
-\frac{\p_r\tilde{\phi}_R}{R'\big(1-H'^2\big)}
+\frac{\gamma_2\,H'\,\p_r\tilde{\psi}}{R'\big(1-H'^2\big)}
\nonumber\\
&-\frac{\p_\theta(\sin\theta\,\tilde{\phi}_\theta)}{R^2(1-H'^2)\sin\theta}
-\frac{\p_\phi\tilde{\phi}_\phi}{R^2(1-H'^2)\sin^2\theta}
+\frac{\gamma_2\,H'^2\,\tilde{\pi}}{1-H'^2}\nonumber\\
&- \frac{ \gamma_2\,H'\tilde{\phi}_R}{1-H'^2}\,,
\end{align}

%%%%%%%%%%%%%%%%%%%%%%%%%%%%%%%%%%%%%%%%%%%%%%%%%%%%%%%%%%%%%%%%%%%%%%%%%%%%%%%%%%%%%%%
%% \bibliography{refs}
\bibliographystyle{apsrev}
\bibliography{Hyp_DF.bbl}{}

\begin{thebibliography}{71}
\expandafter\ifx\csname natexlab\endcsname\relax\def\natexlab#1{#1}\fi
\expandafter\ifx\csname bibnamefont\endcsname\relax
  \def\bibnamefont#1{#1}\fi
\expandafter\ifx\csname bibfnamefont\endcsname\relax
  \def\bibfnamefont#1{#1}\fi
\expandafter\ifx\csname citenamefont\endcsname\relax
  \def\citenamefont#1{#1}\fi
\expandafter\ifx\csname url\endcsname\relax
  \def\url#1{\texttt{#1}}\fi
\expandafter\ifx\csname urlprefix\endcsname\relax\def\urlprefix{URL }\fi
\providecommand{\bibinfo}[2]{#2}
\providecommand{\eprint}[2][]{\url{#2}}

\bibitem[{\citenamefont{Abrahams and Evans}(1992)}]{AbrEva92}
\bibinfo{author}{\bibfnamefont{A.~M.} \bibnamefont{Abrahams}} \bibnamefont{and}
  \bibinfo{author}{\bibfnamefont{C.~R.} \bibnamefont{Evans}},
  \bibinfo{journal}{Phys. Rev. D} \textbf{\bibinfo{volume}{46}},
  \bibinfo{pages}{R4117} (\bibinfo{year}{1992}).

\bibitem[{\citenamefont{Gundlach and Mart{\'i}n-Garc{\'i}a}(2007)}]{GunGar07}
\bibinfo{author}{\bibfnamefont{C.}~\bibnamefont{Gundlach}} \bibnamefont{and}
  \bibinfo{author}{\bibfnamefont{J.~M.} \bibnamefont{Mart{\'i}n-Garc{\'i}a}},
  \bibinfo{journal}{Living Reviews in Relativity} \textbf{\bibinfo{volume}{10}}
  (\bibinfo{year}{2007}),
  \urlprefix\url{http://www.livingreviews.org/lrr-2007-5}.

\bibitem[{\citenamefont{Hilditch et~al.}(2013)\citenamefont{Hilditch,
  Baumgarte, Weyhausen, Dietrich, Br{\"u}gmann, Montero, and
  M{\"u}ller}}]{HilBauWey13}
\bibinfo{author}{\bibfnamefont{D.}~\bibnamefont{Hilditch}},
  \bibinfo{author}{\bibfnamefont{T.~W.} \bibnamefont{Baumgarte}},
  \bibinfo{author}{\bibfnamefont{A.}~\bibnamefont{Weyhausen}},
  \bibinfo{author}{\bibfnamefont{T.}~\bibnamefont{Dietrich}},
  \bibinfo{author}{\bibfnamefont{B.}~\bibnamefont{Br{\"u}gmann}},
  \bibinfo{author}{\bibfnamefont{P.~J.} \bibnamefont{Montero}},
  \bibnamefont{and}
  \bibinfo{author}{\bibfnamefont{E.}~\bibnamefont{M{\"u}ller}},
  \bibinfo{journal}{Phys.Rev.} \textbf{\bibinfo{volume}{D88}},
  \bibinfo{pages}{103009} (\bibinfo{year}{2013}), \eprint{1309.5008}.

\bibitem[{\citenamefont{Hilditch et~al.}(2016)\citenamefont{Hilditch,
  Weyhausen, and Brügmann}}]{HilWeyBru15}
\bibinfo{author}{\bibfnamefont{D.}~\bibnamefont{Hilditch}},
  \bibinfo{author}{\bibfnamefont{A.}~\bibnamefont{Weyhausen}},
  \bibnamefont{and}
  \bibinfo{author}{\bibfnamefont{B.}~\bibnamefont{Brügmann}},
  \bibinfo{journal}{Phys. Rev.} \textbf{\bibinfo{volume}{D93}},
  \bibinfo{pages}{063006} (\bibinfo{year}{2016}), \eprint{1504.04732}.

\bibitem[{\citenamefont{Hilditch}(2015)}]{Hil15}
\bibinfo{author}{\bibfnamefont{D.}~\bibnamefont{Hilditch}}
  (\bibinfo{year}{2015}), \eprint{1509.02071}.

\bibitem[{\citenamefont{Reisswig et~al.}(2009)\citenamefont{Reisswig, Bishop,
  Pollney, and Szilagyi}}]{ReiBisPol09}
\bibinfo{author}{\bibfnamefont{C.}~\bibnamefont{Reisswig}},
  \bibinfo{author}{\bibfnamefont{N.~T.} \bibnamefont{Bishop}},
  \bibinfo{author}{\bibfnamefont{D.}~\bibnamefont{Pollney}}, \bibnamefont{and}
  \bibinfo{author}{\bibfnamefont{B.}~\bibnamefont{Szilagyi}},
  \bibinfo{journal}{Phys. Rev. Lett.} \textbf{\bibinfo{volume}{103}},
  \bibinfo{pages}{221101} (\bibinfo{year}{2009}), \eprint{0907.2637}.

\bibitem[{\citenamefont{Reisswig et~al.}(2013)\citenamefont{Reisswig, Bishop,
  and Pollney}}]{ReiBisPol12}
\bibinfo{author}{\bibfnamefont{C.}~\bibnamefont{Reisswig}},
  \bibinfo{author}{\bibfnamefont{N.~T.} \bibnamefont{Bishop}},
  \bibnamefont{and} \bibinfo{author}{\bibfnamefont{D.}~\bibnamefont{Pollney}},
  \bibinfo{journal}{Gen.Rel.Grav.} \textbf{\bibinfo{volume}{45}},
  \bibinfo{pages}{1069} (\bibinfo{year}{2013}), \eprint{1208.3891}.

\bibitem[{\citenamefont{Taylor et~al.}(2013)\citenamefont{Taylor, Boyle,
  Reisswig, Scheel, Chu, Kidder, and Szilágyi}}]{TayBoyRei13}
\bibinfo{author}{\bibfnamefont{N.~W.} \bibnamefont{Taylor}},
  \bibinfo{author}{\bibfnamefont{M.}~\bibnamefont{Boyle}},
  \bibinfo{author}{\bibfnamefont{C.}~\bibnamefont{Reisswig}},
  \bibinfo{author}{\bibfnamefont{M.~A.} \bibnamefont{Scheel}},
  \bibinfo{author}{\bibfnamefont{T.}~\bibnamefont{Chu}},
  \bibinfo{author}{\bibfnamefont{L.~E.} \bibnamefont{Kidder}},
  \bibnamefont{and}
  \bibinfo{author}{\bibfnamefont{B.}~\bibnamefont{Szilágyi}},
  \bibinfo{journal}{Phys. Rev.} \textbf{\bibinfo{volume}{D88}},
  \bibinfo{pages}{124010} (\bibinfo{year}{2013}), \eprint{1309.3605}.

\bibitem[{\citenamefont{Handmer and Szilagyi}(2015)}]{HanSzi14}
\bibinfo{author}{\bibfnamefont{C.~J.} \bibnamefont{Handmer}} \bibnamefont{and}
  \bibinfo{author}{\bibfnamefont{B.}~\bibnamefont{Szilagyi}},
  \bibinfo{journal}{Class. Quant. Grav.} \textbf{\bibinfo{volume}{32}},
  \bibinfo{pages}{025008} (\bibinfo{year}{2015}), \eprint{1406.7029}.

\bibitem[{\citenamefont{Handmer et~al.}(2015)\citenamefont{Handmer, Szilágyi,
  and Winicour}}]{HanSziWin15}
\bibinfo{author}{\bibfnamefont{C.~J.} \bibnamefont{Handmer}},
  \bibinfo{author}{\bibfnamefont{B.}~\bibnamefont{Szilágyi}},
  \bibnamefont{and} \bibinfo{author}{\bibfnamefont{J.}~\bibnamefont{Winicour}},
  \bibinfo{journal}{Class. Quant. Grav.} \textbf{\bibinfo{volume}{32}},
  \bibinfo{pages}{235018} (\bibinfo{year}{2015}), \eprint{1502.06987}.

\bibitem[{\citenamefont{Handmer et~al.}(2016)\citenamefont{Handmer, Szilágyi,
  and Winicour}}]{HanSziWin16}
\bibinfo{author}{\bibfnamefont{C.~J.} \bibnamefont{Handmer}},
  \bibinfo{author}{\bibfnamefont{B.}~\bibnamefont{Szilágyi}},
  \bibnamefont{and} \bibinfo{author}{\bibfnamefont{J.}~\bibnamefont{Winicour}}
  (\bibinfo{year}{2016}), \eprint{1605.04332}.

\bibitem[{\citenamefont{Winicour}(2012)}]{Win12}
\bibinfo{author}{\bibfnamefont{J.}~\bibnamefont{Winicour}},
  \bibinfo{journal}{Living Rev. Relativity} \textbf{\bibinfo{volume}{15}},
  \bibinfo{pages}{2} (\bibinfo{year}{2012}), \bibinfo{note}{[Online article]},
  \urlprefix\url{http://www.livingreviews.org/lrr-2012-2}.

\bibitem[{\citenamefont{Penrose}(1964)}]{Pen64}
\bibinfo{author}{\bibfnamefont{R.}~\bibnamefont{Penrose}}, in
  \emph{\bibinfo{booktitle}{Relativity, Groups, and Topology (Les Houches,
  France, 1964)}}, edited by
  \bibinfo{editor}{\bibfnamefont{C.}~\bibnamefont{DeWitt}} \bibnamefont{and}
  \bibinfo{editor}{\bibfnamefont{B.}~\bibnamefont{DeWitt}}
  (\bibinfo{publisher}{Gordon and Breach}, \bibinfo{address}{New York},
  \bibinfo{year}{1964}), pp. \bibinfo{pages}{565--584}.

\bibitem[{\citenamefont{Friedrich}(1981{\natexlab{a}})}]{Fri81}
\bibinfo{author}{\bibfnamefont{H.}~\bibnamefont{Friedrich}},
  \bibinfo{journal}{Proc. Roy. Soc. London} \textbf{\bibinfo{volume}{A 375}},
  \bibinfo{pages}{169} (\bibinfo{year}{1981}{\natexlab{a}}).

\bibitem[{\citenamefont{Friedrich}(1981{\natexlab{b}})}]{Fri81a}
\bibinfo{author}{\bibfnamefont{H.}~\bibnamefont{Friedrich}},
  \bibinfo{journal}{Proc. Roy. Soc. London} \textbf{\bibinfo{volume}{A 378}},
  \bibinfo{pages}{401} (\bibinfo{year}{1981}{\natexlab{b}}).

\bibitem[{\citenamefont{Doulis and Frauendiener}(2016)}]{DouFra16}
\bibinfo{author}{\bibfnamefont{G.}~\bibnamefont{Doulis}} \bibnamefont{and}
  \bibinfo{author}{\bibfnamefont{J.}~\bibnamefont{Frauendiener}}
  (\bibinfo{year}{2016}), \eprint{1609.03584}.

\bibitem[{\citenamefont{Zenginoglu}(2008)}]{Zen08}
\bibinfo{author}{\bibfnamefont{A.}~\bibnamefont{Zenginoglu}},
  \bibinfo{journal}{Class. Quant. Grav.} \textbf{\bibinfo{volume}{25}},
  \bibinfo{pages}{195025} (\bibinfo{year}{2008}), \eprint{0808.0810}.

\bibitem[{\citenamefont{Friedrich}(1986)}]{Fri86}
\bibinfo{author}{\bibfnamefont{H.}~\bibnamefont{Friedrich}},
  \bibinfo{journal}{Comm. Math. Phys.} \textbf{\bibinfo{volume}{107}},
  \bibinfo{pages}{587} (\bibinfo{year}{1986}).

\bibitem[{\citenamefont{Garfinkle}(2002)}]{Gar02}
\bibinfo{author}{\bibfnamefont{D.}~\bibnamefont{Garfinkle}},
  \bibinfo{journal}{Phys. Rev. D} \textbf{\bibinfo{volume}{65}},
  \bibinfo{pages}{044029} (\bibinfo{year}{2002}).

\bibitem[{\citenamefont{Moncrief and Rinne}(2009)}]{MonRin08}
\bibinfo{author}{\bibfnamefont{V.}~\bibnamefont{Moncrief}} \bibnamefont{and}
  \bibinfo{author}{\bibfnamefont{O.}~\bibnamefont{Rinne}},
  \bibinfo{journal}{Class.Quant.Grav.} \textbf{\bibinfo{volume}{26}},
  \bibinfo{pages}{125010} (\bibinfo{year}{2009}), \eprint{0811.4109}.

\bibitem[{\citenamefont{Rinne}(2010{\natexlab{a}})}]{Rin09}
\bibinfo{author}{\bibfnamefont{O.}~\bibnamefont{Rinne}},
  \bibinfo{journal}{Class. Quant. Grav.} \textbf{\bibinfo{volume}{27}},
  \bibinfo{pages}{035014} (\bibinfo{year}{2010}{\natexlab{a}}),
  \eprint{0910.0139}.

\bibitem[{\citenamefont{Va{\~n}{\'o}-Vi{\~n}uales
  et~al.}(2015)\citenamefont{Va{\~n}{\'o}-Vi{\~n}uales, Husa, and
  Hilditch}}]{VanHusHil14}
\bibinfo{author}{\bibfnamefont{A.}~\bibnamefont{Va{\~n}{\'o}-Vi{\~n}uales}},
  \bibinfo{author}{\bibfnamefont{S.}~\bibnamefont{Husa}}, \bibnamefont{and}
  \bibinfo{author}{\bibfnamefont{D.}~\bibnamefont{Hilditch}},
  \bibinfo{journal}{Class. Quant. Grav.} \textbf{\bibinfo{volume}{32}},
  \bibinfo{pages}{175010} (\bibinfo{year}{2015}), \eprint{1412.3827}.

\bibitem[{\citenamefont{Va{\~n}{\'o}-Vi{\~n}uales and Husa}(2015)}]{VanHus14}
\bibinfo{author}{\bibfnamefont{A.}~\bibnamefont{Va{\~n}{\'o}-Vi{\~n}uales}}
  \bibnamefont{and} \bibinfo{author}{\bibfnamefont{S.}~\bibnamefont{Husa}},
  \bibinfo{journal}{J. Phys. Conf. Ser.} \textbf{\bibinfo{volume}{600}},
  \bibinfo{pages}{012061} (\bibinfo{year}{2015}), \eprint{1412.4801}.

\bibitem[{\citenamefont{Va{\~n}{\'o}-Vi{\~n}uale}(2015)}]{Van15}
\bibinfo{author}{\bibfnamefont{A.}~\bibnamefont{Va{\~n}{\'o}-Vi{\~n}uale}},
  Ph.D. thesis, \bibinfo{school}{U. Iles Balears, Palma}
  (\bibinfo{year}{2015}), \eprint{1512.00776},
  \urlprefix\url{http://inspirehep.net/record/1407828/files/arXiv:1512.00776.pdf}.

\bibitem[{\citenamefont{Baumgarte and Shapiro}(1998)}]{BauSha98}
\bibinfo{author}{\bibfnamefont{T.~W.} \bibnamefont{Baumgarte}}
  \bibnamefont{and} \bibinfo{author}{\bibfnamefont{S.~L.}
  \bibnamefont{Shapiro}}, \bibinfo{journal}{Phys. Rev. D}
  \textbf{\bibinfo{volume}{59}}, \bibinfo{pages}{024007}
  (\bibinfo{year}{1998}), \eprint{gr-qc/9810065}.

\bibitem[{\citenamefont{Shibata and Nakamura}(1995)}]{ShiNak95}
\bibinfo{author}{\bibfnamefont{M.}~\bibnamefont{Shibata}} \bibnamefont{and}
  \bibinfo{author}{\bibfnamefont{T.}~\bibnamefont{Nakamura}},
  \bibinfo{journal}{Phys. Rev. D} \textbf{\bibinfo{volume}{52}},
  \bibinfo{pages}{5428} (\bibinfo{year}{1995}).

\bibitem[{\citenamefont{Nakamura et~al.}(1987)\citenamefont{Nakamura, Oohara,
  and Kojima}}]{NakOohKoj87}
\bibinfo{author}{\bibfnamefont{T.}~\bibnamefont{Nakamura}},
  \bibinfo{author}{\bibfnamefont{K.}~\bibnamefont{Oohara}}, \bibnamefont{and}
  \bibinfo{author}{\bibfnamefont{Y.}~\bibnamefont{Kojima}},
  \bibinfo{journal}{Prog. Theor. Phys. Suppl.} \textbf{\bibinfo{volume}{90}},
  \bibinfo{pages}{1} (\bibinfo{year}{1987}).

\bibitem[{\citenamefont{Bernuzzi and Hilditch}(2010)}]{BerHil09}
\bibinfo{author}{\bibfnamefont{S.}~\bibnamefont{Bernuzzi}} \bibnamefont{and}
  \bibinfo{author}{\bibfnamefont{D.}~\bibnamefont{Hilditch}},
  \bibinfo{journal}{Phys. Rev. D} \textbf{\bibinfo{volume}{81}},
  \bibinfo{pages}{084003} (\bibinfo{year}{2010}), \eprint{0912.2920}.

\bibitem[{\citenamefont{Zenginoglu}(2007)}]{Zen07}
\bibinfo{author}{\bibfnamefont{A.}~\bibnamefont{Zenginoglu}}, Ph.D. thesis,
  \bibinfo{school}{Potsdam U., Inst. of Math.} (\bibinfo{year}{2007}),
  \eprint{0711.0873},
  \urlprefix\url{https://inspirehep.net/record/766850/files/arXiv:0711.0873.pdf}.

\bibitem[{\citenamefont{Bardeen et~al.}(2011)\citenamefont{Bardeen, Sarbach,
  and Buchman}}]{BarSarBuc11}
\bibinfo{author}{\bibfnamefont{J.~M.} \bibnamefont{Bardeen}},
  \bibinfo{author}{\bibfnamefont{O.}~\bibnamefont{Sarbach}}, \bibnamefont{and}
  \bibinfo{author}{\bibfnamefont{L.~T.} \bibnamefont{Buchman}},
  \bibinfo{journal}{Phys. Rev.} \textbf{\bibinfo{volume}{D83}},
  \bibinfo{pages}{104045} (\bibinfo{year}{2011}), \eprint{1101.5479}.

\bibitem[{\citenamefont{Morales and Sarbach}(2016)}]{MorSar16}
\bibinfo{author}{\bibfnamefont{M.~D.} \bibnamefont{Morales}} \bibnamefont{and}
  \bibinfo{author}{\bibfnamefont{O.}~\bibnamefont{Sarbach}}
  (\bibinfo{year}{2016}), \eprint{1609.05756}.

\bibitem[{\citenamefont{LeFloch and Ma}(2014)}]{LeFYue14}
\bibinfo{author}{\bibfnamefont{P.~G.} \bibnamefont{LeFloch}} \bibnamefont{and}
  \bibinfo{author}{\bibfnamefont{Y.}~\bibnamefont{Ma}} (\bibinfo{year}{2014}),
  \eprint{1411.4910}.

\bibitem[{\citenamefont{LeFloch and Ma}(2015)}]{LeFYue15}
\bibinfo{author}{\bibfnamefont{P.~G.} \bibnamefont{LeFloch}} \bibnamefont{and}
  \bibinfo{author}{\bibfnamefont{Y.}~\bibnamefont{Ma}} (\bibinfo{year}{2015}),
  \eprint{1507.01143}.

\bibitem[{\citenamefont{Lindblom et~al.}(2006)\citenamefont{Lindblom, Scheel,
  Kidder, Owen, and Rinne}}]{LinSchKid05}
\bibinfo{author}{\bibfnamefont{L.}~\bibnamefont{Lindblom}},
  \bibinfo{author}{\bibfnamefont{M.~A.} \bibnamefont{Scheel}},
  \bibinfo{author}{\bibfnamefont{L.~E.} \bibnamefont{Kidder}},
  \bibinfo{author}{\bibfnamefont{R.}~\bibnamefont{Owen}}, \bibnamefont{and}
  \bibinfo{author}{\bibfnamefont{O.}~\bibnamefont{Rinne}},
  \bibinfo{journal}{Class. Quant. Grav.} \textbf{\bibinfo{volume}{23}},
  \bibinfo{pages}{S447} (\bibinfo{year}{2006}), \eprint{gr-qc/0512093}.

\bibitem[{\citenamefont{Zengino{\u g}lu et~al.}(2009)\citenamefont{Zengino{\u
  g}lu, Nunez, and Husa}}]{ZenNunHus08}
\bibinfo{author}{\bibfnamefont{A.}~\bibnamefont{Zengino{\u g}lu}},
  \bibinfo{author}{\bibfnamefont{D.}~\bibnamefont{Nunez}}, \bibnamefont{and}
  \bibinfo{author}{\bibfnamefont{S.}~\bibnamefont{Husa}},
  \bibinfo{journal}{Class. Quant. Grav.} \textbf{\bibinfo{volume}{26}},
  \bibinfo{pages}{035009} (\bibinfo{year}{2009}), \eprint{0810.1929}.

\bibitem[{\citenamefont{Zenginoglu}(2011)}]{Zen10}
\bibinfo{author}{\bibfnamefont{A.}~\bibnamefont{Zenginoglu}},
  \bibinfo{journal}{J. Comput. Phys.} \textbf{\bibinfo{volume}{230}},
  \bibinfo{pages}{2286} (\bibinfo{year}{2011}), \eprint{1008.3809}.

\bibitem[{\citenamefont{Racz and Toth}(2011)}]{RacTot11}
\bibinfo{author}{\bibfnamefont{I.}~\bibnamefont{Racz}} \bibnamefont{and}
  \bibinfo{author}{\bibfnamefont{G.~Z.} \bibnamefont{Toth}},
  \bibinfo{journal}{Class. Quant. Grav.} \textbf{\bibinfo{volume}{28}},
  \bibinfo{pages}{195003} (\bibinfo{year}{2011}), \eprint{1104.4199}.

\bibitem[{\citenamefont{Jasiulek}(2012)}]{Jas12}
\bibinfo{author}{\bibfnamefont{M.}~\bibnamefont{Jasiulek}},
  \bibinfo{journal}{Class. Quant. Grav.} \textbf{\bibinfo{volume}{29}},
  \bibinfo{pages}{015008} (\bibinfo{year}{2012}), \eprint{1109.2513}.

\bibitem[{\citenamefont{Zenginoğlu et~al.}(2014)\citenamefont{Zenginoğlu,
  Khanna, and Burko}}]{ZenKhaBur14}
\bibinfo{author}{\bibfnamefont{A.}~\bibnamefont{Zenginoğlu}},
  \bibinfo{author}{\bibfnamefont{G.}~\bibnamefont{Khanna}}, \bibnamefont{and}
  \bibinfo{author}{\bibfnamefont{L.~M.} \bibnamefont{Burko}},
  \bibinfo{journal}{Gen.Rel.Grav.} \textbf{\bibinfo{volume}{46}},
  \bibinfo{pages}{1672} (\bibinfo{year}{2014}), \eprint{1208.5839}.

\bibitem[{\citenamefont{Harms et~al.}(2013)\citenamefont{Harms, Bernuzzi, and
  Br{\"u}gmann}}]{HarBerBru13}
\bibinfo{author}{\bibfnamefont{E.}~\bibnamefont{Harms}},
  \bibinfo{author}{\bibfnamefont{S.}~\bibnamefont{Bernuzzi}}, \bibnamefont{and}
  \bibinfo{author}{\bibfnamefont{B.}~\bibnamefont{Br{\"u}gmann}},
  \bibinfo{journal}{Class.Quant.Grav.} \textbf{\bibinfo{volume}{30}},
  \bibinfo{pages}{115013} (\bibinfo{year}{2013}), \eprint{1301.1591}.

\bibitem[{\citenamefont{Zengino{\u g}lu and Khanna}(2011)}]{ZenKha11}
\bibinfo{author}{\bibfnamefont{A.}~\bibnamefont{Zengino{\u g}lu}}
  \bibnamefont{and} \bibinfo{author}{\bibfnamefont{G.}~\bibnamefont{Khanna}},
  \bibinfo{journal}{Phys.Rev.} \textbf{\bibinfo{volume}{X1}},
  \bibinfo{pages}{021017} (\bibinfo{year}{2011}), \eprint{1108.1816}.

\bibitem[{\citenamefont{Bernuzzi
  et~al.}(2011{\natexlab{a}})\citenamefont{Bernuzzi, Nagar, and
  Zenginoglu}}]{BerNagZen11}
\bibinfo{author}{\bibfnamefont{S.}~\bibnamefont{Bernuzzi}},
  \bibinfo{author}{\bibfnamefont{A.}~\bibnamefont{Nagar}}, \bibnamefont{and}
  \bibinfo{author}{\bibfnamefont{A.}~\bibnamefont{Zenginoglu}},
  \bibinfo{journal}{Phys.Rev.} \textbf{\bibinfo{volume}{D84}},
  \bibinfo{pages}{084026} (\bibinfo{year}{2011}{\natexlab{a}}),
  \eprint{1107.5402}.

\bibitem[{\citenamefont{Bernuzzi
  et~al.}(2011{\natexlab{b}})\citenamefont{Bernuzzi, Nagar, and
  Zenginoglu}}]{BerNagZen10}
\bibinfo{author}{\bibfnamefont{S.}~\bibnamefont{Bernuzzi}},
  \bibinfo{author}{\bibfnamefont{A.}~\bibnamefont{Nagar}}, \bibnamefont{and}
  \bibinfo{author}{\bibfnamefont{A.}~\bibnamefont{Zenginoglu}},
  \bibinfo{journal}{Phys. Rev.} \textbf{\bibinfo{volume}{D83}},
  \bibinfo{pages}{064010} (\bibinfo{year}{2011}{\natexlab{b}}),
  \eprint{1012.2456}.

\bibitem[{\citenamefont{Barausse et~al.}(2012)\citenamefont{Barausse, Buonanno,
  Hughes, Khanna, O'Sullivan et~al.}}]{BarBuoHug12}
\bibinfo{author}{\bibfnamefont{E.}~\bibnamefont{Barausse}},
  \bibinfo{author}{\bibfnamefont{A.}~\bibnamefont{Buonanno}},
  \bibinfo{author}{\bibfnamefont{S.~A.} \bibnamefont{Hughes}},
  \bibinfo{author}{\bibfnamefont{G.}~\bibnamefont{Khanna}},
  \bibinfo{author}{\bibfnamefont{S.}~\bibnamefont{O'Sullivan}},
  \bibnamefont{et~al.}, \bibinfo{journal}{Phys.Rev.}
  \textbf{\bibinfo{volume}{D85}}, \bibinfo{pages}{024046}
  (\bibinfo{year}{2012}), \eprint{1110.3081}.

\bibitem[{\citenamefont{Taracchini et~al.}(2014)\citenamefont{Taracchini,
  Buonanno, Khanna, and Hughes}}]{TarBuoKha14}
\bibinfo{author}{\bibfnamefont{A.}~\bibnamefont{Taracchini}},
  \bibinfo{author}{\bibfnamefont{A.}~\bibnamefont{Buonanno}},
  \bibinfo{author}{\bibfnamefont{G.}~\bibnamefont{Khanna}}, \bibnamefont{and}
  \bibinfo{author}{\bibfnamefont{S.~A.} \bibnamefont{Hughes}}
  (\bibinfo{year}{2014}), \eprint{1404.1819}.

\bibitem[{\citenamefont{Harms et~al.}(2014)\citenamefont{Harms, Bernuzzi,
  Nagar, and Zenginoglu}}]{HarBerNag14}
\bibinfo{author}{\bibfnamefont{E.}~\bibnamefont{Harms}},
  \bibinfo{author}{\bibfnamefont{S.}~\bibnamefont{Bernuzzi}},
  \bibinfo{author}{\bibfnamefont{A.}~\bibnamefont{Nagar}}, \bibnamefont{and}
  \bibinfo{author}{\bibfnamefont{A.}~\bibnamefont{Zenginoglu}},
  \bibinfo{journal}{Class.Quant.Grav.} \textbf{\bibinfo{volume}{31}},
  \bibinfo{pages}{245004} (\bibinfo{year}{2014}), \eprint{1406.5983}.

\bibitem[{\citenamefont{Harms et~al.}(2015)\citenamefont{Harms,
  Lukes-Gerakopoulos, Bernuzzi, and Nagar}}]{HarLukBer15}
\bibinfo{author}{\bibfnamefont{E.}~\bibnamefont{Harms}},
  \bibinfo{author}{\bibfnamefont{G.}~\bibnamefont{Lukes-Gerakopoulos}},
  \bibinfo{author}{\bibfnamefont{S.}~\bibnamefont{Bernuzzi}}, \bibnamefont{and}
  \bibinfo{author}{\bibfnamefont{A.}~\bibnamefont{Nagar}}
  (\bibinfo{year}{2015}), \eprint{1510.05548}.

\bibitem[{\citenamefont{Penrose}(1963)}]{Pen63}
\bibinfo{author}{\bibfnamefont{R.}~\bibnamefont{Penrose}},
  \bibinfo{journal}{Phys. Rev. Lett.} \textbf{\bibinfo{volume}{10}},
  \bibinfo{pages}{66} (\bibinfo{year}{1963}).

\bibitem[{\citenamefont{Penrose}(1965)}]{Pen65a}
\bibinfo{author}{\bibfnamefont{R.}~\bibnamefont{Penrose}},
  \bibinfo{journal}{Proc. Roy. Soc. Lond.} \textbf{\bibinfo{volume}{A284}},
  \bibinfo{pages}{159} (\bibinfo{year}{1965}).

\bibitem[{\citenamefont{Frauendiener}(1998{\natexlab{a}})}]{Fra98}
\bibinfo{author}{\bibfnamefont{J.}~\bibnamefont{Frauendiener}},
  \bibinfo{journal}{Phys. Rev. D} \textbf{\bibinfo{volume}{58}},
  \bibinfo{pages}{064002} (\bibinfo{year}{1998}{\natexlab{a}}).

\bibitem[{\citenamefont{Frauendiener}(1998{\natexlab{b}})}]{Fra98a}
\bibinfo{author}{\bibfnamefont{J.}~\bibnamefont{Frauendiener}},
  \bibinfo{journal}{Phys. Rev. D} \textbf{\bibinfo{volume}{58}},
  \bibinfo{pages}{064003} (\bibinfo{year}{1998}{\natexlab{b}}).

\bibitem[{\citenamefont{H{\"u}bner}(1999)}]{Hub99}
\bibinfo{author}{\bibfnamefont{P.}~\bibnamefont{H{\"u}bner}},
  \bibinfo{journal}{Class. Quantum Grav.} \textbf{\bibinfo{volume}{16}},
  \bibinfo{pages}{2823} (\bibinfo{year}{1999}).

\bibitem[{\citenamefont{Valiente-Kroon}(2016)}]{Val16}
\bibinfo{author}{\bibfnamefont{J.-A.} \bibnamefont{Valiente-Kroon}},
  \emph{\bibinfo{title}{Conformal Methods in General Relativity}}
  (\bibinfo{publisher}{Cambridge University Press},
  \bibinfo{address}{Cambridge}, \bibinfo{year}{2016}).

\bibitem[{\citenamefont{Calabrese et~al.}(2006)\citenamefont{Calabrese,
  Gundlach, and Hilditch}}]{CalGunHil05}
\bibinfo{author}{\bibfnamefont{G.}~\bibnamefont{Calabrese}},
  \bibinfo{author}{\bibfnamefont{C.}~\bibnamefont{Gundlach}}, \bibnamefont{and}
  \bibinfo{author}{\bibfnamefont{D.}~\bibnamefont{Hilditch}},
  \bibinfo{journal}{Class.Quant.Grav.} \textbf{\bibinfo{volume}{23}},
  \bibinfo{pages}{4829} (\bibinfo{year}{2006}), \eprint{gr-qc/0512149}.

\bibitem[{\citenamefont{Zengino{\u g}lu and Husa}(2008)}]{ZenHus06}
\bibinfo{author}{\bibfnamefont{A.}~\bibnamefont{Zengino{\u g}lu}}
  \bibnamefont{and} \bibinfo{author}{\bibfnamefont{S.}~\bibnamefont{Husa}},
  \bibinfo{journal}{Class. Quantum Grav.} \textbf{\bibinfo{volume}{25}},
  \bibinfo{pages}{19} (\bibinfo{year}{2008}), \eprint{gr-qc/0612161}.

\bibitem[{\citenamefont{Rinne}(2010{\natexlab{b}})}]{Rin10}
\bibinfo{author}{\bibfnamefont{O.}~\bibnamefont{Rinne}},
  \bibinfo{journal}{Class.Quant.Grav.} \textbf{\bibinfo{volume}{27}},
  \bibinfo{pages}{035014} (\bibinfo{year}{2010}{\natexlab{b}}),
  \eprint{0910.0139}.

\bibitem[{\citenamefont{Moncrief}(2000)}]{Mon00}
\bibinfo{author}{\bibfnamefont{V.}~\bibnamefont{Moncrief}},
  \emph{\bibinfo{title}{Conformally regular {ADM} evolution equations}}
  (\bibinfo{year}{2000}), \bibinfo{note}{talk at Santa Barbara,
  \texttt{http://online.itp.ucsb.edu/online/numrel00/moncrief}}.

\bibitem[{\citenamefont{Bieri}(2010)}]{Bie09}
\bibinfo{author}{\bibfnamefont{L.}~\bibnamefont{Bieri}}, \bibinfo{journal}{J.
  Diff. Geom.} \textbf{\bibinfo{volume}{86}}, \bibinfo{pages}{17}
  (\bibinfo{year}{2010}), \eprint{0904.0620}.

\bibitem[{\citenamefont{Wald}(1984)}]{Wal84}
\bibinfo{author}{\bibfnamefont{R.~M.} \bibnamefont{Wald}},
  \emph{\bibinfo{title}{General relativity}} (\bibinfo{publisher}{The
  University of Chicago Press}, \bibinfo{address}{Chicago},
  \bibinfo{year}{1984}), ISBN \bibinfo{isbn}{0-226-87032-4 (hardcover),
  0-226-87033-2 (paperback)}.

\bibitem[{\citenamefont{Sogge}(1995)}]{Sog95}
\bibinfo{author}{\bibfnamefont{C.}~\bibnamefont{Sogge}},
  \emph{\bibinfo{title}{Lectures on nonlinear wave equations}}, no.
  \bibinfo{number}{Bd. 2} in \bibinfo{series}{Monographs in analysis}
  (\bibinfo{publisher}{International Press}, \bibinfo{year}{1995}).

\bibitem[{\citenamefont{{Lindblad} and {Rodnianski}}(2004)}]{LinRod04}
\bibinfo{author}{\bibfnamefont{H.}~\bibnamefont{{Lindblad}}} \bibnamefont{and}
  \bibinfo{author}{\bibfnamefont{I.}~\bibnamefont{{Rodnianski}}},
  \bibinfo{journal}{ArXiv Mathematics e-prints}  (\bibinfo{year}{2004}),
  \eprint{math/0411109}.

\bibitem[{\citenamefont{Trautman}(1958)}]{Tra58}
\bibinfo{author}{\bibfnamefont{A.}~\bibnamefont{Trautman}},
  \bibinfo{journal}{Bulletin of the Polish Academy of Sciences}
  \textbf{\bibinfo{volume}{VI}}, \bibinfo{pages}{403} (\bibinfo{year}{1958}).

\bibitem[{\citenamefont{Br{\"u}gmann}(2013)}]{Bru11}
\bibinfo{author}{\bibfnamefont{B.}~\bibnamefont{Br{\"u}gmann}},
  \bibinfo{journal}{J. Comput. Phys.} \textbf{\bibinfo{volume}{235}},
  \bibinfo{pages}{216} (\bibinfo{year}{2013}), \eprint{1104.3408}.

\bibitem[{\citenamefont{Bugner et~al.}(2015)\citenamefont{Bugner, Dietrich,
  Bernuzzi, Weyhausen, and Br{\"u}gmann}}]{BugDieBer15}
\bibinfo{author}{\bibfnamefont{M.}~\bibnamefont{Bugner}},
  \bibinfo{author}{\bibfnamefont{T.}~\bibnamefont{Dietrich}},
  \bibinfo{author}{\bibfnamefont{S.}~\bibnamefont{Bernuzzi}},
  \bibinfo{author}{\bibfnamefont{A.}~\bibnamefont{Weyhausen}},
  \bibnamefont{and}
  \bibinfo{author}{\bibfnamefont{B.}~\bibnamefont{Br{\"u}gmann}}
  (\bibinfo{year}{2015}), \eprint{1508.07147}.

\bibitem[{\citenamefont{Alcubierre et~al.}(2001)\citenamefont{Alcubierre,
  Brandt, Br{\"u}gmann, Holz, Seidel, Takahashi, and Thornburg}}]{AlcBraBru99}
\bibinfo{author}{\bibfnamefont{M.}~\bibnamefont{Alcubierre}},
  \bibinfo{author}{\bibfnamefont{S.~R.} \bibnamefont{Brandt}},
  \bibinfo{author}{\bibfnamefont{B.}~\bibnamefont{Br{\"u}gmann}},
  \bibinfo{author}{\bibfnamefont{D.}~\bibnamefont{Holz}},
  \bibinfo{author}{\bibfnamefont{E.}~\bibnamefont{Seidel}},
  \bibinfo{author}{\bibfnamefont{R.}~\bibnamefont{Takahashi}},
  \bibnamefont{and}
  \bibinfo{author}{\bibfnamefont{J.}~\bibnamefont{Thornburg}},
  \bibinfo{journal}{Int. J. Mod. Phys. D} \textbf{\bibinfo{volume}{10}},
  \bibinfo{pages}{273} (\bibinfo{year}{2001}), \eprint{gr-qc/9908012},
  \urlprefix\url{http://ejournals.worldscientific.com.sg/ijmpd/10/1003/S0218271801000834.html}.

\bibitem[{\citenamefont{Pretorius}(2005)}]{Pre04}
\bibinfo{author}{\bibfnamefont{F.}~\bibnamefont{Pretorius}},
  \bibinfo{journal}{Class. Quant. Grav.} \textbf{\bibinfo{volume}{22}},
  \bibinfo{pages}{425} (\bibinfo{year}{2005}), \eprint{gr-qc/0407110}.

\bibitem[{\citenamefont{Bizon and Zenginoglu}(2009)}]{BizZen08}
\bibinfo{author}{\bibfnamefont{P.}~\bibnamefont{Bizon}} \bibnamefont{and}
  \bibinfo{author}{\bibfnamefont{A.}~\bibnamefont{Zenginoglu}},
  \bibinfo{journal}{Nonlinearity} \textbf{\bibinfo{volume}{22}},
  \bibinfo{pages}{2473} (\bibinfo{year}{2009}), \eprint{0811.3966}.

\bibitem[{\citenamefont{Frauendiener}(1998{\natexlab{c}})}]{Fra98b}
\bibinfo{author}{\bibfnamefont{J.}~\bibnamefont{Frauendiener}}
  (\bibinfo{year}{1998}{\natexlab{c}}), \eprint{gr-qc/9806103}.

\bibitem[{\citenamefont{Buchman et~al.}(2009)\citenamefont{Buchman, Pfeiffer,
  and Bardeen}}]{BucPfeBar09}
\bibinfo{author}{\bibfnamefont{L.~T.} \bibnamefont{Buchman}},
  \bibinfo{author}{\bibfnamefont{H.~P.} \bibnamefont{Pfeiffer}},
  \bibnamefont{and} \bibinfo{author}{\bibfnamefont{J.~M.}
  \bibnamefont{Bardeen}}, \bibinfo{journal}{Phys.Rev.}
  \textbf{\bibinfo{volume}{D80}}, \bibinfo{pages}{084024}
  (\bibinfo{year}{2009}), \eprint{0907.3163}.

\bibitem[{\citenamefont{Schinkel
  et~al.}(2014{\natexlab{a}})\citenamefont{Schinkel, Macedo, and
  Ansorg}}]{SchPanAns13}
\bibinfo{author}{\bibfnamefont{D.}~\bibnamefont{Schinkel}},
  \bibinfo{author}{\bibfnamefont{R.~P.} \bibnamefont{Macedo}},
  \bibnamefont{and} \bibinfo{author}{\bibfnamefont{M.}~\bibnamefont{Ansorg}},
  \bibinfo{journal}{Class. Quant. Grav.} \textbf{\bibinfo{volume}{31}},
  \bibinfo{pages}{075017} (\bibinfo{year}{2014}{\natexlab{a}}),
  \eprint{1310.4699}.

\bibitem[{\citenamefont{Schinkel
  et~al.}(2014{\natexlab{b}})\citenamefont{Schinkel, Ansorg, and
  Panosso~Macedo}}]{SchPanAns13a}
\bibinfo{author}{\bibfnamefont{D.}~\bibnamefont{Schinkel}},
  \bibinfo{author}{\bibfnamefont{M.}~\bibnamefont{Ansorg}}, \bibnamefont{and}
  \bibinfo{author}{\bibfnamefont{R.}~\bibnamefont{Panosso~Macedo}},
  \bibinfo{journal}{Class. Quant. Grav.} \textbf{\bibinfo{volume}{31}},
  \bibinfo{pages}{165001} (\bibinfo{year}{2014}{\natexlab{b}}),
  \eprint{1301.6984}.

\end{thebibliography}

%%%%%%%%%%%%%%%%%%%%%%%%%%%%%%%%%%%%%%%%%%%%%%%%%%%%%%%%%%%%%%%%%%%%%%%%%%%%%%%%%%%%%%%

\end{document}